\documentclass[a4paper,12pt]{article}
\usepackage[]{inputenc,graphicx,amsmath,bbm,latexsym,textcomp,amssymb, mathrsfs,appendix}
\usepackage{color, multicol,multirow}
\usepackage{subfigure,float, afterpage}  
\usepackage{hyperref}
\usepackage[margin=1.15in,top=1.1in]{geometry}

\newcommand{\beq}{\begin{equation}}
\newcommand{\eeq}{\end{equation}}
\newcommand{\bea}{\begin{eqnarray}}
\newcommand{\eea}{\end{eqnarray}}
\newcommand{\Z}{\mathbb{Z}}
\def\sv{\left\langle\sigma v\right\rangle}

\def\nn{\nonumber}
\def\Eqn#1{Eq.\ (\ref{#1})}

\let\jnfont=\rm
\def\NPB#1,{{\jnfont Nucl.\ Phys.\ B }{\bf #1},}
\def\PLB#1,{{\jnfont Phys.\ Lett.\ B }{\bf #1},}
\def\EPJC#1,{{\jnfont Eur.\ Phys.\ Jour.\ C }{\bf #1},}
\def\PRD#1,{{\jnfont Phys.\ Rev.\ D }{\bf #1},}
\def\PRL#1,{{\jnfont Phys.\ Rev.\ Lett.\ }{\bf #1},}
\def\MPLA#1,{{\jnfont Mod.\ Phys.\ Lett.\ A }{\bf #1},}
\def\JPG#1,{{\jnfont J.\ Phys.\ G }{\bf #1},}
\def\CTP#1,{{\jnfont Commun.\ Theor.\ Phys.\ }{\bf #1},}
\def\JHEP#1,{{\jnfont JHEP \ }{\bf #1},}
\def\NPPS#1,{{\jnfont Nucl.\ Phys.\ Proc.\ Suppl.\ }{\bf #1},}
\def\CPC#1,{{\jnfont Computl.\ Phys.\ Commun.\ }{\bf #1},}
\def\CPL#1,{{\jnfont Chin.\ Phys.\ Lett. }{\bf #1},}
\def\AJS#1,{{\jnfont Astrophys.\ J.\ Suppl. }{\bf #1},}
\def\PR#1,{{\jnfont Phys.\ Rept. }{\bf #1},}
\def\AP#1,{{\jnfont Astropart.\ Phys. }{\bf #1},}
\def\EPL#1,{{\jnfont Europhys.\ Lett. }{\bf #1},}
\def\FP#1,{{\jnfont Fortsch.\ Phys. }{\bf #1},}
\def\JCAP#1,{{\jnfont JCAP \ }{\bf #1},}

\def\etc{ {\em etc.\ }}
\def\ie{ {\em i.e.,\ }}

\def\etal{\!{\em et al.\ }}

\begin{document}
 \begin{center}
{\Large{\bf  \color{black}
A Possible Explanation of Low Energy $\gamma$-ray Excess from Galactic Centre and Fermi Bubble
by a Dark Matter Model with Two Real Scalars}}\\
\vspace {1.0cm}

{\bf Kamakshya Prasad Modak}$^\ast$\footnote{kamakshya.modak@saha.ac.in},
{\bf Debasish Majumdar}$^\ast$\footnote{debasish.majumdar@saha.ac.in},
{\bf Subhendu Rakshit}$^\dagger$\footnote{rakshit@iiti.ac.in} \\
	
\vskip 0.15in
{\it
$^\ast${Astroparticle Physics and Cosmology Division, Saha Institute of Nuclear Physics, \\
1/AF Bidhannagar, Kolkata 700064, India.} \\

$^\dagger${Discipline of Physics, Indian Institute of Technology Indore,   \\
IET-DAVV Campus, Indore 452017, India.}\\
}
\vspace{0.40cm}
\end{center}
\begin{abstract}
We promote the idea of multi-component Dark Matter (DM) to explain results from both direct and indirect detection experiments. In these models as contribution of each DM candidate to relic abundance is summed up to meet WMAP/Planck measurements of $\Omega_{\rm DM}$, these candidates have larger annihilation cross-sections compared to the single-component DM models. This results in larger $\gamma$-ray flux in indirect detection experiments of DM. We illustrate this fact by introducing an extra scalar to the popular single real scalar DM model. We also present detailed calculations for the vacuum stability bounds, perturbative unitarity and triviality constraints on this model.
As direct detection experimental results still show some conflict, we kept our options open, discussing different scenarios with different DM mass zones. In the framework of our model we make an interesting observation:  The existing direct detection experiments like CDMS~II, CoGeNT, CRESST~II, XENON~100 or LUX together with the observation of excess low energy $\gamma$-ray from Galactic Centre and Fermi Bubble by FGST already have the capability to distinguish between different DM halo profiles.
\end{abstract}

\newpage

\section{Introduction}

The overwhelming cosmological and astrophysical evidences have now
 established the existence of an unknown non-luminous matter present
 in the universe in enormous amount, namely the dark matter~(DM).
Experiments like Wilkinson Microwave Anisotropy Probe (WMAP)~\cite{wmap},
 BOSS~\cite{boss} or more recently Planck~\cite{planck} measure the baryonic fraction precisely
 to consolidate the fact that this non-baryonic DM
 constitutes around $\sim$ 26.5\% of the content of the universe.
  The particle nature of DM candidate is still unknown.
The relic density of dark matter deduced from cosmological observations
 mentioned above tends to
 suggest that most of the DM could be made of weakly interacting massive
 particles (WIMPs)~\cite{jungman,griest,bertone,murayama}
 and they are non-relativistic or cold in nature. This calls for an extension of the standard model~(SM) of particle physics.
 Many such extensions have been suggested in the literature in the framework of supersymmetry, extra dimensions, axion {\em etc.} 
Models such as Kaluza Klein \cite{kaluza}, inert triplet
\cite{triplet} or supersymmetry breaking models like mAMSB \cite{Modak:2012wk} predict very massive
DM whereas models like SMSSM \cite{smssm}, axion \cite{axion} predict DM of lower mass.
Phenomenology of simpler extensions of SM like fermionic DM model \cite{sing_ferm} or inert doublet model \cite{idm}
has been elaborately studied.
Amongst all such options, extending the scalar sector is particularly interesting because of its simplicity.

The minimal extension with a single gauge singlet real scalar stabilised by a $\Z_2$ symmetry in the context of dark matter was proposed by Silveira and Zee in Ref.~\cite{scalar_singlet1} and then it was extensively
studied in the literature~\cite{veltman}~-~\cite{Holz:2001cb}. In Ref.~\cite{scalar_singlet2} the singlet scalar DM model has been discussed with a global U(1) symmetry.


Amongst the non-minimal extensions, a DM model where SM is extended by a complex singlet scalar has been considered in Refs.~\cite{Barger:2008jx, Barger:2010yn, Gonderinger:2012rd}.
A DM model with two real scalars has been discussed in Refs.~\cite{2singlets_a, 2singlets_b}, 
where one scalar is protected by a $\Z_2$ symmetry, but the $\Z_2$ symmetry protecting the other 
one spontaneously breaks. In all these non-minimally extended models there is, however, only one DM candidate.  

A pertinent question to ask at this stage is that, if our visible sector is enriched with so many particles, why the DM should be composed of only one component? We therefore intend to discuss in this paper a model with two DM candidates. In some earlier works~\cite{brooks} the idea of multicomponent dark matter has been discussed in details. We extend the SM with two gauge singlet real scalars protected by unbroken $\Z_2$ symmetries. As mentioned in the abstract, the advantage of such a multi-component DM model is that the DM annihilation can be enhanced so that one can expect spectacular signals in the indirect detection experiments. Because of this simple fact, the thermal averaged annihilation cross-sections in this model can enjoy enhancement upto two orders of magnitude compared to the models with one real scalar. As a result, in the present model with two real scalars we can make an attempt to explain both direct and indirect detection DM experimental observations, which was {\em not} possible with the model with a single real scalar.

Direct detection DM experiments can detect DM by measuring
 the recoil energy of a target nucleon of detecting material in case a DM  particle 
 happens to scatter off such nucleons. Experiments like CDMS~\cite{cdms_science, cdms_2013},
 DAMA~\cite{dama}, CoGeNT~\cite{cogent} or
 CRESST~\cite{cresst} present their results indicating allowed zones in the scattering cross-section -- DM mass plane. 
 These experiments seem to prefer low dark matter masses $\sim$ 10 GeV. Some earlier works on $\sim$ 10 GeV DM mass have been done~\cite{Kyae:2013qna,Belanger:2013tla}. XENON~100~\cite{xenon2011, xenon2012}, however, did not observe any potential DM event contradicting claims of the earlier experiments and has presented an upper bound on DM-nucleon scattering cross-section for various DM masses. Recent findings by LUX~\cite{LUX2013} have fortified claims by XENON~100 collaboration.

 The indirect detection of DM involves detecting the
 particles and their subsequent decay products, produced due to
 DM annihilations. Huge concentration of DM are expected at the centre of gravitating
 bodies such as the Sun or the galactic centre~(GC) as they can capture DM particles over time.

 The region in and around the GC are looked for detecting
 the dark matter annihilation products such as $\gamma$, $\nu$ {\em etc.}
 Fermi Gamma Space Telescope (FGST), operated from mid of '08, have been looking for
 the gamma ray from the GC~\cite{fermilat}.
 The low energy gamma ray from GC shows some bumpy structures around a few GeV which cannot be
 properly explained by known astrophysics. A plausible explanation of such a non-power law spectrum is provided by DM annihilations~\cite{hooperplb, boyarsky}.

 The emission of gamma rays from Fermi Bubble may also be partially caused by DM 
 annihilations.
 The Fermi Bubble is a lobular structure of gamma ray emission zone
 both upward and downward from the galactic plane
 has been discovered recently by Fermi's Large Area Telescope~\cite{Su:2010qj}.
 The lobes spread up to a few kpc above and below
 the galactic plane and emit gamma ray with energy extending
 from a few GeV to about a hundred GeV. The gamma emission is supposed to
 be produced from the inverse Compton scattering~(ICS) of cosmic ray electrons.
 But more involved study of this emission reveals
 that while the spectra from
 the high galactic latitude region can be explained by ICS taking into consideration cosmic electron
 distribution, it cannot satisfactorily explain the emission from the lower latitudes.
 The $\gamma$-ray flux from possible DM annihilation in the galactic halo
 may help explain this apparent anomaly~\cite{Huang:2013pda, Huang:2013apa, hooperfermibubble}.

As mentioned earlier, in this work, the proposed model with two gauge singlet real scalars protected by a 
$\Z_2\times\Z_2'$ symmetry is confronted with the experimental findings of both direct and indirect DM experiments. 
As direct DM experimental results are contradictory, we keep an open mind in considering them. We find constraints on the parameter space from these results and then try to constrain them further imposing the requirement of producing the relic density of the DM candidates consistent with Planck observations. We then choose benchmark points from this constrained parameter space to explain results from indirect detection experiments from DM annihilation taking into consideration different DM halo profiles.

The paper is organised as follows. In Sec.~\ref{S:theory} we discuss the theoretical framework of our proposed model. 
 The theoretical constraints from vacuum stability, perturbative unitarity, triviality and experimental constraints from the invisible branching ratio of the Higgs boson have been discussed in Sec.~\ref{S:constraints}. 
  The next section contains the relevant relic density calculations. The model is confronted with direct detection experiments and Planck observations in Sec.~\ref{S:Direct}. Explanation of the observed excess of $\gamma$-ray from GC and Fermi bubble by our model is studied in Sec.~\ref{S:Indirect}. We conclude in Sec.~\ref{S:conc}.

\section{Theoretical Framework}
\label{S:theory}

 We propose a model where two real scalar singlets ($S$ and $S'$)
are added to the standard model.

The general form of the renormalisable scalar potential is then given by,
\begin{eqnarray}
 V(H, S, S') &=& \frac{m^2}{2} H^{\dag} H  +\frac{\lambda}{4} (H^{\dag} H)^2  \nonumber\\
&& +\frac{\delta_1}{2} H^{\dag} HS + \frac{\delta_2}{2} H^{\dag} HS^2 + \frac{\delta_1 m^2}{2\lambda} S
   + \frac{k_2}{2} S^2 + \frac{k_3}{3} S^3 + \frac{k_4}{4} S^4 \nonumber\\
&& + \frac{\delta'_1}{2} H^{\dag} HS' + \frac{\delta'_2}{2} H^{\dag} H{S'}^2 + \frac{\delta'_1 m^2}{2\lambda} S' + \frac{k'_2}{2} {S'}^2 + \frac{k'_3}{3} {S'}^3 + \frac{k'_4}{4} {S'}^4\nonumber \\
&&  + \frac{\delta''_2}{2}H^{\dag} H S'S + \frac{k''_2}{2} SS' + \frac{1}{3}(k^a_3 SSS' + k^b_3 SS'S')\nonumber \\
&& + \frac{1}{4}(k^a_4 SSS'S' + k^b_4 SSSS' + k^c_4 SS'S'S') \,\,\, ,
\label{gen_potential}
\end{eqnarray}
where $H$ is the ordinary (SM) Higgs doublet. In the above $\delta$'s denote the
couplings between the singlets and the Higgs and
$k$'s are the couplings between these singlets themselves.

The stability of DM particles is achieved by imposing a discreet symmetry $\Z_2$ onto the Lagrangian. Depending on whether $S$ and $S'$ are odd under the same $\Z_2$ or not, we discuss two scenarios for completeness.

\subsection[]{Lagrangian Invariant under $\Z_2\times \Z_2$}\label{ss:z2}

If only $S$ and $S'$ are odd under the same $\Z_2$, and the rest of the particles are even,
\bea
\left ( \begin{array}{c} S \\ S'
\end{array} \right ) &\xrightarrow{\Z_2\times \Z_2}&
\left ( \begin{array}{c} -S \\ -S'
\end{array} \right )\,\, ,
\label{basis}
\eea
some parameters of the potential vanish:
\begin{equation}
 \delta_1 = k_3 = \delta'_1 = k'_3 = k^a_3 = k^b_3 = 0 \,\,\, ,
\end{equation}
so that the scalar potential~(\ref{gen_potential}) reduces to the following,
\begin{eqnarray}
 V(H, S, S') &=& \frac{m^2}{2} H^{\dag} H +  \frac{\lambda}{4} (H^{\dag} H)^2 \nonumber\\
 && + \frac{\delta_2}{2} H^{\dag} HS^2  + \frac{k_2}{2} S^2  + \frac{k_4}{4} S^4 \nonumber\\
 && + \frac{\delta'_2}{2} H^{\dag} H{S'}^2  + \frac{k'_2}{2} {S'}^2 + \frac{k'_4}{4} {S'}^4 \nonumber\\
&&  + \frac{\delta''_2}{2}H^{\dag} H S'S + \frac{k''_2}{2} SS' \nonumber\\
&& + \frac{1}{4}(k^a_4 SSS'S' + k^b_4 SSSS' + k^c_4 SS'S'S') \,\,\, .
\label{z2z2_pot}
\end{eqnarray}

After the spontaneous symmetry breaking the mass matrix for $S$ and $S'$ is given by
\begin{equation}
M_{SS'} = \left ( \begin{array}{cc}
                   k_2 + \delta_2 v^2/2 & \delta''_2 v^2/4 + k''_2/2  \\
                   \delta''_2 v^2/4 + k''_2/2&  k'_2 + \delta'_2 v^2/2
                   \end{array} \right )
	\equiv \left ( \begin{array}{cc}
                   M_{11} & M_{12}  \\
                   M_{12} & M_{22}
                   \end{array} \right )\,\,.\nonumber\label{mssp}\\
\end{equation}
$\frac{v}{\sqrt{2}}$ denotes the vacuum expectation value of
the Higgs.
After diagonalisation the masses of the physical eigenstates $S_1$ and $S_2$ are given by
\bea
M^2_{S_1} &=& \cos^2\theta M_{11} + \sin^2\theta M_{22} + 2\cos\theta \sin\theta M_{12}\,\,\,  \label{Ms1} \\
M^2_{S_2} &=& \cos^2\theta M_{22} + \sin^2\theta M_{11} - 2\cos\theta \sin\theta M_{12} \label{Ms2}\,\,\, ,
\eea
where
\begin{equation}
\tan2\theta = \frac{2 M_{12}}{M_{11}-M_{22}}\,\, .
\end{equation}

\subsection[]{Lagrangian Invariant under $\Z_2\times\Z_2'$}\label{ss:z2z2}

If $S$ and $S'$ are stabilised by different discrete symmetries,
\begin{equation}
 S \xrightarrow{\Z_2} -S\,\, ~{\rm and}~ \,\, \\
 S' \xrightarrow{\Z_2'} -S' \,\,\, ,
 \label{bases}
\end{equation}
$\delta''_2=k''_2=k^b_4=k^c_4=0$, so that the scalar potential~(\ref{z2z2_pot}) further reduces to
\begin{eqnarray}
 V(H, S, S') &=& \frac{m^2}{2} H^{\dag} H +  \frac{\lambda}{4} (H^{\dag} H)^2 \nonumber \\
&& + \frac{\delta_2}{2} H^{\dag} HS^2   + \frac{k_2}{2} S^2  + \frac{k_4}{4} S^4  \nonumber \\
&&+ \frac{\delta'_2}{2} H^{\dag} H{S'}^2  + \frac{k'_2}{2} {S'}^2 + \frac{k'_4}{4} {S'}^4\nonumber \\
&&+ \frac{1}{4}k^a_4 SSS'S' \,\,\,  .
\label{vhssp}
\end{eqnarray}

After spontaneous symmetry breaking
the respective masses of $S$ and $S'$ are given by
\begin{eqnarray}
 M^2_S &=& k_2 + \frac{\delta_2 v^2}{2} \\
 M^2_{S'} &=& k_2' + \frac{\delta_2' v^2}{2} \,\,\, .
\end{eqnarray}
The four beyond SM parameters determining
the masses of the scalars are $k_2$, $k_2'$, $\delta_2$ and $\delta_2'$.

In both $\Z_2\times\Z_2$ and $\Z_2\times\Z_2'$ cases if $SS\leftrightarrow S'S'$ scattering processes can be avoided, the model can give rise to a two-component DM scenario. However, as the later case has fewer number of beyond SM parameters, in the following we will restrict ourselves only to the $\Z_2\times\Z_2'$ invariant Lagrangian.

\section{Constraints on Model Parameters}\label{S:constraints}
The extra scalars present in the model modifies the scalar potential. Hence it is prudent to revisit constraints emanating from vacuum stability conditions and triviality of the Higgs potential. Perturbative unitarity can also get affected by these scalars. Limits on the invisible decay width of Higgs from LHC severely restricts such models. In the following we elaborate on these constraints.

\subsection{Vacuum Stability Conditions}\label{ss:vac}

Calculating the exact vacuum stability conditions for any model is generally difficult.
However, for many dark matter models
the quartic part of the scalar potential can be expressed as quadratic form
($\lambda_{ab} S_{a}^{2} S_{b}^{2}$) with the squares of real fields as single entity.
Lagrangian respecting $\Z_{2}$ symmetry which ensures the stability of scalar dark matter
have the terms which can be expressed like that.
The scalar potential of our proposed model can also be expressed in a similar form
as above because of preservation of $\Z_{2}$ symmetry.
The criteria for copositivity allow one to derive
properly the analytic vacuum stability conditions for the such matrix $\lambda_{ab}$
from which sufficient conditions for vacuum stability can be obtained.\footnote{
Derivation of the necessary and sufficient conditions for the model
is much simpler with copositivity than with the other used formalisms.}

The necessary conditions for a symmetric matrix $A$ of order 3 to be
copositive are given by~\cite{copositive,Hadeler198379,Chang1994113,Chakrabortty:2013mha},
\begin{equation}
\begin{split}
  a_{11} &\geqslant 0, a_{22} \geqslant 0, a_{33} \geqslant 0,
  \\
  \bar{a}_{12} &= a_{12} + \sqrt{ a_{11} a_{22} } \geqslant 0,
  \\
  \bar{a}_{13} &= a_{13} + \sqrt{ a_{11} a_{33} } \geqslant 0,
  \\
  \bar{a}_{23} &= a_{23} + \sqrt{ a_{22} a_{33} } \geqslant 0,
\end{split}
\label{eq:A:3:copos:1:2:crit}
\end{equation}
and
\begin{equation}
  \sqrt{a_{11} a_{22} a_{33}} + a_{12} \sqrt{a_{33}} + a_{13} \sqrt{a_{22}} + a_{23} \sqrt{a_{11}}
  + \sqrt{2 \bar{a}_{12} \bar{a}_{13} \bar{a}_{23}} \geqslant 0.
\label{eq:A:3:copos:3:crit}
\end{equation}
The last criterion given in \Eqn{eq:A:3:copos:3:crit} is a simplified form of the two conditions
(Eqs. (\ref{eq:A:3:copos:3:crit:new}) and (\ref{eq:A:3:copos:3:crit:det})) below
\begin{align}
  \sqrt{a_{11} a_{22} a_{33}}
  + a_{12} \sqrt{a_{33}} + a_{13} \sqrt{a_{22}} + a_{23} \sqrt{a_{11}} &\geqslant 0,
  \label{eq:A:3:copos:3:crit:new}
  \\
  \det A = a_{11} a_{22} a_{33}
  - (a_{12}^2 a_{33} + a_{13}^2 a_{22} + a_{11} a_{23}^2)
  + 2 a_{12} a_{13} a_{23} &\geqslant 0,
  \label{eq:A:3:copos:3:crit:det}
\end{align}
where one \emph{or} the other inequality has to be satisfied~\cite{Hadeler198379}\footnote{The criterion, $\det A \geqslant 0$ is a part of well known
Sylvester's criterion for positive semidefiniteness.}.
The conditions Eq.~(\ref{eq:A:3:copos:1:2:crit}) impose that the
three $2 \times 2$ principal submatrices of $A$ are copositive.

The matrix of quartic couplings $\Lambda$ in the $(h^{2}, S^{2}, S'^{2})$ basis
for the potential Eq.~(\ref{vhssp}) is given by

\begin{equation}
  4 \, \Lambda
  =
  \begin{pmatrix}
    \lambda & \delta_{2} & \delta'_{2}
    \\
    \delta_{2} & k_4
    & \frac{k_4^a}{2}
    \\
    \delta_2'  & \frac{k_4^a}{2}
    & k_{4}'
  \end{pmatrix}.
    \label{eq:lambdas:SM:singlet}
\end{equation}
Copositivity criteria of Eqs.~(\ref{eq:A:3:copos:1:2:crit}) and
(\ref{eq:A:3:copos:3:crit}) yield the necessary and sufficient vacuum stability conditions,
\begin{equation}
\begin{split}
  \lambda \geqslant & ~0, k_4 \geqslant 0, k_4' \geqslant 0,
  \\
  \delta_2 &+ \sqrt{ \lambda k_4 } \geqslant 0,
  \\
  \delta_2' &+ \sqrt{ \lambda k_4' } \geqslant 0,
  \\
  k_4^a &+ \sqrt{ k_4 k_4' } \geqslant 0,
\end{split}
\label{vac}
\end{equation}
and
\beq
 \sqrt{\lambda k_4 k_4'} + \delta_2\sqrt{k_4'} + \delta_2' \sqrt{k_4} + 2 k_4'\sqrt{\lambda}
+ \sqrt{(\delta_2 + \sqrt{ \lambda k_4 })(\delta_2' + \sqrt{ \lambda k_4' })(k_4^a + \sqrt{ k_4 k_4' })}
\geqslant 0.
\label{vsz2z2}
\eeq

The conditions of Eqs.~(\ref{vac}) and~(\ref{vsz2z2}) simply determine the vacuum stability bounds on
our model. We restrict the parameter space by these conditions for later calculation.

\subsection{Perturbative Unitarity Bounds}\label{ss:uni}

The potential of the $\Z_2\times\Z_2'$ model is bounded
from below if \Eqn{vac} and \Eqn{vsz2z2}
are simultaneously satisfied. Then,
$\lambda,k_4>0$ and $\delta_2>0$ or\begin{equation}
\delta_2^{2}<k_4\lambda\qquad\textrm{for }\delta_2<0.\label{eq:2}\end{equation}
The Higgs mechanism generates
a mass of $M_{H}^{2}=\frac{1}{2}\lambda v^{2}$ for the Higgs and also contributes
to the mass of the $S$ particle
\begin{eqnarray}
 M^2_S &=& k_2 + \frac{\delta_2 v^2}{2} \\
 M^2_{S'} &=& k_2' + \frac{\delta_2' v^2}{2} \,\,\, ,
\end{eqnarray}
For $\left\langle H\right\rangle =(0,v/\sqrt{2})$
and $\left\langle S\right\rangle =0$, $\left\langle S'\right\rangle =0$ be a local minimum
we should have $M_{S}^{2}>0$ and $M_{S'}^{2}>0$ are required. This is
also a global minimum as long as $k_2>-\frac{1}{2}v^{2}\sqrt{k_4\lambda}$
and $k_2'>-\frac{1}{2}v^{2}\sqrt{k_4'\lambda}$
\cite{Burgess:2000yq}. The potential of the scalar sector after electroweak symmetry breaking
in the unitary gauge can be written as,
\begin{eqnarray}
V_{SS'H} &=& \frac{\lambda}{4}H^{4} + \frac{m^2}{4}H^{2} + \frac{m^2 v}{2}H
     +v\lambda H^{3} + \frac{3 v^2\lambda}{2}H^{2} + v^3\lambda H \nonumber \\
    &&+\frac{\delta_2}{2}H^{2} S^{2} +  v\delta_2 H S^{2} + \frac{v^2\delta_2}{2} S^{2}
    + \frac{k_2}{2} S^{2} + \frac{k_4}{4} S^{4}
    + \frac{\delta_2'}{2}H^{2} S'^{2} \nonumber \\
    &&+ v\delta_2' H S'^{2} + \frac{v^2\delta_2'}{2} S'^{2}
    + \frac{k'_2}{2} S^{2} + \frac{k'_4}{4} S'^{4} + \frac{k_4^a}{4} S^{2} S'^{2}\,\,\,.
\label{eq:VhS}
\end{eqnarray}
After that tree-level perturbative unitarity \cite{lee} to scalar
elastic scattering processes have been applied in this model (\Eqn{eq:VhS}). The zeroth
partial wave amplitude,
\begin{equation}
a_{0}=\frac{1}{32\pi}\sqrt{\frac{4p_{f}^{\rm CM}p_{i}^{\rm CM}}{s}}
\int_{-1}^{+1}T_{2\rightarrow2}d\cos\!\theta
\label{eq:partw}\end{equation}
must satisfy the condition $\left|{\rm Re}(a_{0})\right|\leq\frac{1}{2}$ \cite{Cynolter:2004cq}.
In the above, $s$ is the
centre of mass (CM) energy, $p_{i,f}^{\rm CM}$ are the initial and final
momenta in CM system and $T_{2\rightarrow2}$ denotes the matrix element
for $2\rightarrow 2$ processes with $\theta$ being the incident angle between
two incoming particles.

The possible two particle states are $HH,HS,SS,S'S',S'S,HS'$ and the
scattering processes include many possible diagrams such as
$HH\rightarrow HH$, $SS\rightarrow SS$, $HS\rightarrow HS$, $HH\rightarrow SS$,
$SS\rightarrow HH$, $S'S'\rightarrow HH$, $S'S'\rightarrow SS$,
$HS'\rightarrow HS'$, $S'S\rightarrow S'S$, $HH\rightarrow S'S'$,
$SS\rightarrow S'S'$, $S'S'\rightarrow S'S'$.
The matrix elements ($T_{2\rightarrow2}$) for the above $2\rightarrow 2$ processes are
calculated from the tree level Feynman diagrams for corresponding
scattering and given by,


\begin{eqnarray}
T_{HH\rightarrow HH}&=&3\frac{M_{H}^{2}}{v^{2}}\left(1+3M_{H}^{2}\left(\frac{1}{s-M_{H}^{2}}
+\frac{1}{t-M_{H}^{2}}+\frac{1}{u-M_{H}^{2}}\right)\right)\,\,\, ,\label{eq:thh}\\
T_{SS\rightarrow SS}&=&6k_4+\delta_2\left(\frac{\delta_2 v^{2}}{s-M_{H}^{2}}
+\frac{\delta_2 v^{2}}{t-M_{H}^{2}}+\frac{\delta_2 v^{2}}{u-M_{H}^{2}}\right)\,\,\, ,\label{eq:tss}\\
T_{SS\rightarrow HH}&=&\delta_2\left(1+3M_{H}^{2}\frac{1}{s-M_{H}^{2}}
+\delta_2 v^{2}\left(\frac{1}{t-M_{S}^{2}}+\frac{1}{u-M_{S}^{2}}\right)\right)\,\,\, ,\label{eq:tsshh}\\
T_{HS\rightarrow HS}&=&\delta_2\left(1+v^{2}\left(\frac{\delta_2}{s-M_{S}^{2}}
+\frac{3\lambda}{t-M_{H}^{2}}+\frac{\delta_2}{u-M_{S}^{2}}\right)\right)\,\,\, ,\label{eq:tsh}\\
T_{S' S'\rightarrow HH}&=&\delta_2'\left(1+3M_{H}^{2}\frac{1}{s-M_{H}^{2}}
+\delta_2' v^{2}\left(\frac{1}{t-M_{S}^{2}}+\frac{1}{u-M_{S}^{2}}\right)\right)\,\,\, ,\\
T_{S' S'\rightarrow SS}&=&k_4^a+ \left(\frac{\delta_2\delta_2' v^2}{s-M_H^2}\right)\,\,\, ,\\
T_{HS'\rightarrow HS'}&=&\delta_2'\left(1+v^{2}\left(\frac{\delta_2'}{s-M_{S'}^{2}}
+\frac{3\lambda}{t-M_{H}^{2}}+\frac{\delta_2'}{u-M_{S'}^{2}}\right)\right)\,\,\, ,\\
T_{S' S\rightarrow S' S}&=&k_4^a + \left(\frac{\delta_2\delta_2' v^2}{t-M_H^2}\right)\,\,\, ,\\
T_{S' S'\rightarrow S' S'}&=&6k_4'+\delta_2'\left(\frac{\delta_2' v^{2}}{s-M_{H}^{2}}
+\frac{\delta_2' v^{2}}{t-M_{H}^{2}}+\frac{\delta_2 v^{2}}{u-M_{H}^{2}}\right)\,\,\, .
\end{eqnarray}

Now using \Eqn{eq:partw}, we have calculated the partial wave amplitude for each
of the scattering processes and the coupled amplitude can be written as a matrix
form,
\begin{equation}
\begin{split}
\mathbb{M} = \left(\begin{array}{cccccc}
a_{0}^{HH\rightarrow HH} & a_{0}^{HH\rightarrow SS} & a_{0}^{HH\rightarrow SH}
& a_{0}^{HH\rightarrow S' S'} & a_{0}^{HH\rightarrow S' S} & a_{0}^{HH\rightarrow S' H} \\
a_{0}^{SS\rightarrow HH} & a_{0}^{SS\rightarrow SS} & a_{0}^{SS\rightarrow SH}
& a_{0}^{SS\rightarrow S' S'} & a_{0}^{SS\rightarrow S' S} & a_{0}^{SS\rightarrow S' H}\\
a_{0}^{SH\rightarrow HH} & a_{0}^{SH\rightarrow SS} & a_{0}^{SH\rightarrow SH}
& a_{0}^{SH\rightarrow S' S'} & a_{0}^{SH\rightarrow S' S} & a_{0}^{SH\rightarrow S' H}\\
a_{0}^{S' S'\rightarrow HH} & a_{0}^{S' S'\rightarrow SS} & a_{0}^{S' S'\rightarrow SH}
& a_{0}^{S' S'\rightarrow S' S'} & a_{0}^{S' S'\rightarrow S' S} & a_{0}^{S' S'\rightarrow S' H}\\
a_{0}^{S' S\rightarrow HH} & a_{0}^{S' S\rightarrow SS} & a_{0}^{S' S\rightarrow SH}
& a_{0}^{S' S\rightarrow S' S'} & a_{0}^{S' S\rightarrow S' S} & a_{0}^{S' S\rightarrow S' H}\\
a_{0}^{S' H\rightarrow HH} & a_{0}^{S' H\rightarrow SS} & a_{0}^{S' H\rightarrow SH}
& a_{0}^{S' H\rightarrow S' S'} & a_{0}^{S' H\rightarrow S' S} & a_{0}^{S' H\rightarrow S' H}
\end{array}\right) & \\
\begin{array}{c}
\\\longrightarrow\\
{\scriptstyle s\gg M_{H}^{2},M_{S}^{2},M_{S'}^{2}}\end{array}\frac{1}{16\pi}\left(\begin{array}{cccccc}
3\lambda & \delta_2 & 0 & \delta_2' & 0 & 0\\
\delta_2 & 6 k_4 & 0 & k_4^a & 0 & 0\\
0 & 0 & 2\delta_2 & 0 & 0&0\\
\delta_2' & k_4^a & 0 & 6 k_4' & 0 & 0 \\
0 & 0 & 0 & 0& k_4^a & 0 \\
0 & 0 & 0 & 0 & 0 & 2\delta_2'
\end{array}\right).
&
\end{split}
\end{equation}

Requiring $\left|{\rm Re}(a_{0})\right|\leq\frac{1}{2}$ for each individual
process above we obtain
\begin{eqnarray}
{\rm for}\,\,\, HH\rightarrow HH &  & M_{H}\leq\sqrt{\frac{8\pi}{3}}v\label{eq:Hunit1}\,\,\, ,\\
{\rm for}\,\,\, HS\rightarrow HS\textrm{ and }HH\rightarrow SS &  & |\delta_2|\leq8\pi\label{eq:k8pi}\,\,\, ,\\
{\rm for}\,\,\, SS\rightarrow SS &  & k_4\leq\frac{8}{6}\pi\label{eq:lambda8pi}\,\,\, , \\
{\rm for}\,\,\, S' S'\rightarrow S' S' &  & k_4'\leq\frac{8}{6}\pi\label{eq:lambdap8pi}\,\,\, ,\\
{\rm for}\,\,\, S' S'\rightarrow SS\textrm{ and }S' S\rightarrow S' S &  & |k_4^a|\leq8\pi\label{eq:k4a8pi}\,\,\, ,\\
{\rm for}\,\,\, HS'\rightarrow HS'\textrm{ and }HH\rightarrow S' S' &  & |\delta_2'|\leq8\pi\label{eq:kp8pi}\,\,\, .
\end{eqnarray}

\subsection{Triviality Bound}\label{ss:tri}

The requirement for `triviality bound' on any model is guaranteed by one of the conditions
that the renormalization group evolution should not
push the quartic coupling constant of such models (say, $\lambda$)
to infinite value up to the ultraviolet cut-off scale $\Lambda$ of the model. 
This requires that Landau pole of the Higgs boson should be in higher scale than $\Lambda$.

Therefore to check the triviality in our model, namely the two scalar singlet model with $\Z_2\times\Z_2'$
symmetry, we have to solve the renormalization group (RG) evolution equations for all the running
parameters of this model. We have chosen only one-loop contribution in determining the beta functions
for our model.
The RG equations for
the couplings in the model, namely,
$\lambda$, $k_2$, $k_2'$, $k_4$, $k_4'$, $k_4^a$, $\delta_2$,
$\delta_2'$ are thus obtained at one-loop level as
\hspace{-5mm}
\begin{eqnarray}
16 \pi^2 \frac{d\delta_2}{dt}
&=& 4\delta_{2}^2 +\delta_2'k_4^a+
\delta_2 (2 \gamma_h+6k_4+3\lambda),
\label{eq:RGEdelta2}
 \\ \nonumber \\
16 \pi^2 \frac{d\delta_2'}{dt}
&=& 4\delta_{2}'^2 +\delta_2 k_4^a+
\delta_2' (2 \gamma_h+6k_4'+3\lambda),
 \\ \nonumber \\
16 \pi^2 \frac{d\lambda}{dt}
&=&6\lambda^2+4 \lambda \gamma_h -24y_t^4
\nonumber \\ && 
+\frac{3}{2}(g_1^4+2 g_1^2g_2^2+3g_2^4)
+2\delta_{2}^2+2\delta_{2}'^2,\label{eq:RGEH} \\
16 \pi^2 \frac{d k_2}{dt} 
&=& 2 m^2 \delta_2 + 6 k_2 k_4 + k^a_4\delta'_2, \label{eq:RGEk2} \\ \nonumber \\
16 \pi^2 \frac{d k_2'}{dt}
&=& 2 m^2 \delta_2' + 6 k_2' k_4' + k^a_4\delta_2, \label{eq:RGEk2p} \\ \nonumber \\
16 \pi^2 \frac{d k_4}{dt}
&=& 18 k_4^2 + \frac{1}{2}{k_4^a}^2+2\delta_2^2, \label{eq:RGE1}
\end{eqnarray}

\begin{eqnarray}
16 \pi^2 \frac{d k_4'}{dt}
&=& 18 k_4'^2 + \frac{1}{2}{k_4^a}^2+2\delta_2'^2,
\\ \nonumber  \\
16 \pi^2 \frac{d k_4^a}{dt}
&=& 4 {k_4^a}^2 + 6 k_4^a (k_4+k_4')
+4\delta_2\delta_2',
\label{eq:RGE3}
\end{eqnarray}
where $t=\log(\mu/M)$. In \Eqn{eq:RGEdelta2} $-$ \Eqn{eq:RGE3}
$\mu$ denotes the renormalization scale and $M$ is an
arbitrary scale.
Here $\gamma_h=-(9/4) g_2^2-(3/4)g_1^2+3y_t^2$ and $g_1$, $g_2$, $y_t$
are ${\rm U}(1)_{\rm Y}$, ${\rm SU}(2)_{\rm L}$ gauge couplings and top Yukawa coupling,
respectively.
In our calculation, the RG equations for
gauge and top Yukawa couplings are also taken into
account. We have taken the initial condition, $y_t(\mu=m_t)=\sqrt{2}m_t
(1+4\alpha_s(m_t)/3\pi)^{-1}/v$ for running of top Yukawa coupling, 
where $m_t=171~{\rm GeV}$ and
$\alpha_s(m_t)$ is strong coupling at the scale of $\mu=m_t$
\cite{Casas:1994qy}

We have solved all the RG equations given above and checked the
consistency of all the quartic couplings within the suitably
chosen scale of the theory. 
For the initialisation, we have taken
$\lambda_{init}$ corresponding to the recent value of Higgs
mass, 126 GeV.
The other initial values of parameters in our model
should have been chosen from the allowed region of parameter space.
\begin{figure}[h]
\begin{center}
\subfigure{
\includegraphics[width=2.7in,height=1.3in]{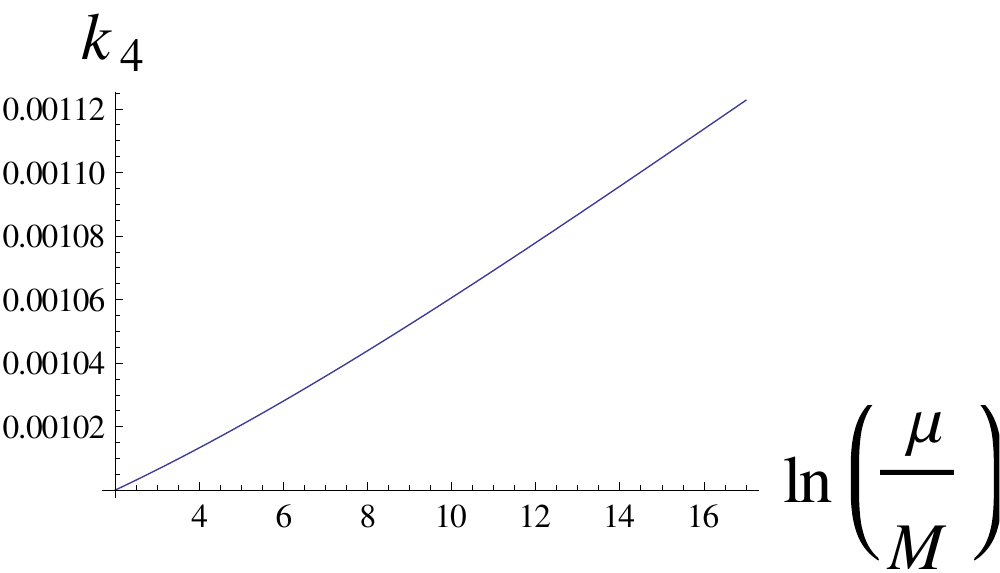}}
\subfigure{
\includegraphics[width=2.7in,height=1.3in]{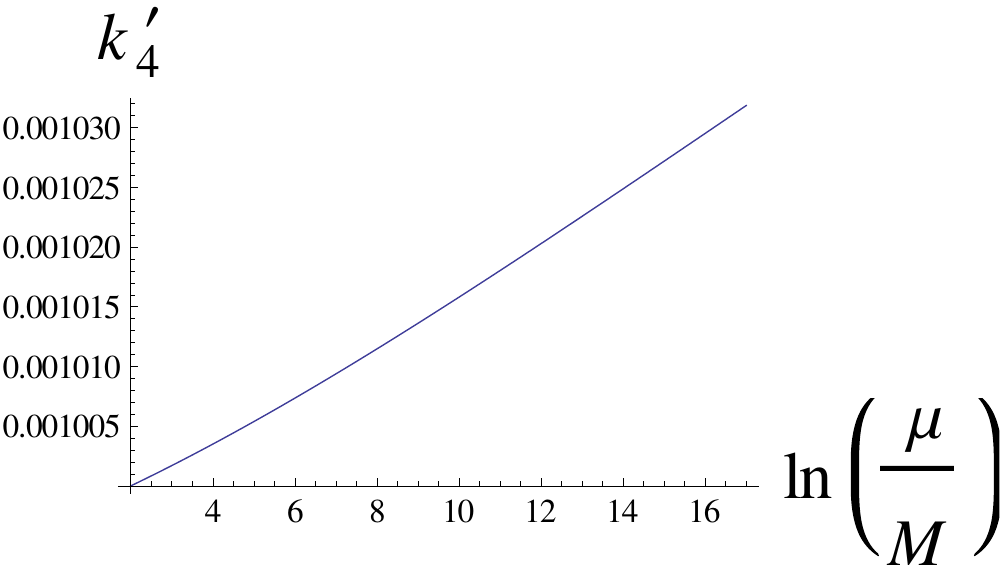}}
\subfigure{
\includegraphics[width=2.7in,height=1.3in]{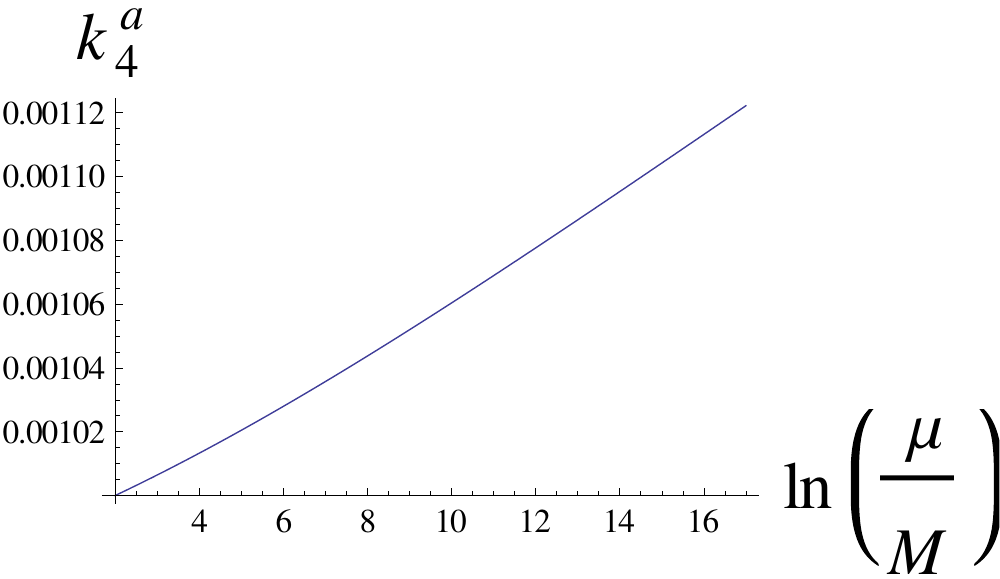}}
\subfigure{
\includegraphics[width=2.7in,height=1.3in]{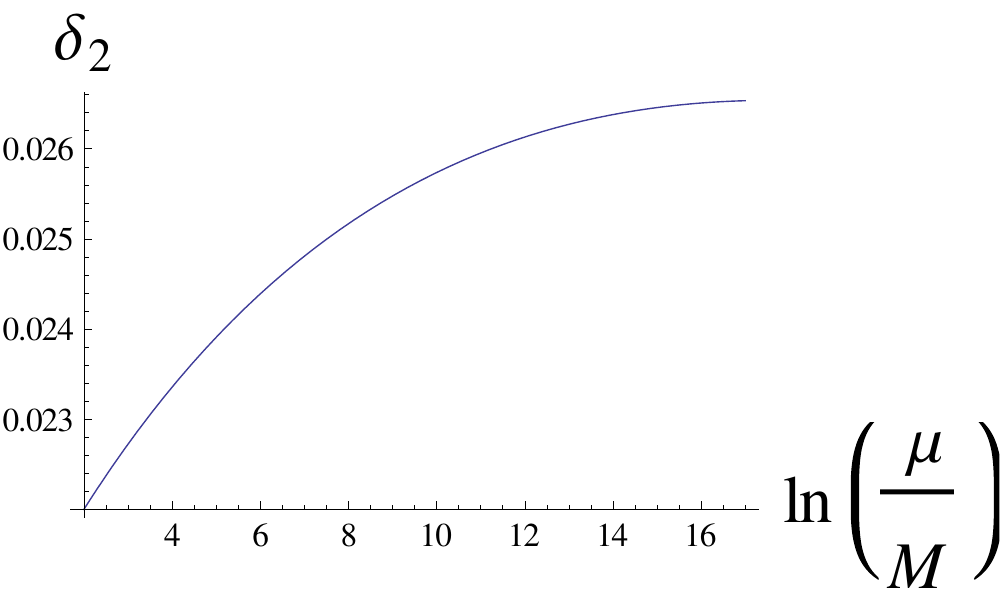}}
\subfigure{
\includegraphics[width=2.7in,height=1.3in]{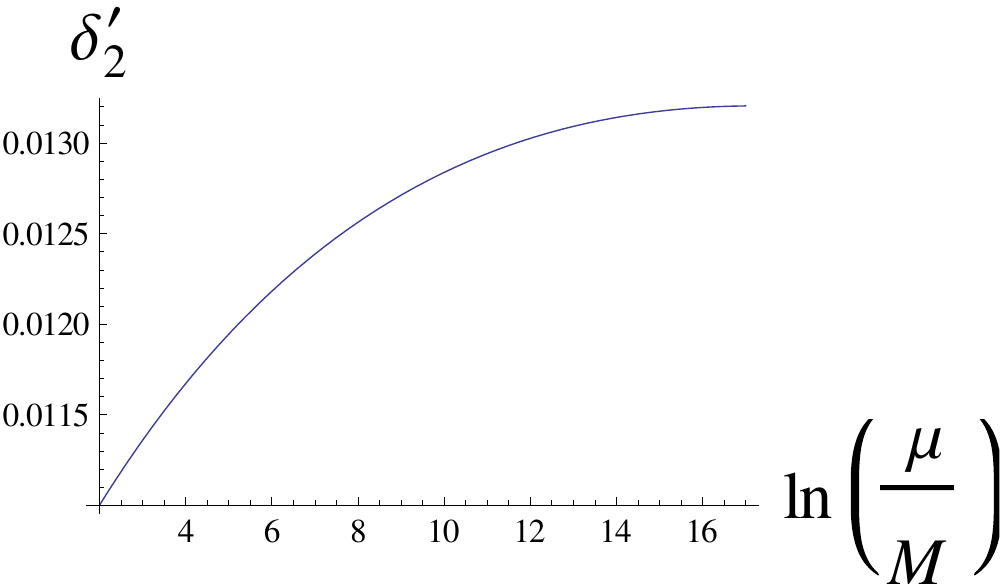}}
\subfigure{
\includegraphics[width=2.7in,height=1.3in]{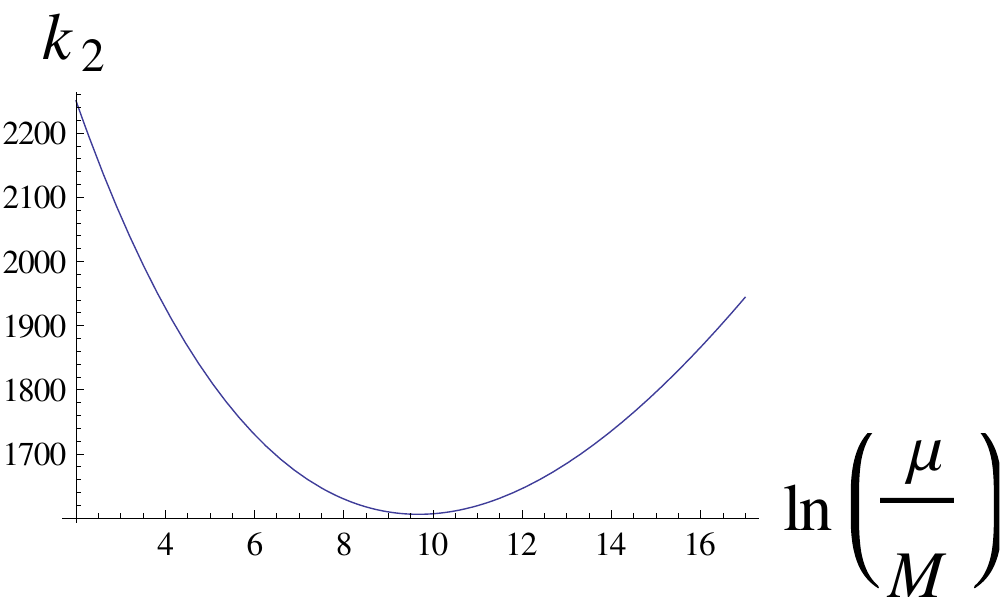}}
\subfigure{
\includegraphics[width=2.7in,height=1.3in]{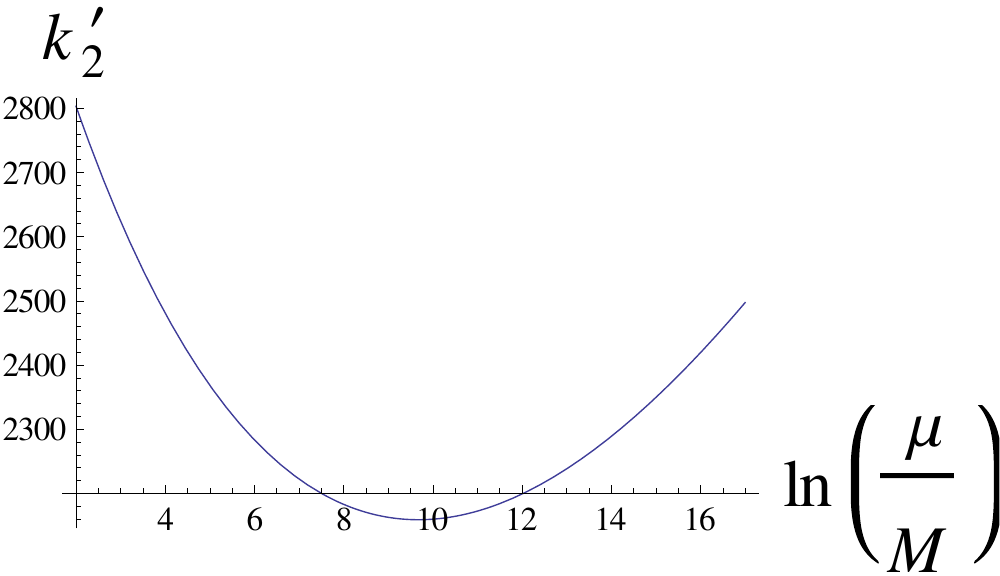}}
\subfigure{
\includegraphics[width=2.7in,height=1.3in]{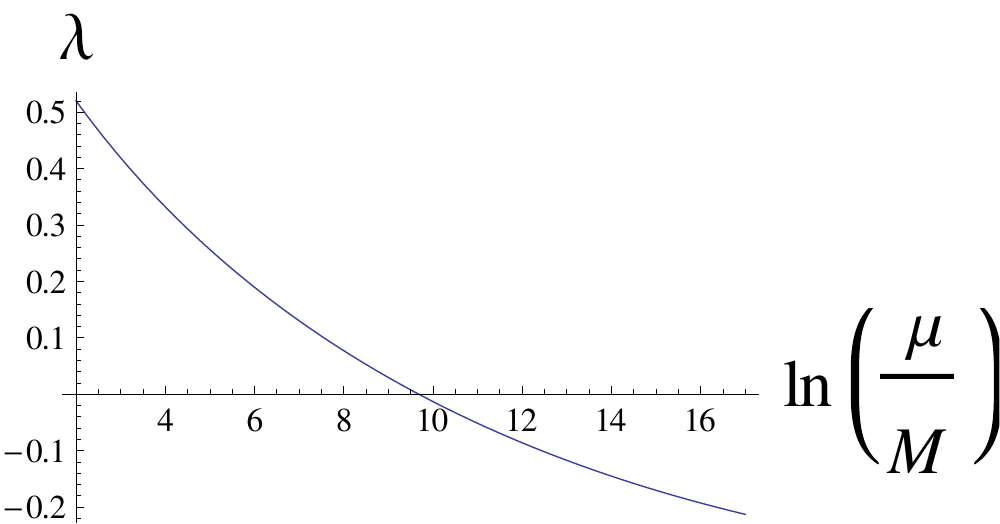}}
\caption{\label{fig:rge}\textit{Plot showing the variation of different
couplings present in the framework of two scalar singlet model 
with $\Z_2\times\Z_2'$ symmetry with different
energy scales.} }
\end{center}
\end{figure}
In Fig.~\ref{fig:rge} the variation of
different parameters ($k_4, k_4', k_4^a, k_2, k_2', \delta_2, \delta_2', \lambda$) with scale
used in this model is shown. 
The benchmark point 4A of Table~\ref{table_XENON100} has been chosen
for assigning initial values in the evaluation of the running
of various couplings.
The variation of mass of each scalar $S$ (or $S'$)
with energy scale can be obtained from the plots as it is 
determined by couplings $k_2$ and $\delta_2$ (or $k_2$ and $\delta'_2$).
Although we have solved the RG equation for $\delta_2$ and $\delta_2'$,
the influence of $\delta_2$ and $\delta_2'$ on $\lambda$ is very
small as we can see from \Eqn{eq:RGEH} that the RG equation of $\lambda$
is deviated from the SM RG equation only by the almost smooth term
$(2{\delta_2}^2 + 2{\delta_2'}^2)$. 
But the allowed region for `triviality bound' for a given
Higgs mass shrinks as the
term, $(2{\delta_2}^2 + 2{\delta_2'}^2)$ starts growing.

\subsection{Constraints from Invisible Higgs Decay Width}\label{S:lhc}

If kinematically allowed, Higgs boson can decay to $SS$ or $S'S'$. Such invisible decay channels are severely restricted by the present data from Large Hadron Collider (LHC). The branching fraction
\begin{equation}
  \mathcal{B}(H\rightarrow {\rm inv}) = \frac{\Gamma_{\rm inv}}{\Gamma_{\rm SM} + \Gamma_{\rm inv}}
 \end{equation}
is bounded at 95\%~CL to be less than 19\% by the global fits to the Higgs data keeping Higgs to fermion couplings fixed to their SM values. If such a restriction is lifted and additional particles are allowed in the loops the bound get relaxed to $\mathcal{B}(H\rightarrow {\rm inv})<38\%$~\cite{Belanger:2013xza}. $\Gamma_{\rm SM}$ denotes the SM Higgs decay width and  $\Gamma_{\rm inv}$ is the invisible Higgs decay width, which in our model is given by~\cite{Guo:2010hq},
  \begin{equation}
     \Gamma_{\rm inv} = \frac{v^2}{32 \pi M_H}\left(\delta_2^2 \sqrt{1 - \frac{4 M_S^2}{M_H^2}} +
     \delta_2'^2 \sqrt{1 - \frac{4 M^2_{S'}}{M_H^2}}\right) \,\,\, .
  \label{width_inv}
  \end{equation}
The benchmark point 4A in Table~\ref{table_XENON100}, consistent with the XENON~100 direct detection results, gives  $\mathcal{B}(H\rightarrow {\rm inv}) \sim 0.26$ which at present is allowed at 95\% CL~\cite{Chpoi:2013wga, Kang:2013zba, Belanger:2013kya, Djouadi:2013qya,Banerjee:2012xc}.
 However as we intend to interpret the low mass regions of dark matter claimed to be probed by several other dark matter direct
 search experiments (CDMS~II, CRESST, CoGeNT {\em etc.}) along with indirect searches (low energy $\gamma$-ray
 from Fermi Bubble and Galactic Centre), in some cases $M_S, M_{S'} \le M_H/2$ and the Higgs boson decays invisibly to $SS$ or $S'S'$, with a $\mathcal{B}(H\rightarrow {\rm inv})$ disfavoured by the LHC observations. This is a well known problem with all such models, where DM annihilation is mediated by the SM Higgs. It can be circumvented by adding extra degrees of freedom lighter than DM particles~\cite{zupan}. In that case it is possible to delineate the dominant Higgs-mediated annihilation channels from the diagrams contributing to the invisible Higgs decay. Such an elaborate model building issue is out of the scope of the present work as we feel given the non-observation of low mass DM events by XENON~100 and LUX, it is still premature to rely on CDMS~II and other experiments betting on the low mass DM. For quantitative estimations we have chosen the Higgs to be the $126$~GeV SM Higgs as a benchmark.

\section{Calculation of Relic Abundance }\label{S:relic}

In order to calculate the relic abundance for the dark matter candidates in the present
formalism, we need to solve the relevant Boltzmann equations.

In presence of one singlet scalar $S$, the Boltzmann eqn. is given by
\bea
\frac{dn_S}{dt} + 3Hn_S = -\langle\sigma v\rangle (n^2_S - n_{S_{\rm eq}}^2)\,\, ,
\label{eqn1}
\eea
where $n_S$  and $n_{S_{\rm eq}}$ are the number density and
equilibrium number density of the singlet scalar, $S$.
$\langle\sigma v\rangle$ is
the thermal averaged annihilation cross-section of dark matter
annihilating to SM particles.

Defining dimensionless quantities $Y=\frac{n}{e}$ and $x=\frac{M_{S}}{T}$, where
$e$ is the total entropy density, \Eqn{eqn1} can be written in the form,
\begin{equation}
\frac{dY}{dx} = -\left( \frac{45}{\pi}G \right )^{-1/2}\frac{g_*^{1/2}M_S}{x^2}
\langle \sigma v \rangle (Y^2-Y_{eq}^2)\,\,\, ,
\label{8}
\end{equation}
where $g_*$ is the degrees of freedom.
The relic density $Y_0$ (value of $Y$ at $T = T_0$) is obtained by
integrating \Eqn{8} from initial $x$ value, $x_0 = M_S/T_0$
to final $x$ value, $x_f = M_S/T_f$,
where $T_0$ and $T_f$ are the present photon temperature (2.726$^o$ K)
and freeze-out temperature respectively.

The relic density of a dark matter candidate, $S$,
in the units of critical energy density,
$\rho_{cr}=3H^2/8\pi G$, can be expressed as
\begin{equation}
 \Omega_{S}=\frac{M_{S}\,n}{\rho_{cr}}=\frac{M_{S}\,e_0\, Y_0}{\rho_{cr}}\,\,\, ,
\end{equation}
where $e_0$ is the present entropy density evaluated at $T_0$.
It follows that knowing $Y_0$, we can compute the relic density of the
dark matter candidate
from the relation~\cite{gondolo},
\bea
\Omega_S h^2 &=& 2.755\times10^8 \, \frac{M_{S}\:({\rm in~GeV})}{\rm GeV}\,Y_0\,\,.
\label{omega_s}
\eea
In \Eqn{omega_s}
$h=100\,{\rm km}\,\,{\rm sec}^{-1}{\rm Mpc}^{-1}$ is the Hubble constant.
The Planck survey provides the
constraints on the dark matter density $\Omega_{\rm DM} h^2$
from precision measurements of anisotropy of cosmic microwave background radiation as
\begin{equation}
 0.1165 < \Omega_{\rm DM}h^2 < 0.1227\,\,\, ,
\end{equation}
consistent with the previous WMAP measurement $ 0.1093 < \Omega_{\rm DM}h^2 < 0.1183$.

In our model with two real scalars, both $S$ and $S'$ contribute to the relic density.
Their individual contributions can be obtained by solving the following
coupled Boltzmann equations,
\bea
\frac{dn_S}{dt} + 3Hn_S &=& -\langle\sigma v\rangle_{SS\rightarrow XX}
(n_S^2-n_{S_{\rm eq}}^2) - \langle\sigma v\rangle_{SS\rightarrow S'S'}
\left(n_S^2-\frac{n_{S_{\rm eq}}^2}{n_{S'_{\rm eq}}^2}n_{S'}^2\right)
\label{be1}\\
\frac{dn_{S'}}{dt} + 3Hn_{S'} &=& -\langle\sigma v\rangle_{S'S'\rightarrow XX}
(n_{S'}^2-n_{S'_{\rm eq}}^2) - \langle\sigma v\rangle_{S'S'\rightarrow SS}
\left(n_{S'}^2-\frac{n_{S'_{\rm eq}}^2}{n_{S_{\rm eq}}^2}n_S^2\right)\,\, ,
\label{be2}
\eea
where $X$ stands for a standard model particle.
Thus, first term on the right hand side of Eqs.~(\ref{be1}) and (\ref{be2})
take care of the contribution of annihilation to SM particles whereas the
second terms of both the equations denote the contribution of
the self-scattering of the two scalars.

In the very early universe, both of the scalars are in thermal and chemical
equilibrium. But as the universe expands, the temperature falls resulting
some species to be decoupled from the universe plasma and contribute to the relic density.
The heavier scalar decouples earlier than the lighter one.
But today they are both freezed out giving rise to a total contribution
in relic abundance that is probed by WMAP/Planck or other cosmological observations.
Hence the total relic abundance ($\Omega_{\rm DM}$) due to both of the scalars is
the sum of their individual contributions ($\Omega_S$ and $\Omega_{S'}$),
\beq
\Omega_{\rm DM} = 
\Omega_S + \Omega_{S'}\, .
\label{omega}
\eeq
If the self-scattering cross-section between the two scalars is small
compared to that of the scalars going to standard model particles, such that
\beq
\langle\sigma v\rangle_{SS\rightarrow XX},
\langle\sigma v\rangle_{S'S'\rightarrow XX} >> \langle\sigma v\rangle_{SS\leftrightarrow S'S'} \,\,\, ,
\eeq
then the coupled Boltzmann equations in Eqs.~(\ref{be1}) and (\ref{be2}) can be treated
as two decoupled equations each one of which describes the evolution of each of the scalars
independently. In the present work we have ensured this condition by taking the masses of $S$ and $S'$ close enough\footnote{A situation like this can be realised by assuming that $M_S$ and $M_{S'}$ are degenerate at some high scale and then at low scale the degeneracy is slightly lifted due to some hidden sector physics.} so that $\langle\sigma v\rangle_{SS\leftrightarrow S'S'}$ is negligible from phase space considerations. We have then used  {\tt micrOMEGAs}~\cite{micrOMEGAs2, micrOMEGAs3} to calculate $\Omega_S$ and $\Omega_{S'}$.

The thermally-averaged values of cross-section ($\sv$)
for the annihilation channels of dark matter to Standard Model particles ($DM + DM \rightarrow SM + SM$)
can be expressed as (\cite{Guo:2010hq}, \cite{gondolo}),
\beq
\sv = \frac{x}{16 M_{S}^5 K_{2}^2(x)}
\int_{4M_{S}^2}^{\infty} ds\, K_{1}\left(\frac{\sqrt{s}}{T}\right) \sqrt{s-4M_{S}^2} \, \, \hat{\sigma}(s)\, \, ,
\eeq
with
\beq
\hat{\sigma}(s) = 2 \sqrt{s(s-4M_{S}^2)} \,  \sigma (s) \,\, , \nn
\eeq
where $x=M_{S}/T$. $K_{i}(x)$ denote the $i$th order modified Bessel function of second kind.
$\sigma(s)$ is the normalized annihilation cross-section of dark matter for $DM + DM \rightarrow SM + SM$
processes.
For non-relativistic dark matter $\sv$ can be approximated as
$\sv \sim  \hat{\sigma} (4 M_{S}^2)/ 4 M_{S}^2 $.

\begin{figure}[h]
\centering
\includegraphics[width=10cm, keepaspectratio=true]{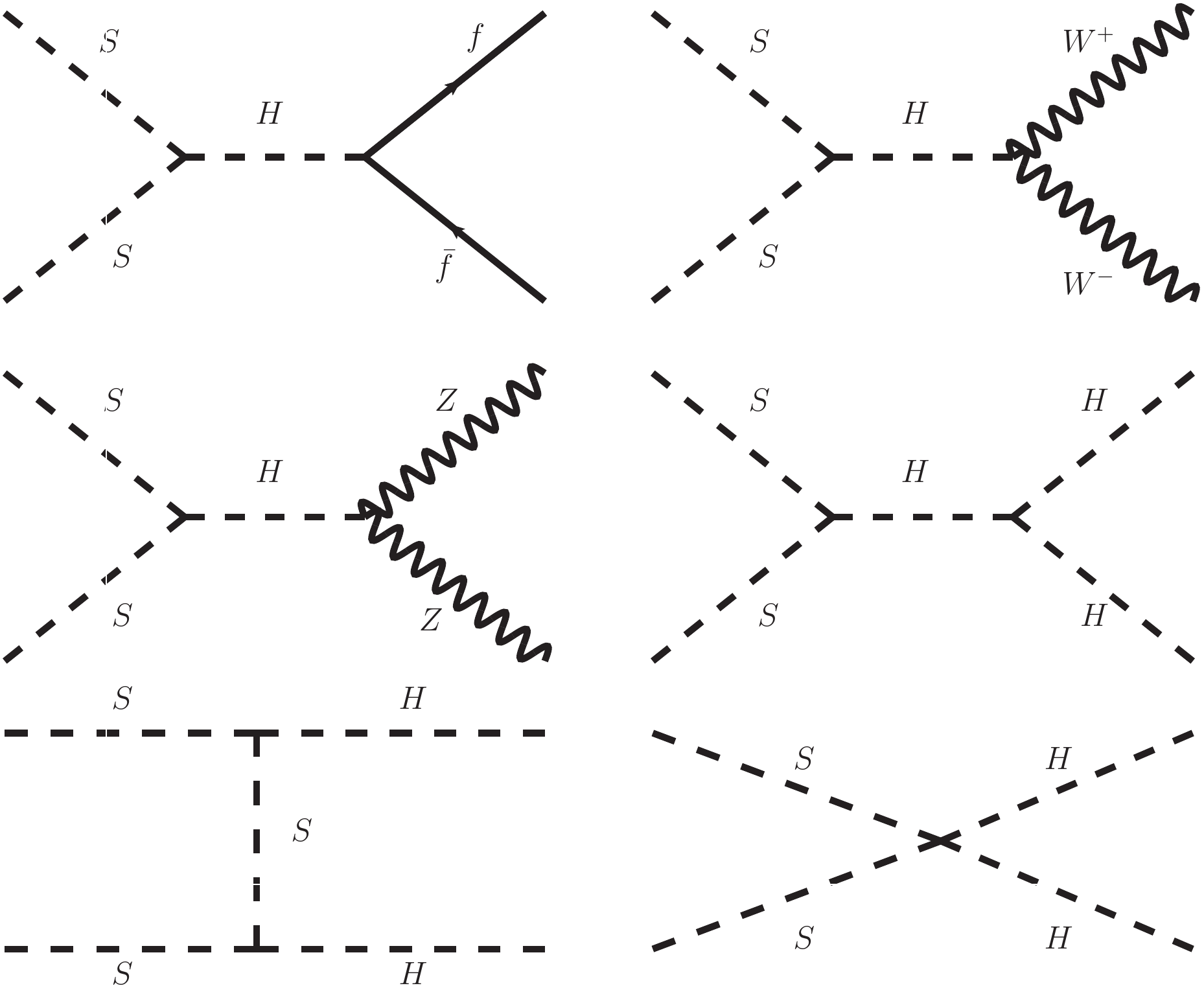}
\caption{\textit{Tree level Feynman diagrams of DM pair annihilation to a pair
of fermion and anti-fermion, $W^+ W^-$, $ZZ$ and Higgs. Similar annihilation channels exist for both $S$ and $S'$.}
\label{ann_diag}}
\end{figure}

The Feynman diagrams for
singlet scalar ($S$ or $S'$)
pair annihilation into SM particles in the unitary gauge are shown in Fig.~\ref{ann_diag}. The corresponding expressions for cross-sections can be found in Refs.~\cite{Burgess:2000yq, Guo:2010hq}.

\section{Constraining the Parameter Space with Dark Matter Direct Detection Experiments and Planck Survey}\label{S:Direct}

\begin{figure}[h]
\begin{center}
\includegraphics[width=4cm, keepaspectratio=true]{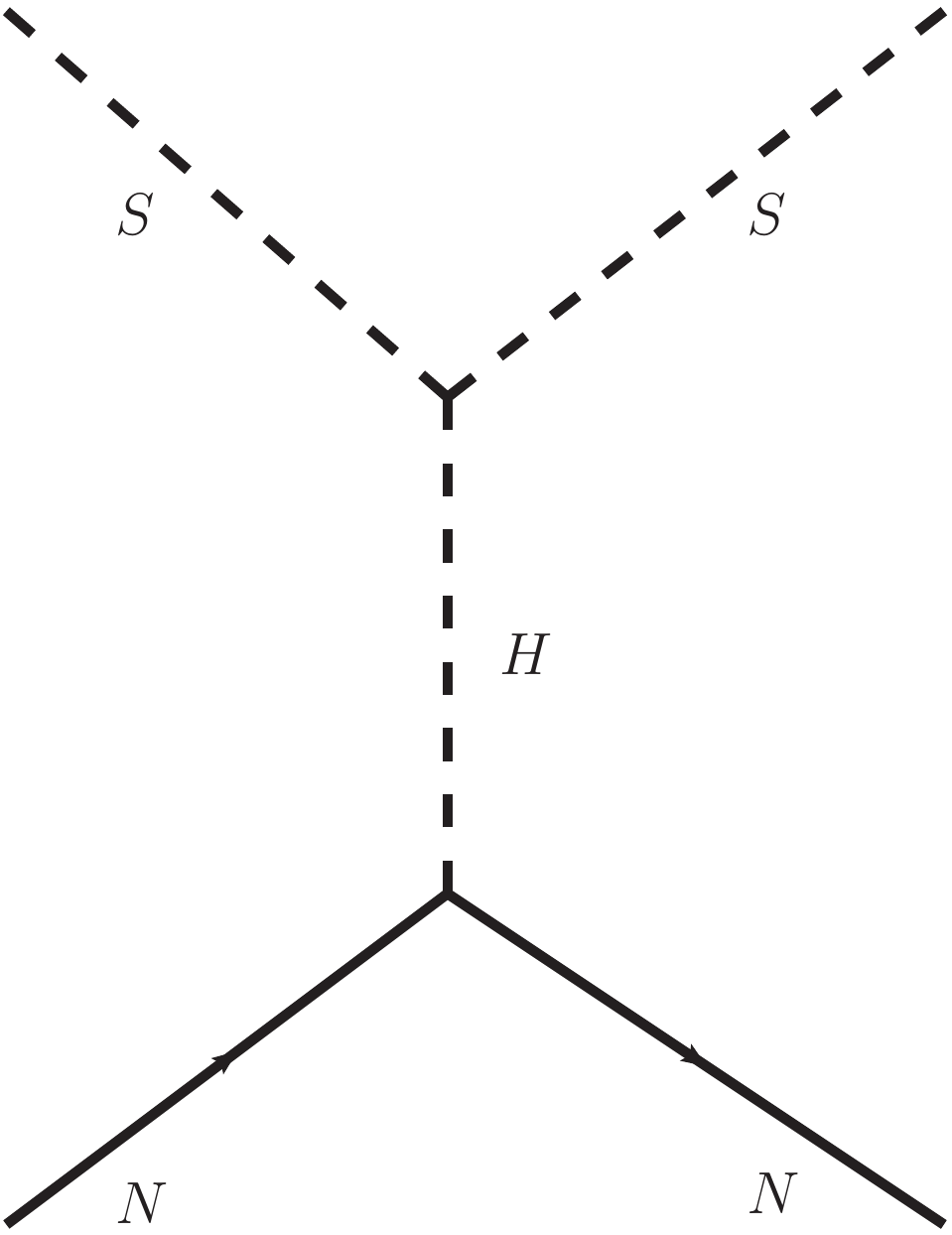}
\caption{\label{fig:fig_dd}
\textit{Lowest order Feynman diagram for singlet scalar--nucleus elastic scattering via Higgs mediation. A similar diagram exists for $S'$ as well.}}
\end{center}
\end{figure}

The dark matter particles $S$ and $S'$ can interact with the nuclei of the active material (see Fig.~\ref{fig:fig_dd}) in the direct detection experiments and leave their signature in form of a recoiled nucleus.
These experiments have indicated some preferred or excluded zones in the mass of DM vs. DM-nucleon cross-section plane. If we can express these cross-sections in terms of the parameters of our model, we can translate the results obtained from direct detection experiments into some preferred or excluded region in the parameter space of our model comprising of $\delta_2, M_S, \delta '_2$ and  $M_{S'}$. The requirement of producing the right relic abundance will further restrict the allowed parameter space.

As presented in Appendix~\ref{App-A}, the expressions for singlet scalar-nucleon elastic scattering cross-section are rather involved. But
for all practical purposes \Eqn{cs_nucleon} can be approximated as~\cite{Burgess:2000yq}
\begin{eqnarray}
\sigma^{\rm SI}_{\rm nucleon}
&=&
{(\delta_2)}^2
{\left(\frac{100 ~{\rm GeV}}{M_H ({\rm in~GeV})}\right)}^4
{\left(\frac{50 ~{\rm GeV}}{M_S ({\rm in~GeV})}\right)}^2
(5 \times 10^{-42} {\rm cm^2}) .
\label{eqcross2}
\end{eqnarray}
Similar expression works for $S'$ as well.

Empowering ourselves with the expression of cross-section in terms of model parameters, we will now go ahead in constraining the model parameter space from direct detection experimental results and relic density requirements from Planck survey. We have broadly explored three DM mass ranges. CDMS~II and CoGeNT vouch for low ($\sim 10$~GeV) mass DM. CRESST~II data prefer a relatively higher mass zone ($\sim 25$~GeV), in addition to the low zone. XENON~100 and LUX provide exclusion regions from non-observation of any interesting event. Only high mass DM ($> 50$~GeV) are consistent with these two experiments and Planck data. We will now elaborate more on them in the following.

\subsection{Constraints from CDMS~II and Planck data}

\begin{figure}[h!]
\begin{center}
\includegraphics[width=4.0in,keepaspectratio=true,angle=0]{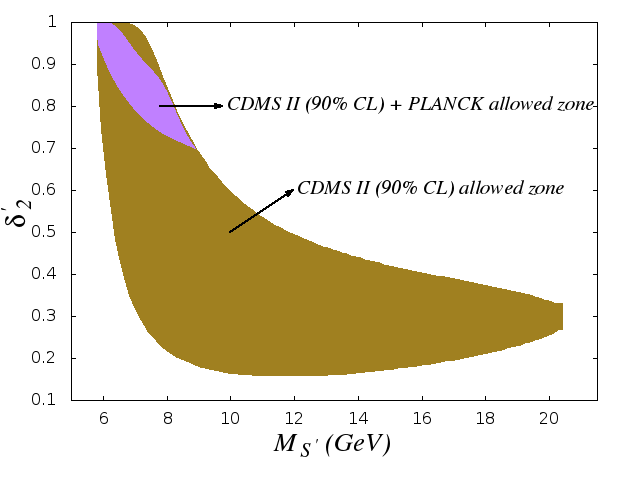}
\caption{\label{cdms2delta2} \textit{The hatched olive region is the parameter space allowed in the $M_{S'}$ -- $\delta '_2$ (or $M_{S} -\delta_2$) plane by CDMS~II 90\%CL data. Choosing ($M_S=8.6$~GeV, $\delta_2= 0.45$) in the $M_S -\delta_2$ plane, the only parameter space allowed in the $M_{S'} -\delta '_2$ by Planck data is shown as the blue shaded region.}}
\end{center}
\end{figure}

 CDMS collaboration has recently reported observation of three WIMP events and provided a preferred contour in the mass of DM -- SI scattering cross-section plane, with the maximum
 likelihood point at a mass of $8.6$~GeV with cross-section $1.9\times 10^{-41}$ cm$^2$. This value of cross-section
 corresponds to $\delta_2\simeq 0.45$ for $M_S=8.6$~GeV. Likewise assuming only $S$ (or $S'$) responsible for CDMS~II findings, the contour in the mass of DM -- SI scattering cross-section plane can now be translated to the $M_S$ -- $\delta_2$ (or $M_{S'}$ -- $\delta '_2$) plane as shown in Fig.~\ref{cdms2delta2} by the hatched region.

\begin{table}[h!]
 \caption{Benchmark point consistent with CDMS~II contour and Planck data}
 \centering
 \vskip 2 mm
 \begin{tabular}{|c| c | c | c | c | c |}
  \hline
 &$M_S$ (or $M_{S'}$) & $\delta_2$ (or $\delta_2'$)
& $\sigma^{\rm SI}$  & $\langle\sigma v\rangle$ & Contribution  \\
 &(GeV) & & ($\times10^{-41}$~cm$^{2}$)  & ($\times 10^{-26}$~cm$^3$/s) & in $\Omega_{\rm S} ~({\rm or} ~\Omega_{\rm S'} )$\\
\hline \hline
 \parbox[b]{6mm}{\multirow{2}{*}[10pt]{\rotatebox{90}{~~~~Benchmark point 1}}} &    8.6    &  0.45  &  $1.9$ &  $3.2$ $\begin{cases} 2.6 ~(b\bar b) & \\0.4 ~(c\bar c) & \\0.2 ~(l\bar l) &\end{cases}$ &
     $\begingroup{\renewcommand{\arraystretch}{1.2}\begin{array}{c} 81\% \\12\% \\7\% \end{array}}\endgroup $\\
     \cline{2-6}
   & 6.7 &  0.82  & $9.9$  & $8.3$ $\begin{cases} 6.3 ~(b\bar b) & \\1.2 ~(c\bar c) & \\7.2 ~(l\bar l) &\end{cases}$ &
   $\begingroup{\renewcommand{\arraystretch}{1.2}\begin{array}{c} 77\% \\15\% \\8\% \end{array}}\endgroup $\\
    \hline
\end{tabular}
  \label{table_CDMS}
\end{table}

 However in our model both $S$ and $S'$ contribute to the relic density and participate in direct detection experiments. So we fix $\Omega_S$ by choosing the point corresponding to the maximum likelihood point ($M_S=8.6$~GeV, $\delta_2\simeq 0.45$) in the $M_S$ -- $\delta_2$ plane. This restricts $\Omega_{S'}$ from Planck constraints on relic abundance. Only a small part of the previously allowed region in the $M_{S'}$ -- $\delta '_2$ plane (shown as the blue shaded region) can now account for such $\Omega_{S'}$. We then complete choosing our benchmark point (see Table \ref{table_CDMS}) by taking $M_{S'}=6.7$~GeV, $\delta '_2\simeq 0.82$ from this blue region. This point in the parameter space is thus consistent with both CDMS~II and Planck observations.

\subsection{Constraints from CoGeNT and Planck data}
 
CoGeNT dark matter direct detection experiment predicts dark matter particle with
 a mass roughly $\sim$ 7-11 GeV and elastic scattering cross-section with nucleon
 which is $\sim 10^{-41}$--$10^{-40}$ cm$^2$. 
\begin{figure}[h!]
\begin{center}
\includegraphics[width=4.0in,keepaspectratio=true,angle=0]{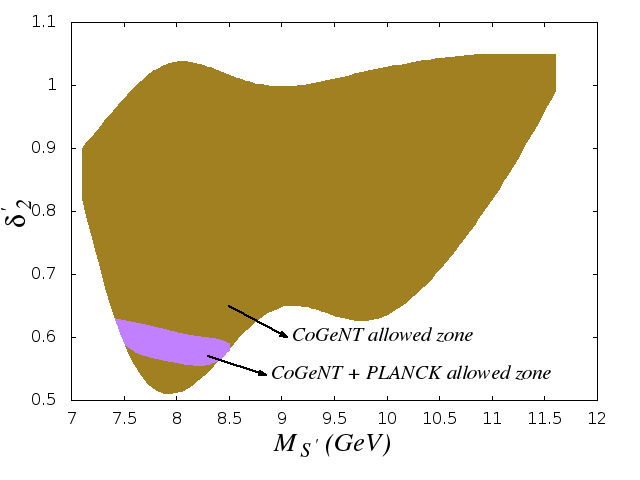}
\caption{\label{cogentdelta2} \textit{The hatched olive region is the parameter space allowed in the $M_{S'}$ -- $\delta '_2$ (or $M_{S}$ -- $\delta_2$) plane by CoGeNT data. Choosing ($M_S=7.8$~GeV, $\delta_2= 0.56$) in the $M_S$ -- $\delta_2$ plane, the only parameter space allowed in the $M_{S'}$ -- $\delta '_2$ by Planck data is shown as the blue shaded region.}}
\end{center}
\end{figure}
The other direct detection experiments like
 DAMA/LIBRA or
 CRESST~II have also reported signals nearly in that zone which is not consistent
 with the known background sources. 
Also the spectrum of the events reported by
 experiments like CRESST~II and
 CoGeNT are consistent with each other~\cite{hooper2011} and possibly attribute
 to dark matter of mass $\sim 10$ GeV.

 With CoGeNT preferred zone we can do similar analysis
 as we did with CDMS~II. If we are to explain CoGeNT findings with either $S$ (or $S'$), we find the olive hatched zone in the $M_S$ -- $\delta_2$ (or $M_{S'}$ -- $\delta '_2$) plane (see Fig.~\ref{cogentdelta2}). We choose a point ($M_S=7.8$~GeV, $\delta_2\simeq 0.56$) in the $M_S$ -- $\delta_2$ plane. This fixes $\Omega_S$. Planck results then restrict $\Omega_{S'}$, which then can be reproduced by a tiny region in the $M_{S'}$ -- $\delta '_2$ plane (shown as blue shaded region). We then take $M_{S'}=8.2$~GeV, $\delta '_2\simeq 0.61$ from this region to complete the CoGeNT benchmark point presented in Table~\ref{table_CoGeNT}.
 \begin{table}[h!]
 \caption{Benchmark point consistent with CoGeNT and Planck data}
 \centering
 \vskip 2 mm
 \begin{tabular}{|c| c | c | c | c | c |}
  \hline
 &$M_S$ (or $M_{S'}$) & $\delta_2$ (or $\delta_2'$)
& $\sigma^{\rm SI}$  & $\langle\sigma v\rangle$ & Contribution  \\
 &(GeV) & & ($\times10^{-41}$~cm$^{2}$)  & ($\times 10^{-26}$~cm$^3$/s) & in $\Omega_{\rm S} ~({\rm or} ~\Omega_{\rm S'} )$\\
\hline \hline
  \parbox[b]{6mm}{\multirow{2}{*}[10pt]{\rotatebox{90}{~~~~Benchmark point 2}}} &   7.8    &    0.56  &  $9.0$ &  $4.6$ $\begin{cases} 3.7 ~(b\bar b) & \\0.6 ~(c\bar c) & \\0.4 ~(l\bar l) &\end{cases}$ &
     $\begingroup{\renewcommand{\arraystretch}{1.2}\begin{array}{c} 80\% \\12.5\% \\7.5\% \end{array}}\endgroup $\\
     \cline{2-6}
   & 8.2 &  0.61     & $9.8$  & $5.7$ $\begin{cases} 4.6 ~(b\bar b) & \\0.7 ~(c\bar c) & \\0.4 ~(l\bar l) &\end{cases}$ &
   $\begingroup{\renewcommand{\arraystretch}{1.2}\begin{array}{c} 81\% \\12\% \\7\% \end{array}}\endgroup $\\
    \hline
\end{tabular}
  \label{table_CoGeNT}
\end{table}

\subsection{Constraints from CRESST~II and Planck data} \label{ss:cresst}

 We have analysed the $1\sigma$ contour of the CRESST~II data.
 We could have chosen $2\sigma$
 region of CRESST~II data as well. But as the low mass part of the $2\sigma$ contour
 has crossover with CDMS~II and CoGeNT low mass regions, the outcome is expected to be similar to these experiments. We rather prefer to work with DM of higher mass $\sim 25$~GeV.
\begin{figure}[h!]
\begin{center}
\includegraphics[width=4.0in,keepaspectratio=true,angle=0]{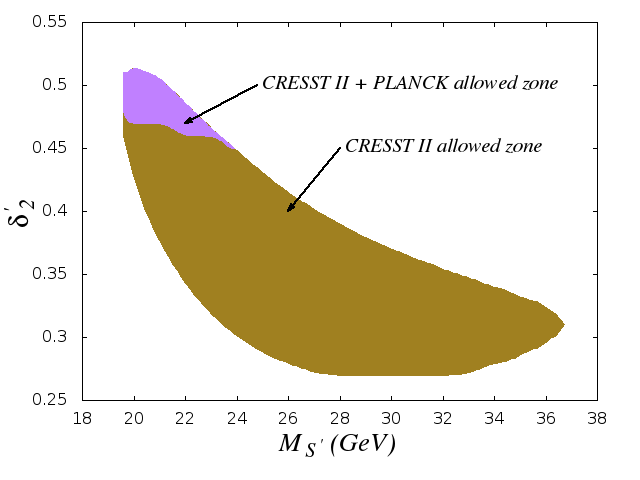}
\caption{\label{cresst2delta2} \textit{The hatched olive region is the parameter space allowed in the $M_{S'}$ -- $\delta '_2$ (or $M_{S}$ -- $\delta_2$) plane by CRESST~II $1\sigma$ contour. Choosing ($M_S=25.3$~GeV, $\delta_2= 0.36$) in the $M_S$ -- $\delta_2$ plane, the only parameter space allowed in the $M_{S'}$
-- $\delta '_2$ plane by Planck data is shown as the blue shaded region.}}
\end{center}
\end{figure}
 CRESST~II $1\sigma$ contour can similarly be translated to the hatched region in the $M_S$ -- $\delta_2$ (or $M_{S'}$ -- $\delta '_2$) plane (see Fig.~\ref{cresst2delta2}). The best fit point for the  CRESST~II $1\sigma$ contour (mass of DM = $25.3$~GeV, $\sigma^{\rm SI}=1.6\times 10^{-42}$ cm$^2$) corresponds to the point ($M_S=25.3$~GeV, $\delta_2\simeq 0.36$) in the $M_S$ -- $\delta_2$ plane. Choice of this point fixes $\Omega_S$. The relic abundance constraint from Planck survey then restricts $\Omega_{S'}$, which in turn limits the allowed parameter space in the $M_{S'}$ -- $\delta '_2$ plane to the tip (shown as the blue zone in Fig.~\ref{cresst2delta2}) of the CRESST~II allowed zone. We then choose  $M_{S'}=23.3$~GeV, $\delta '_2\simeq 0.47$ from this region to complete the CoGeNT benchmark point presented in Table~\ref{table_CRESST}.
 \begin{table}[h!]
 \caption{Benchmark Point consistent with CRESST~II $1\sigma$ contour and Planck data}
 \centering
 \vskip 2 mm
 \begin{tabular}{|c| c | c | c | c | c |}
  \hline
 &$M_S$ (or $M_{S'}$) & $\delta_2$ (or $\delta_2'$)
& $\sigma^{\rm SI}$  & $\langle\sigma v\rangle$ & Contribution  \\
 &(GeV) & & ($\times10^{-41}$~cm$^{2}$)  & ($\times 10^{-26}$~cm$^3$/s) & in $\Omega_{\rm S} ~({\rm or} ~\Omega_{\rm S'} )$\\
\hline \hline
  \parbox[b]{6mm}{\multirow{2}{*}[10pt]{\rotatebox{90}{~~~~Benchmark point 3}}} &   25.3    &    0.36  &  $1.6$ &  $3.0$ $\begin{cases} 2.4 ~(b\bar b) & \\0.4 ~(c\bar c) & \\0.2 ~(l\bar l) &\end{cases}$ &
     $\begingroup{\renewcommand{\arraystretch}{1.2}\begin{array}{c} 81\% \\12\% \\7\% \end{array}}\endgroup $\\
     \cline{2-6}
   & 23.3 &  0.47     & $3.2$  & $4.8$ $\begin{cases} 3.9 ~(b\bar b) & \\0.6 ~(c\bar c) & \\0.3 ~(l\bar l) &\end{cases}$ &
   $\begingroup{\renewcommand{\arraystretch}{1.2}\begin{array}{c} 81\% \\12\% \\7\% \end{array}}\endgroup $\\
    \hline
\end{tabular}
  \label{table_CRESST}
\end{table}

\subsection[]{Constraints from XENON~100 and Planck data}
\begin{figure}[h]
 \begin{center}
 \subfigure[]{
 \includegraphics[width=2.9in,height=2.2in, angle=0]{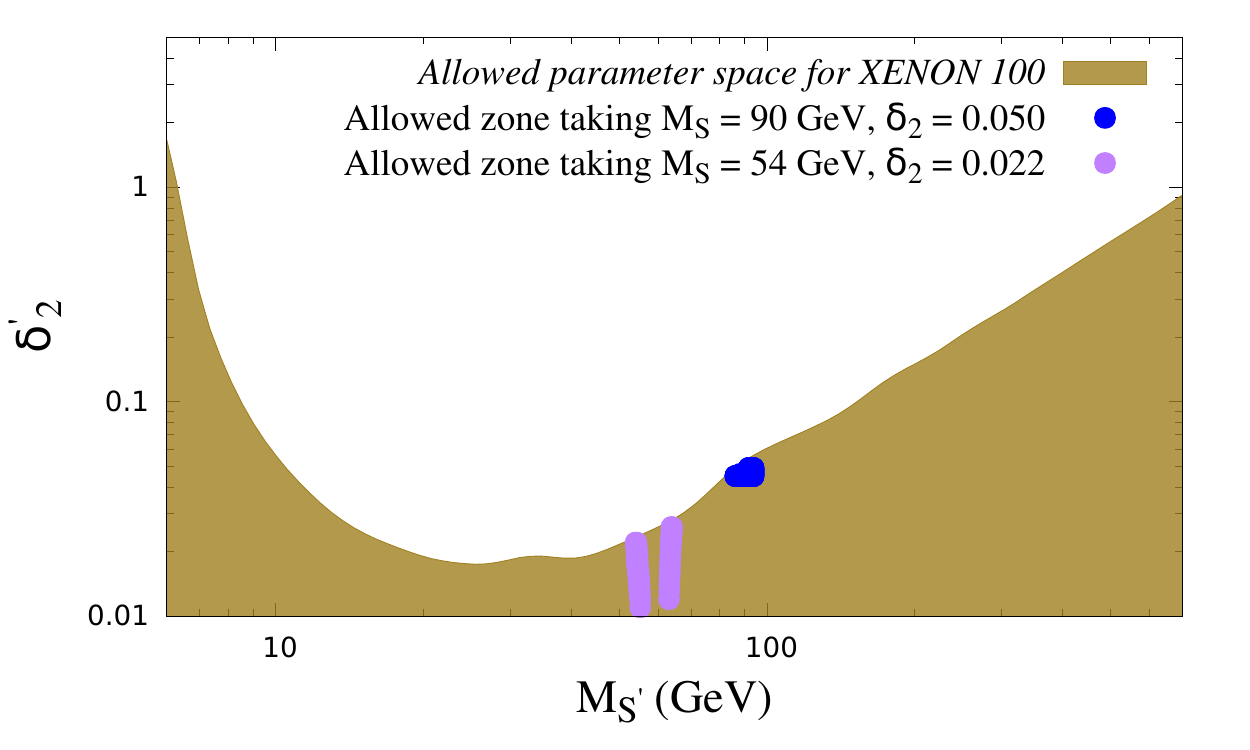}}
 \subfigure[]{
 \includegraphics[width=2.9in,height=2.55in, angle=0]{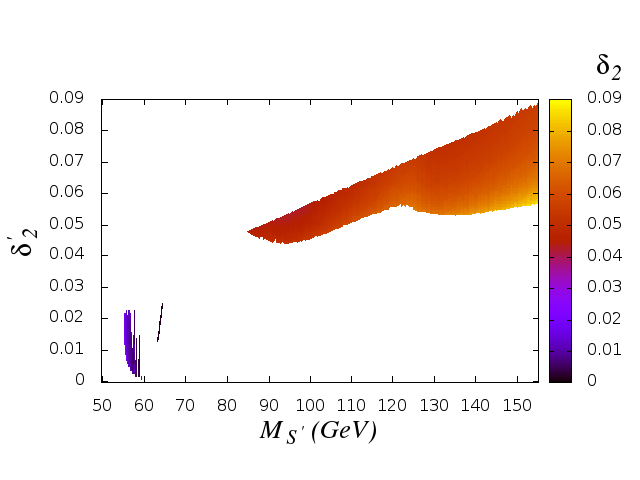}}
 \caption{\label{fig:xenon100relic} \textit{In pane (a) the solid olive region represents the parameter space allowed in the $M_S -\delta_2$ (or $M_{S'} - \delta '_2$) plane by XENON~100. Pane (b) shows the parameter space allowed by XENON~100 keeping $M_S - M_{S'}=2$~GeV. Choosing $M_S=54$~GeV, $\delta_2\simeq 0.022$ ($M_S=90$~GeV, $\delta_2\simeq 0.05$) and keeping the DM mass difference less than 2~GeV, we now denote in pane (a) the parameter space consistent with both XENON~100 and Planck observations.} }
 \end{center}
 \end{figure}

XENON~100 collaboration did not observe any prospective signal of DM. From this non-observation
they have set an upper bound on spin independent scattering cross-section $\sigma^{\rm SI}$ for various dark matter masses.
In the context of our model this translates to an allowed region in the $M_S -\delta_2$ (or $M_{S'} - \delta '_2$) plane, indicated as the olive region in Fig~\ref{fig:xenon100relic}(a).  Here we assume only $S$ (or $S'$) participate to have a conservative estimate. We need to find the parameter space suitable for producing DM relic abundance compatible with Planck observations. Our strategy here is a bit different from earlier experiments which indicated some allowed zone in the scattering cross-section -- mass of DM plane. 

For $M_S - M_{S'}=2$~GeV we scan the parameter space spanned by $M_{S'}$, $\delta_2$ and $\delta '_2$. The outcome is presented in Fig~\ref{fig:xenon100relic}(b). We see that there is a small island of allowed parameter space at around $M_{S'}\sim 55$~GeV and a continuum from $\sim 85$~GeV onwards. From each of these two regions we will now choose a benchmark point. 

First we  choose ($M_S=54$~GeV, $\delta_2\simeq 0.022$). This fixes $\Omega_S$. This in turn restricts $\Omega_{S'}$ from Planck data. To reproduce this $\Omega_{S'}$ window, we do a parameter scan in the $M_{S'} - \delta '_2$ plane imposing the constraint $M_S - M_{S'}\leq 2$~GeV. We thus get the zone (shown as blue dots in Fig~\ref{fig:xenon100relic}(a)) consistent with both XENON~100 and Planck observations. We thus get benchmark point 4A  as given in Table~\ref{table_XENON100}.

Then choosing ($M_S=90$~GeV, $\delta_2\simeq 0.05$) and proceeding as above leads to the benchmark point 4B as shown in Table~\ref{table_XENON100}.
\begin{table}[h]
 \caption{Benchmark Points consistent with XENON~100 and Planck data}
 \centering
 \vskip 2 mm
 \begin{tabular}{|c| c | c | c | l | c |}
  \hline
 & $M_S$ (or $M_{S'}$) & $\delta_2$ (or $\delta_2'$)
& $\sigma^{\rm SI}$  & $~~~~~~~~~~\langle\sigma v\rangle$ & Contribution  \\
 &(GeV) & & ($\times10^{-45}$~cm$^{2}$)  & ~~~~~($\times 10^{-26}$~cm$^3$/s) & in $\Omega_{\rm S} ~({\rm or} ~\Omega_{\rm S'} )$\\
\hline \hline
 \parbox[b]{6mm}{\multirow{2}{*}[10pt]{\rotatebox{90}{Benchmark point 4A}}} &    54.0    &    0.022  &  $1.4$ &  $9.94$ $\begin{cases} 7.83 ~(b\bar b) & \\1.29 ~(c\bar c) & \\0.82 ~(l\bar l) &\end{cases}$ &
     $\begingroup{\renewcommand{\arraystretch}{1.2}\begin{array}{c} 78.8\% \\13.2\% \\8.2\% \end{array}}\endgroup $\\
     \cline{2-6}
    & 56.0 &  0.011     & $0.8$  & $3.95$ $\begin{cases} 3.11 ~(b\bar b) & \\0.52 ~(c\bar c) & \\0.32 ~(l\bar l) &\end{cases}$ &
   $\begingroup{\renewcommand{\arraystretch}{1.2}\begin{array}{c} 78.7\% \\13.2\% \\8.1\% \end{array}}\endgroup $\\
    \hline
   \parbox[b]{6mm}{\multirow{2}{*}[-10pt]{\rotatebox{90}{Benchmark point 4B}}} &    90.0    &    0.050  &  $2.6$ &  $4.32$ $\begin{cases}  4.29~(W^+ W^-) & \\0.024 ~(b\bar b) & \\0.004 ~(c\bar c) & \\0.002 ~(l\bar l) &\end{cases}$ &
     $\begingroup{\renewcommand{\arraystretch}{1.2}\begin{array}{c} 99.3\% \\0.55\% \\0.10\% \\ 0.05\% \end{array}}\endgroup $\\
     \cline{2-6}
    & 92.0 &  0.045     & $2.0$  & $3.78$ $\begin{cases} 3.28 ~(W^+ W^-) & \\0.48 ~(Z Z) & \\0.016 ~(b\bar b) & \\0.003 ~(c\bar c) & \\0.002 ~(l\bar l) &\end{cases}$ &
   $\begingroup{\renewcommand{\arraystretch}{1.2}\begin{array}{c} 86.6\% \\12.8\% \\0.43\% \\0.08\% \\0.05\% \end{array}}\endgroup $\\
    \hline
 \end{tabular}
  \label{table_XENON100}
\end{table}

\subsection[]{Constraints from LUX and Planck data} 
\begin{figure}[h]
 \begin{center}
 \subfigure[]{
 \includegraphics[width=2.9in,height=2.2in, angle=0]{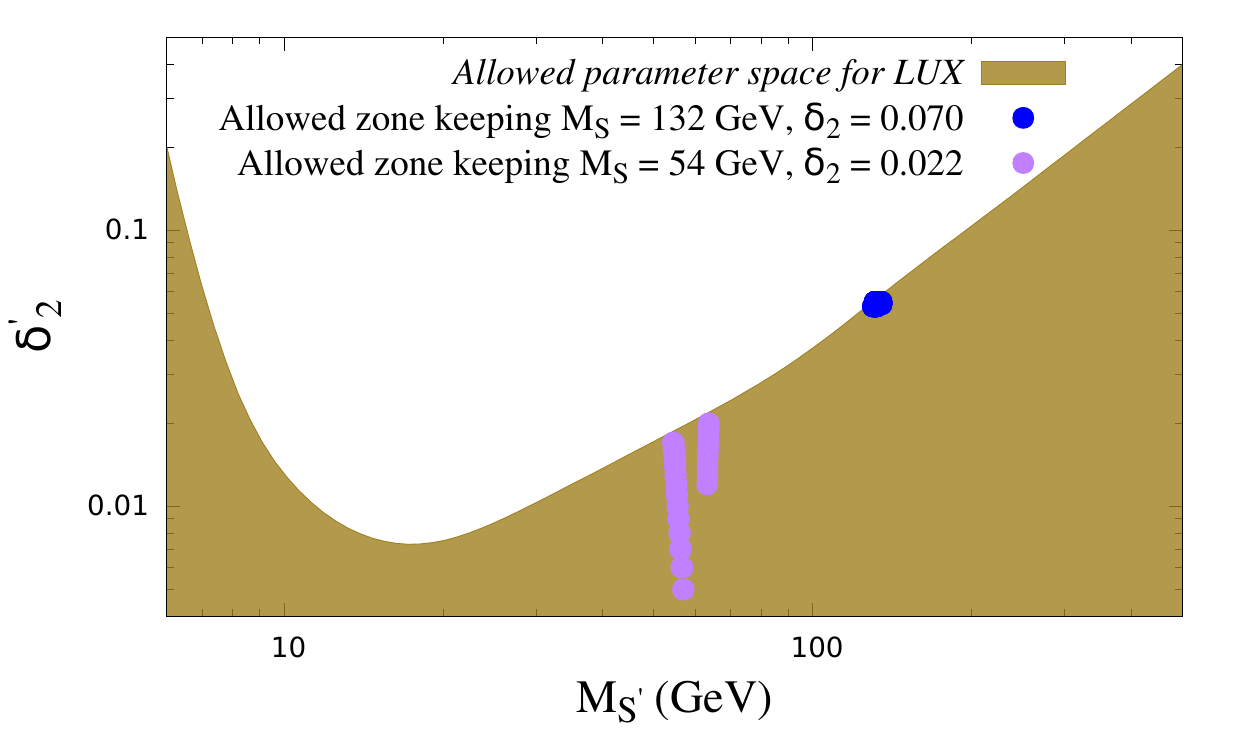}}
\subfigure[]{
 \includegraphics[width=2.9in,height=2.55in, angle=0]{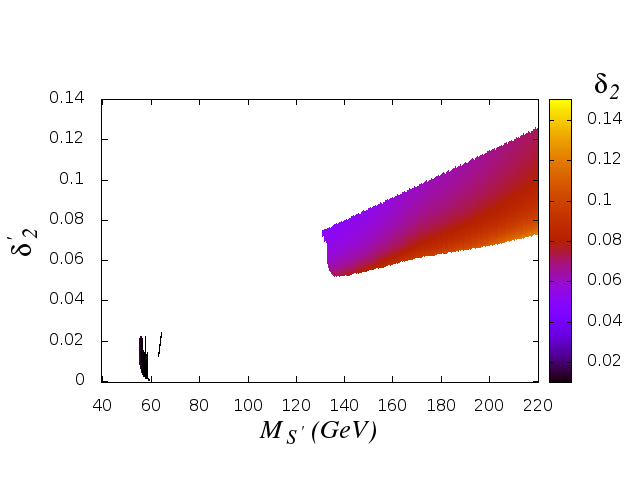}}
 \caption{\label{fig:luxrelic} \textit{In pane (a) the solid olive region represents the parameter space allowed in the $M_S -\delta_2$ (or $M_{S'} - \delta '_2$) plane by LUX. Pane (b) shows the parameter space allowed by XENON~100 keeping $M_S - M_{S'}=2$~GeV. Choosing $M_S=54$~GeV, $\delta_2\simeq 0.022$ ($M_S=132$~GeV, $\delta_2\simeq 0.07$) and keeping the DM mass difference less than 2~GeV, we now denote in pane (a) the parameter space consistent with both LUX and Planck observations.} }
 \end{center}
 \end{figure}

LUX collaboration has recently published their results which confirm to XENON~100 findings. 
As we have done for XENON~100, we show the allowed zones by LUX and Planck observations in Fig.~\ref{fig:luxrelic}. We see that similar to XENON~100, here also we get an island in the parameter space around $M_S\sim 57$~GeV. But the continuum starts around 135~GeV. For the $M_S\sim 57$~GeV point the phenomenology will be similar to XENON~100. However high DM masses $\sim 135$~GeV can not reproduce the morphological features of indirect detection observations.  So we will not discuss the LUX allowed parameter space any further in this work.

To compare with the literature which attempts explaining the experimental observations assuming certain branching fractions of Higgs to SM particles we have denoted in Tables~\ref{table_CDMS}---\ref{table_XENON100} the relevant branching ratios for the chosen benchmark points. 
The main feature of our benchmark points is that in our case the braching ratios are determined from the model precisely whereas the previous 
analysis has been performed assuming certain branching fractions.
Still the experimental data can be confronted remarkably well. 

\section{Confronting Indirect Dark Matter Detection Experiments}\label{S:Indirect}
The region surrounding the Milky Way is rich in astrophysics and is assumed to have a high
 density of dark matter. This region is promising for better
 understanding of the properties of dark matter, as no other astrophysical source or region
 is as accessible as the galactic centre (GC). The Fermi Gamma Ray Space Telescope (FGST) has been
 employed to survey the high luminous gamma ray emission between $\sim$ 50 MeV to $\sim$ 100 GeV.

 DM distribution follows a density
 function, $\rho (\vec{r})$, with $\vec{r}$ is the position
 vector from the centre of the galaxy. Several such DM halo profiles are available in the literature.
 We choose some representative cuspy to flat profiles for our numerical estimations. These are presented in the Appendix~\ref{App-C}.
 For a particular DM halo profile one can calculate the photon flux due to DM annihilation using \Eqn{gammaflux2} in Appendix~\ref{App-B}.

 We now discuss observations of excess $\gamma$-ray emission from GC and low latitude of Fermi Bubble, which does not appear to have ``standard" origins, but can be understood in the light of DM annihilation. In particular we show that the morphological features of these observations can be explained by our proposed model with parameter spaces consistent with DM direct detection experiments and DM relic abundance constraints from Planck survey. We will point out in subsequent discussions that the uncertainties involved in understanding the significance of such astrophysical observations are quite substantial. So at this point we do not intend to fit the data, but rather limit ourselves to reproduce the morphological features of the observations in terms of DM annihilation by our model.

\subsection{Explanation of Excess Gamma Ray Emission from Galactic Centre}
 The low energy (few GeV) $\gamma$-ray data from galactic
 centre region observed by Fermi telescope give a hint of a low mass dark matter.
 In this section we have discussed the phenomenology of gamma ray from the annihilation of
 dark matter from galactic centre in this present formalism. We have
 computed $\gamma$-ray flux from the singlet scalars in this model constrained by CDMS~II and CoGeNT
 experiments and finally comparison with the observed $\gamma$-ray flux has been done.

 Detailed studies on spectral and morphological features of the gamma rays from the
 galactic centre region has been studied in Refs.~\cite{hooper2011,hooperplb}. In Ref.~\cite{hooper2011},
 the spectrum of the
 gamma ray emission from the region that encompasses $5^{o}$ surrounding the galactic centre
 is studied after subtracting
 the known sources from the data of the Fermi Second Source Catalog \cite{fermi_source_catalog} and
 disc emission template. The main reason of the disc template emission is the gamma ray
 produced from the neutral pion decay which is outcome of cosmic ray interaction with
 gas. Though inverse Compton and Bremsstrahlung can also contribute. Assuming the gas distribution
 to be of the following form,
 \begin{eqnarray}
\rho_{\rm gas} &\propto& e^{-|z|/z_{\rm sc}(R)}, \,\,\,\,\,\,\,\,\,\,\,\,\,\,\,\,\,\,\,\,\,\,\,\,\, {{\rm for}\,\,\,\, R < 7\, {\rm kpc}}, \\
\rho_{\rm gas} &\propto& e^{-|z|/z_{\rm sc}(R)} \,  e^{-R/R_{\rm sc}},\,\,\,\,\, {{\rm for}\,\,\,\, R > 7\, {\rm kpc}},\nonumber
\end{eqnarray}
 where ($z$, $R$) denotes the location relative to the GC in cylindrical coordinates.
 The chosen values of $R_{\rm sc}=3.15$ kpc
 and $z_{\rm sc}(R) = 0.1 + 0.00208 \times (R/{\rm kpc})^2$~\cite{kalberla, nakanishi} kpc 
 as these values are best suitable to fit for the observational data of 21-cm H line surveys
 which is the conventional tool used to probe the density of neutral hydrogen.
 The flux of gamma rays from pion decay is estimated by integrating this distribution over
 the line-of-sight and it is found to be in good agreement wih the observed morphology of the diffuse emission 
 The residual gamma ray spectrum is brighter for $\gamma$-energy range from 300 MeV to 10 GeV and drops by
 order of magnitude beyond 10 GeV. From the morphological characteristics of this residual
 gamma ray emission from the central region of our galaxy, it has been shown in~\cite{hooper2011,boyarsky},
 that below 300 MeV the residual gamma ray could originate from
 a point-like source but at higher energies it could originate from spatially extended
 components or may be from annihilating dark matter. Also, if the very
 high energetic portion of the residual gamma ray emission from the galactic centre is analysed,
 the spectral shape is found to match fairly well with the gamma emission from galactic
 ridge. The galactic ridge is an inner region of galaxy extending up to a width of
 5$^o$ galactic latitude and $\pm 40^o$ galactic longitude containing huge amount of
 white dwarfs \cite{gehrels}. The standard convention is that the high energy cosmic nucleons interact
 with molecular cloud in the ridge and pions are produced in huge amount which
 subsequently decay to high energy gamma. Therefore, the residual emission from GC considered here
 can be assumed to contain low energy tail of ridge emission.
 Also very low energy part of the residual spectrum is supposed to be dominated by
 the point source~\cite{chernyakova,boyarsky} which loses its dominance at above GeV scale.

 However there are some astrophysical propositions that can
 morphologically explain the gamma-ray flux structure
 from the inner part of galactic centre. These include millisecond pulsar population~\cite{abazajian},
 central supermassive black hole~\cite{hooperplb, chernyakova} \etc\, which can explain this
 spatially extended gamma ray distribution feature from GC.

 {\em Super-massive black holes} can also accelerate both electrons and cosmic ray protons.
 These accelerated electrons then produce gamma ray from inverse Compton scattering
 and can be accounted for any unresolved gamma ray emission from galactic centre region. But
 these electrons produce $\gamma$-ray in TeV-scale~\cite{supermassive_bh}
 which may in principle explain the high energy gamma ray from galactic centre observed
 by different experiments like HESS, HAWC. Hence this type of mechanism
 cannot fully account for the FGST data for low energy gamma rays.
 But the cosmic ray protons accelerated
 by the black hole can produce pions through the interaction with interstellar gas.
 Decay of these pions yield gamma rays of lower energy. This scenario may
 partially explain the FGST residual emission feature as there appear a lot of astrophysical
 parameters like ISM gas distribution or unknown diffusion coefficient for proton
 propagation through ISM gas \etc which are not fully understood.

\begin{figure}[h!]
\begin{center}
\subfigure[Isothermal]{
\includegraphics[width=2.1in,height=2.9in, angle=-90]{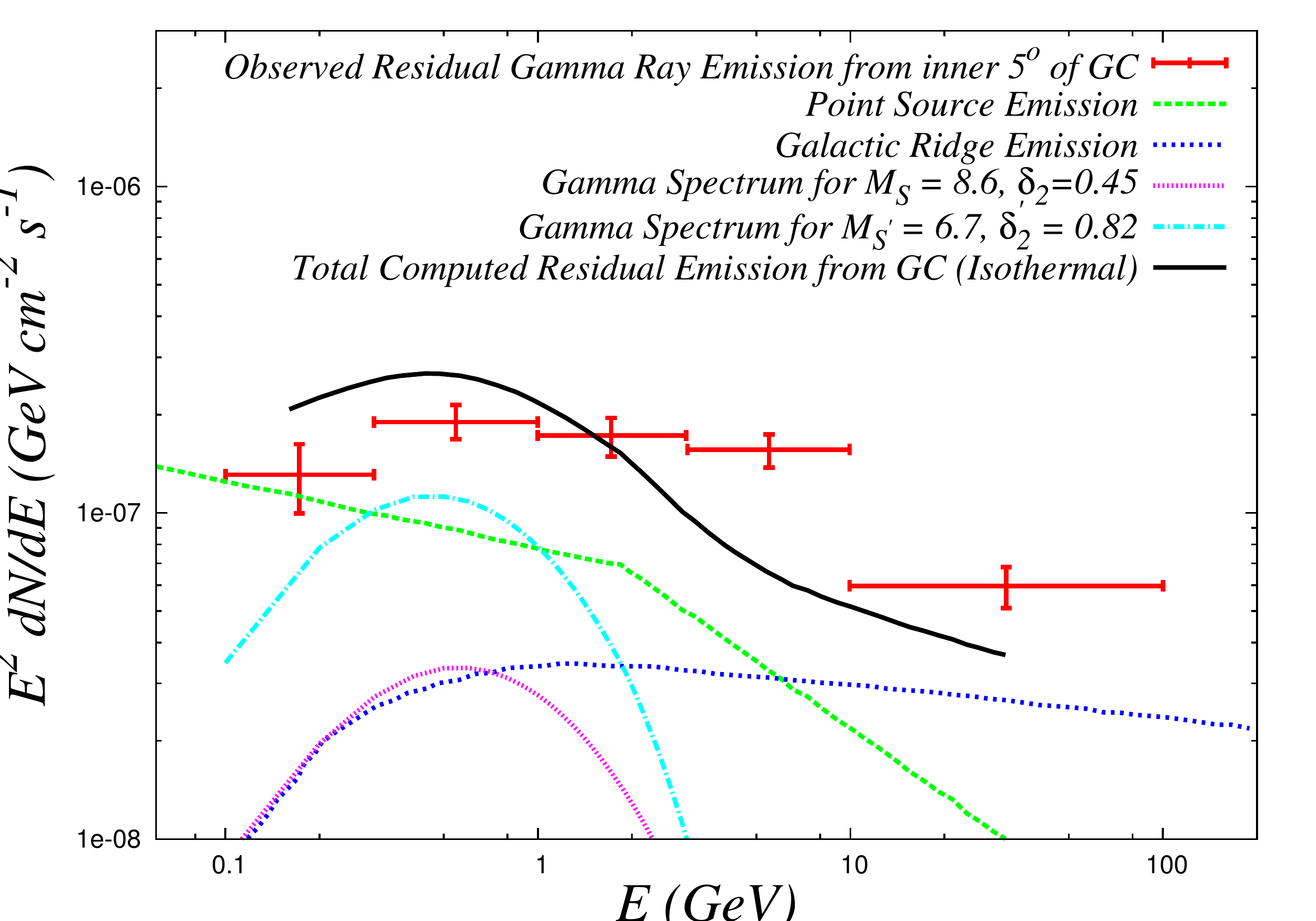}}
\subfigure[NFW]{
\includegraphics[width=2.1in,height=2.9in, angle=-90]{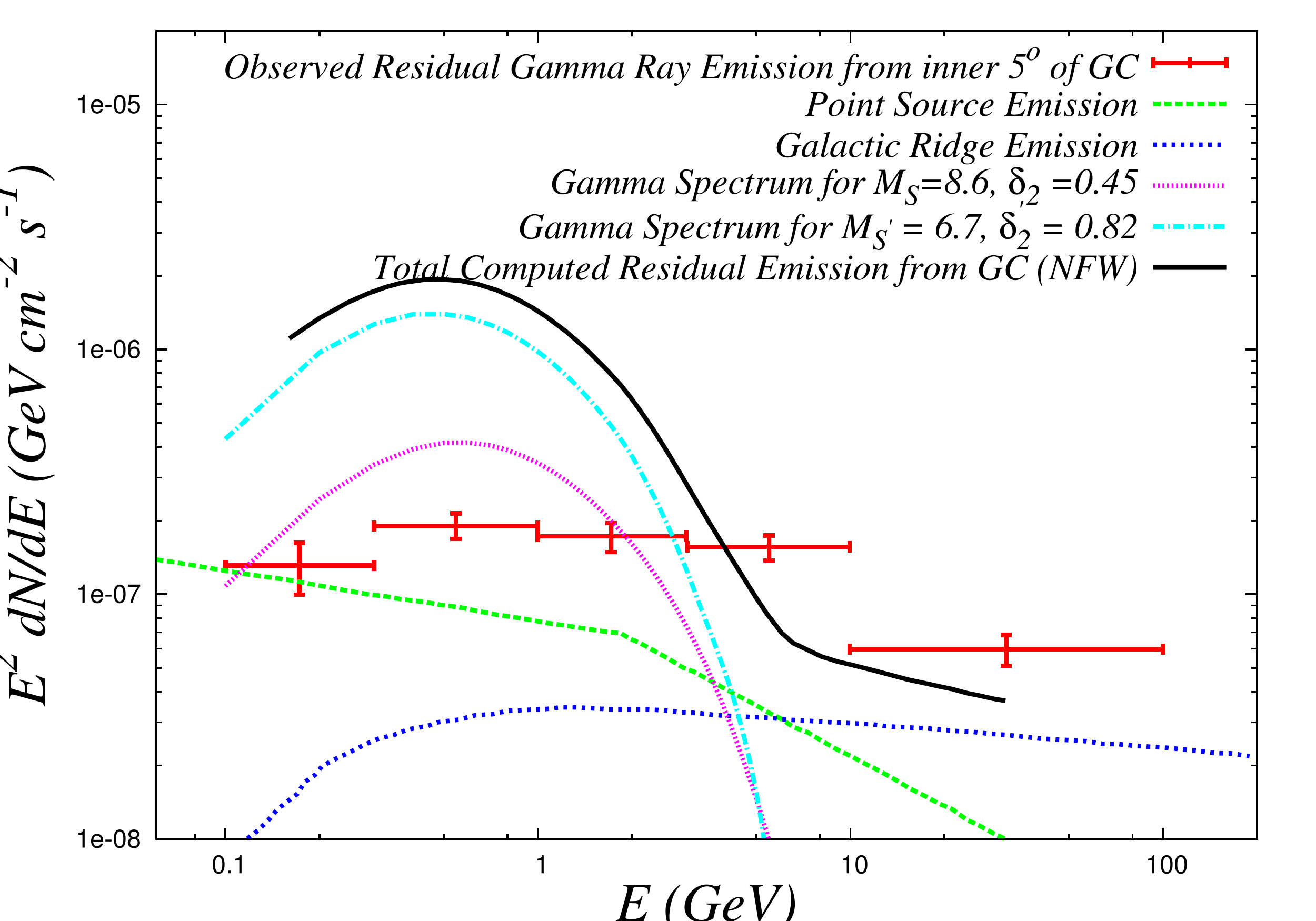}}
\subfigure[Einasto]{
\includegraphics[width=2.1in,height=2.9in, angle=-90]{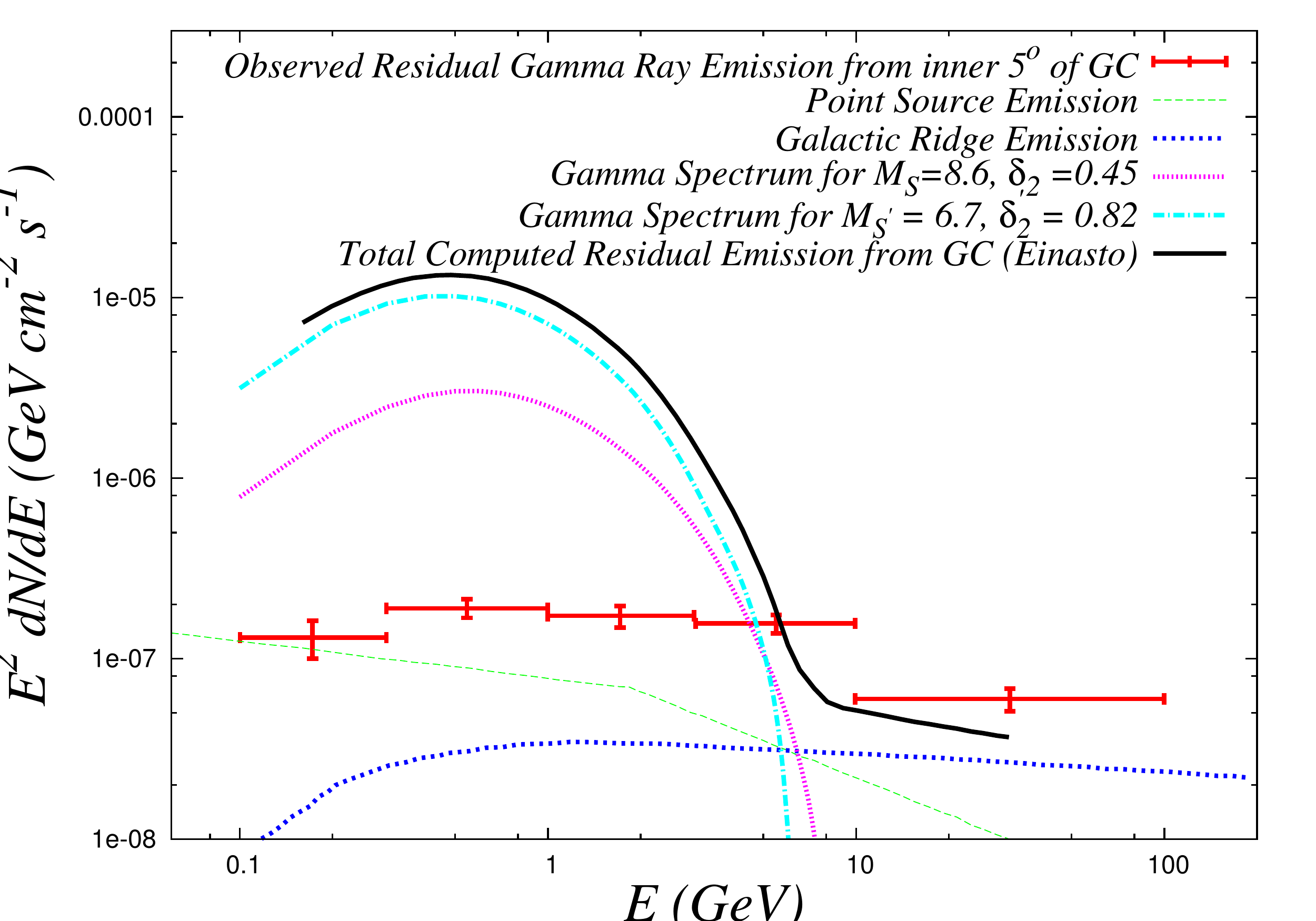}}
\subfigure[Moore]{
\includegraphics[width=2.1in,height=2.9in, angle=-90]{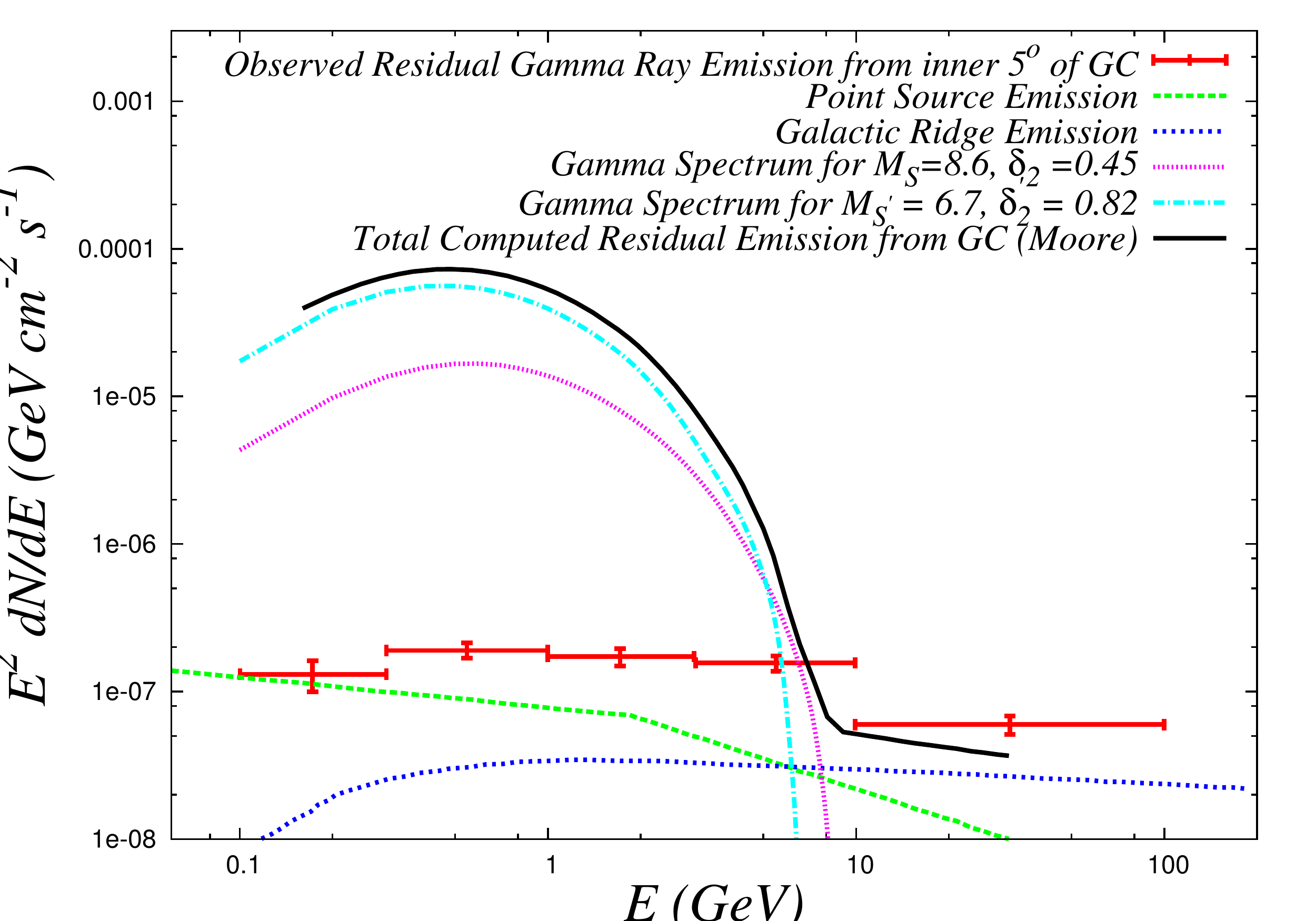}}
\caption{\label{fig:gammafromGC_cdms} \textit{Residual $\gamma$-ray flux from the inner 5$^o$ of
galactic centre. The red data points represent the observed flux. Point source and galactic ridge emissions are represented by the light green dashed and blue dotted lines respectively.  DM annihilation in our model is calculated for benchmark point 1 consistent with CDMS~II and Planck data (see Table~\ref{table_CDMS}). $SS$-annihilation is calculated for $M_S=8.6$~GeV and $\delta_2=0.45$ and is denoted by the dotted violet line. For $S'S'$ annihilation we use $M_{S'}=6.7$~GeV and $\delta'_2=0.82$ and is represented by the dash-dotted cyan line. Total calculated residual $\gamma$-ray flux is denoted by the solid black curve.
Each sub-figure is calculated for different DM halo profiles, as indicated in their respective captions.
} }
\end{center}
\end{figure}

 The other astrophysical objects, the gamma rays from which may yield spectra similar to 
 that observed by FGST data are {\em Millisecond pulsars}. 
 The spectra from the millisecond pulsars are hard in nature beyond
 a few GeV, \ie it falls off with much rapidity after a few GeV.
 his tend to indicate that surrounding the galactic centre, there may be other
 millisecond pulsars in considerable numbers which are still to be probed experimentally. But there are
 discrepancies which immediately contradict this scenario. From the FGST's
 first pulsar catalog, the spectral index of gamma from pulsars is centred at 1.38 but
 a much harder spectrum for the average pulsar is required to match
 the observed gamma spectrum. Although a very few pulsars have certainly a very
 hard spectral index that can be accounted for the residual emission~\cite{abazajian}
 below 10 GeV but to fit the gamma flux of the bumpy spectral shape
 one needs to have larger number of these types of pulsars which is not present
 in Fermi pulsar catalog. The globular clusters, rich in gamma pulsars
 have also been studied to measure the spectral index but here too, the data do not favour
 very hard spectral nature. In order to comply the angular distribution
 pattern of the emission, the pulsar density should decrease very rapidly
 along the outward radial distance but the significance of such rapidity has not been
 found from astrophysical data. 
 From all the above discussions, one may conclude the fact that some
 different mechanism is required to explain this bumpy spectral shape
 of the residual emission from galactic centre observed by FGST.

\begin{figure}[h!]
\begin{center}
\subfigure[Isothermal]{
\includegraphics[width=2.1in,height=2.9in, angle=-90]{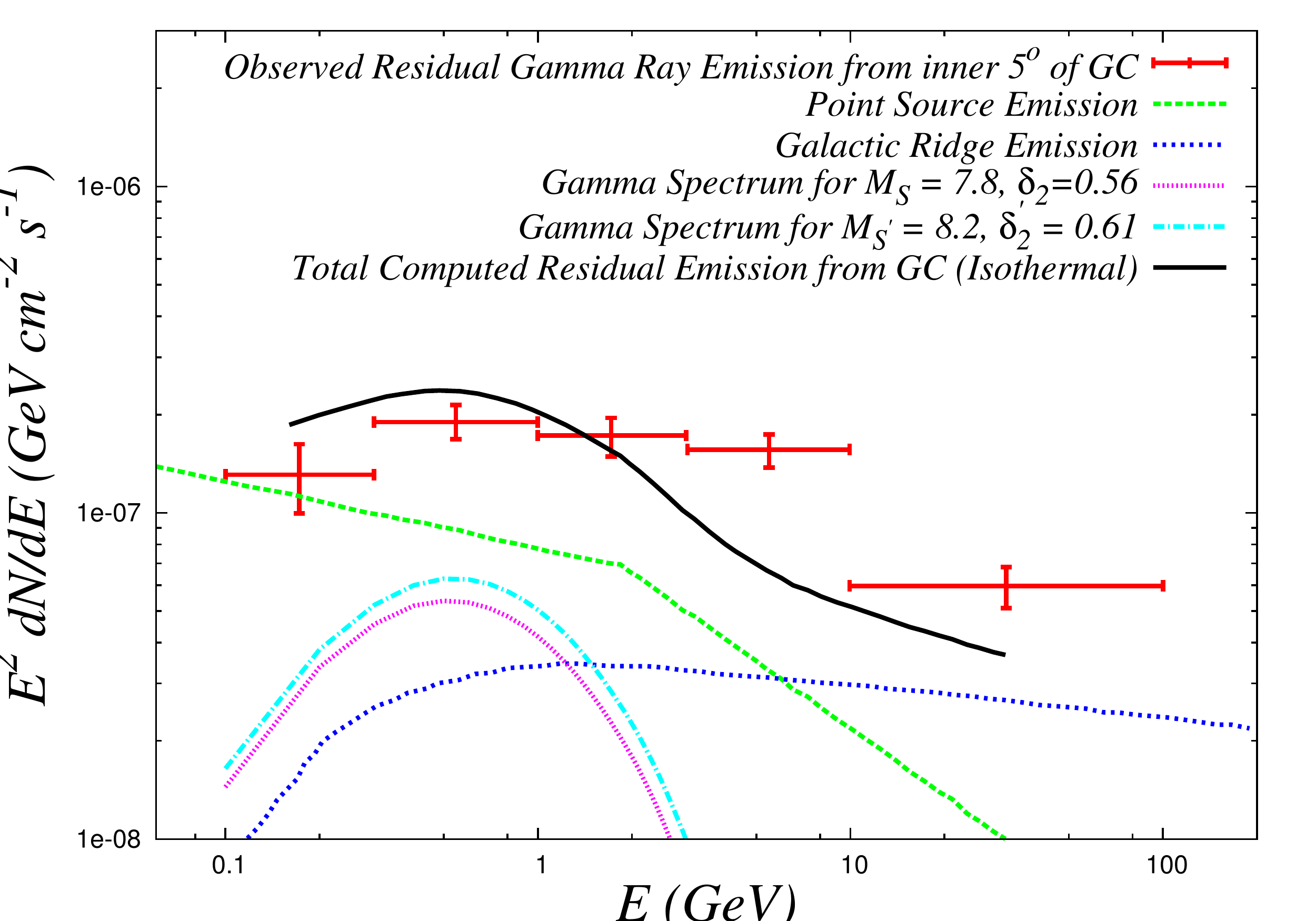}}
\subfigure[NFW]{
\includegraphics[width=2.1in,height=2.9in, angle=-90]{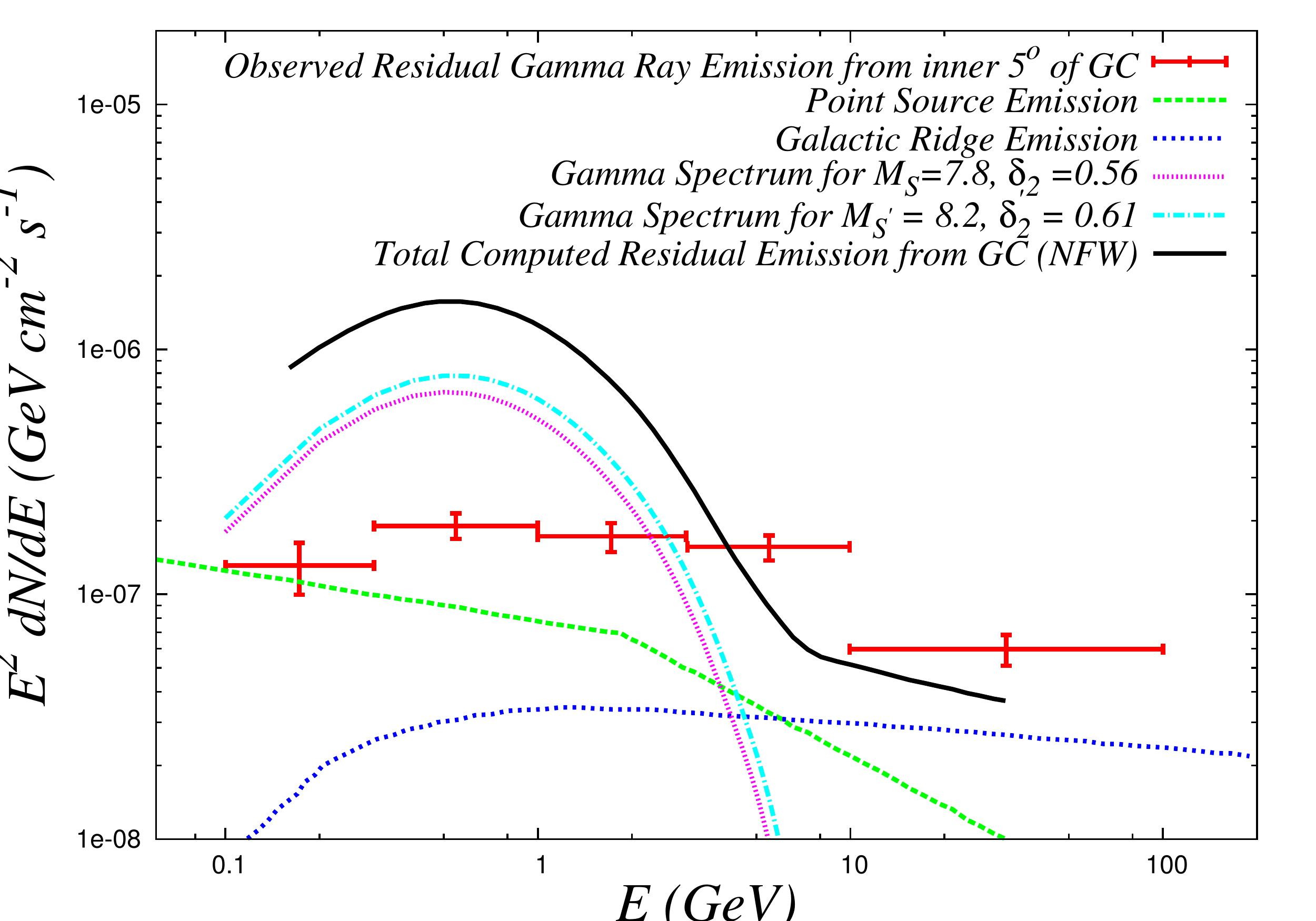}}
\subfigure[Einasto]{
\includegraphics[width=2.1in,height=2.9in, angle=-90]{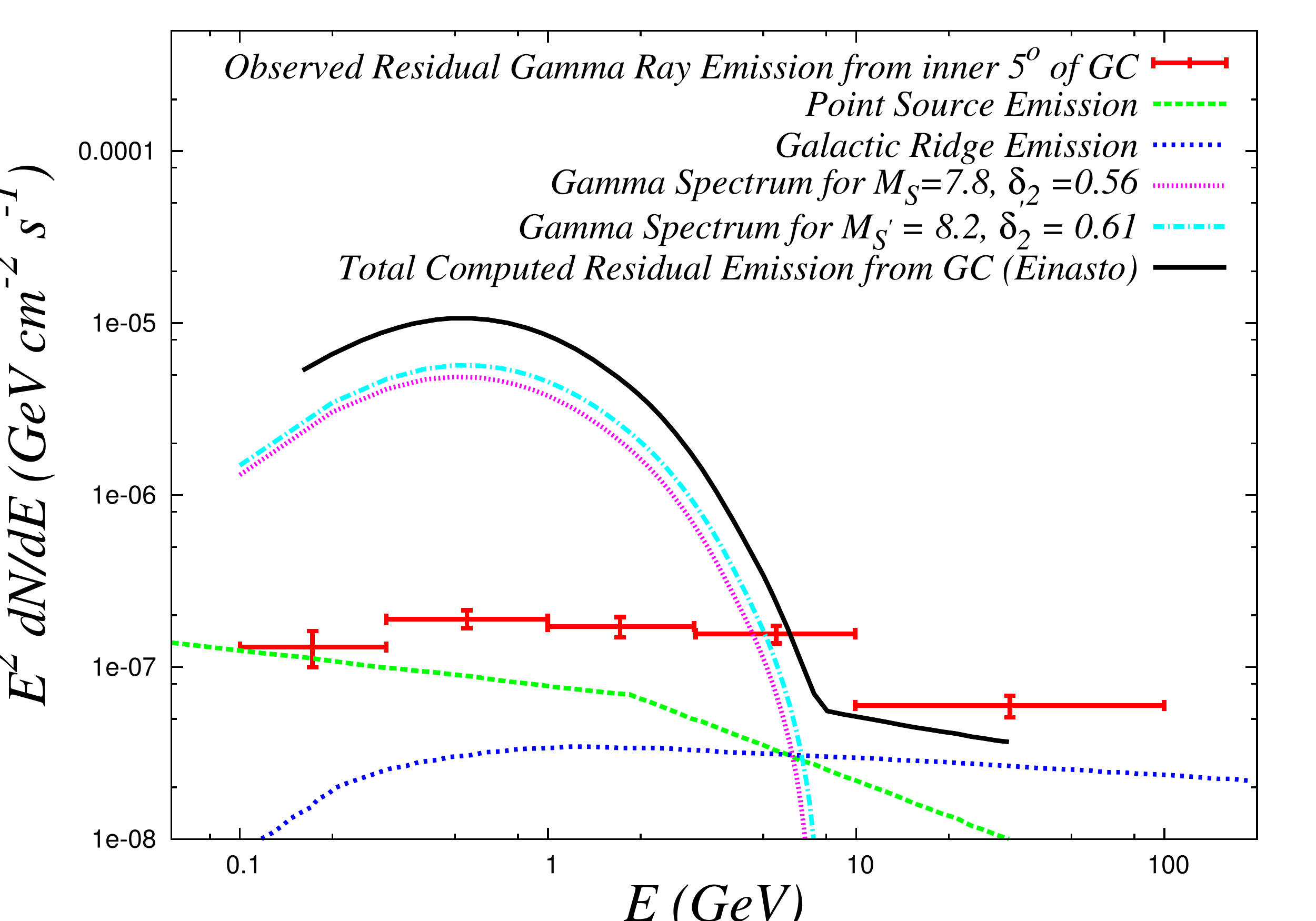}}
\subfigure[Moore]{
\includegraphics[width=2.1in,height=2.9in, angle=-90]{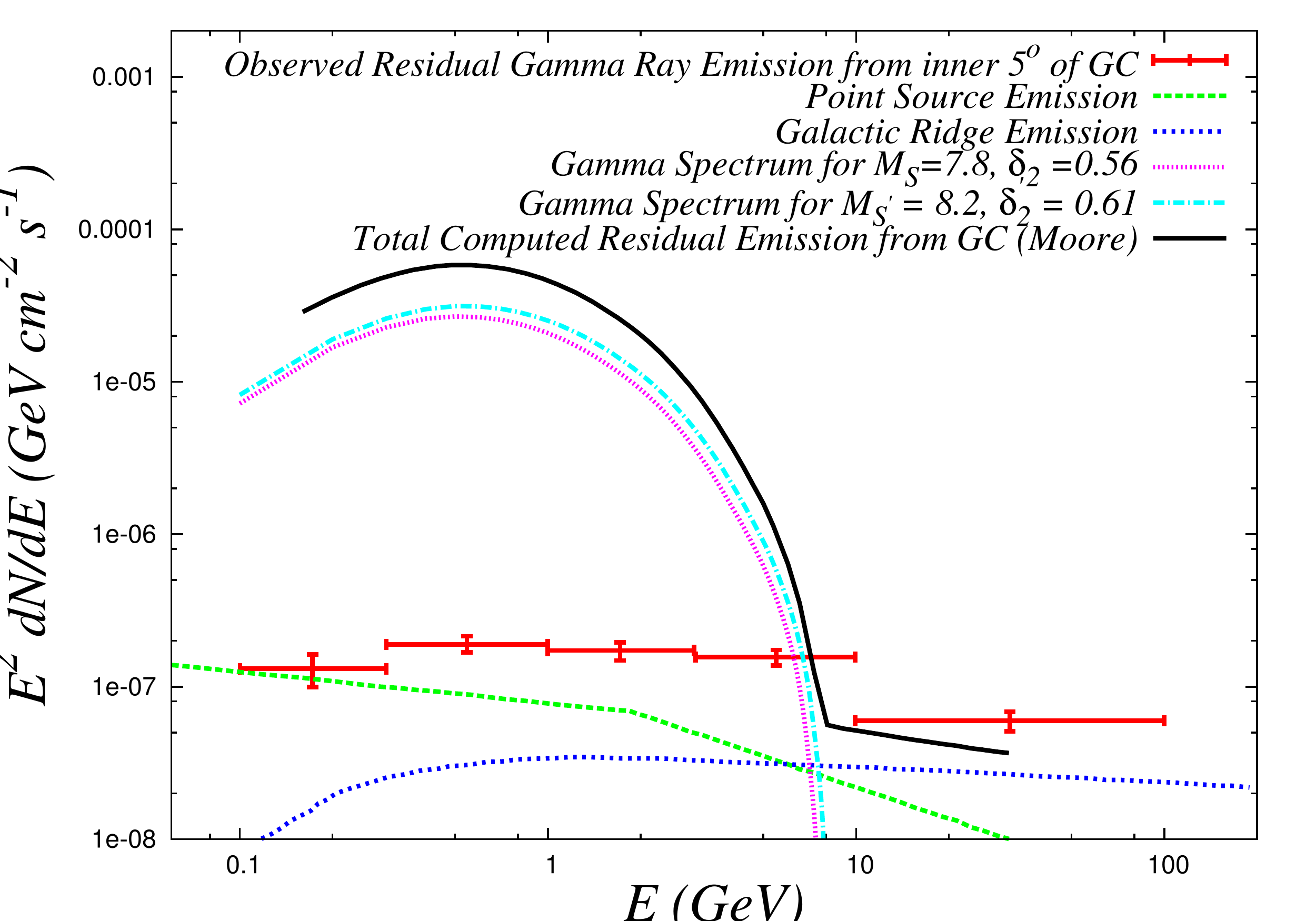}}
\caption{\label{fig:gammafromGC_cogent} \textit{Residual $\gamma$-ray flux from the inner 5$^o$ of
galactic centre.  DM annihilation is calculated for benchmark point 2 consistent with CoGeNT and Planck data (see Table~\ref{table_CoGeNT}). $SS$-annihilation is calculated for $M_S=7.8$~GeV and $\delta_2=0.56$.  For $S'S'$ annihilation we use $M_{S'}=8.2$~GeV and $\delta'_2=0.61$. Notations are same as in Fig.~\ref{fig:gammafromGC_cdms}.
} }
\end{center}
\end{figure}
\begin{figure}[h!]
\begin{center}
\subfigure[Isothermal]{
\includegraphics[width=2.1in,height=2.9in, angle=-90]{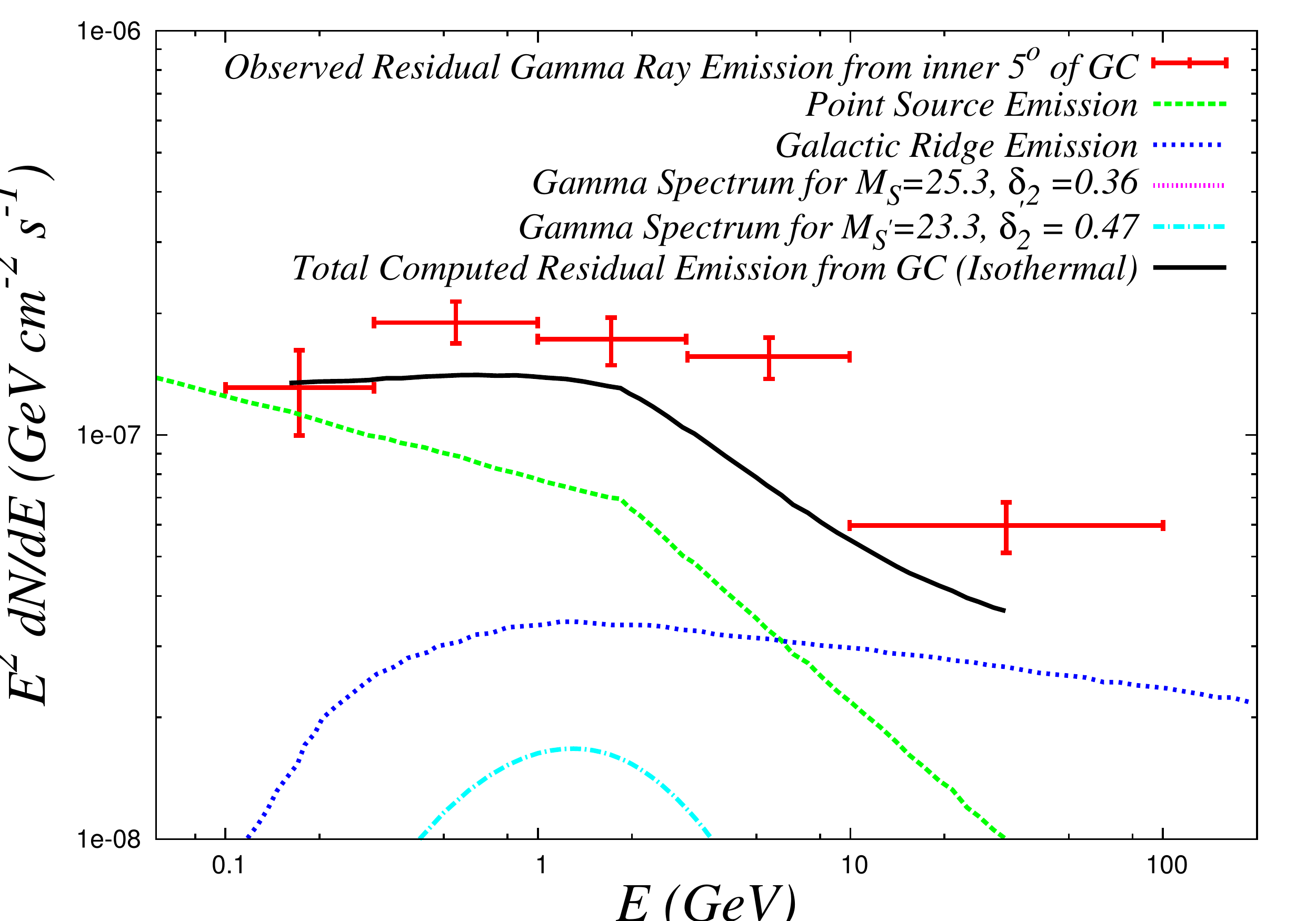}}
\subfigure[NFW]{
\includegraphics[width=2.1in,height=2.9in, angle=-90]{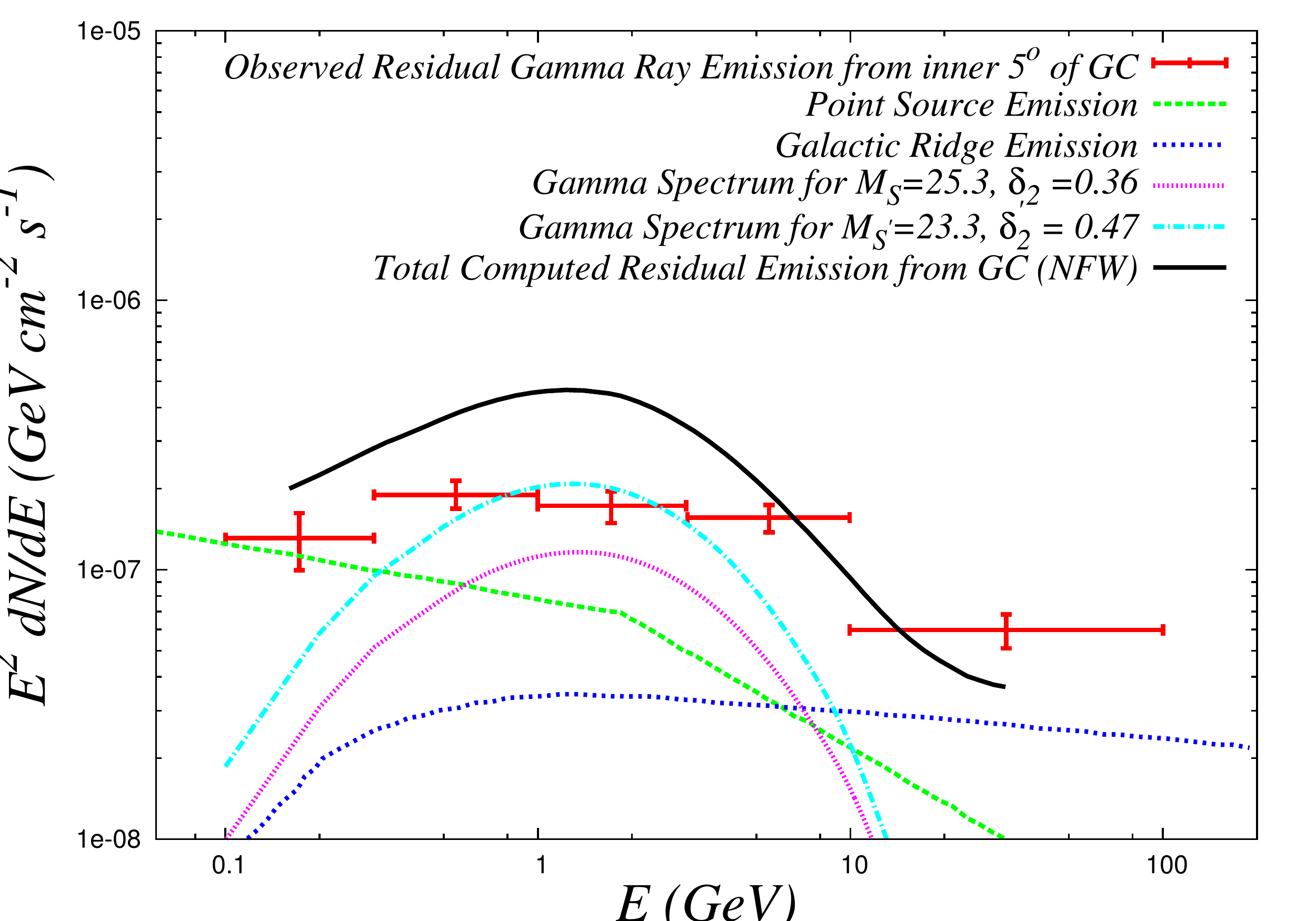}}
\subfigure[Einasto]{
\includegraphics[width=2.1in,height=2.9in, angle=-90]{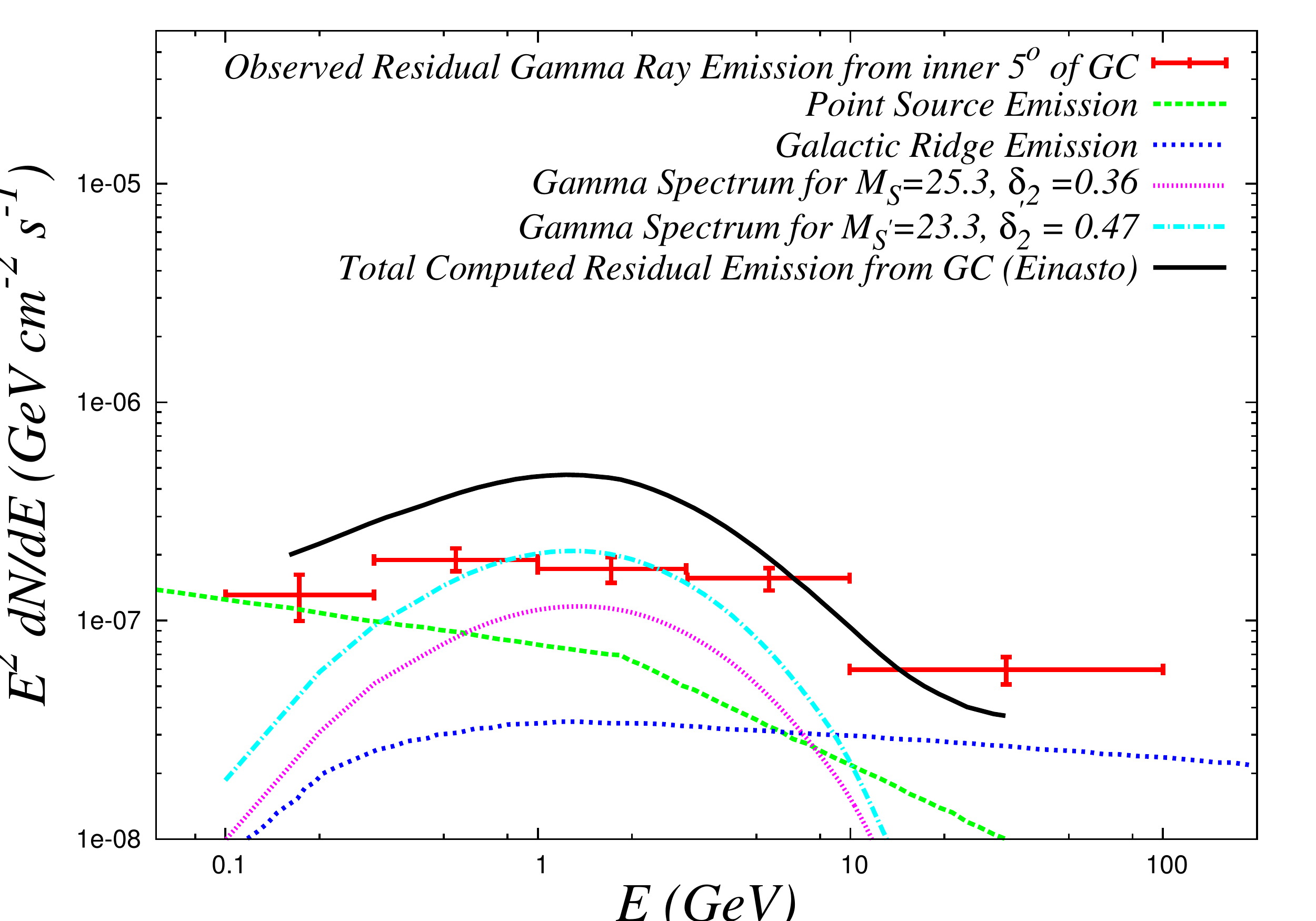}}
\subfigure[Moore]{
\includegraphics[width=2.1in,height=2.9in, angle=-90]{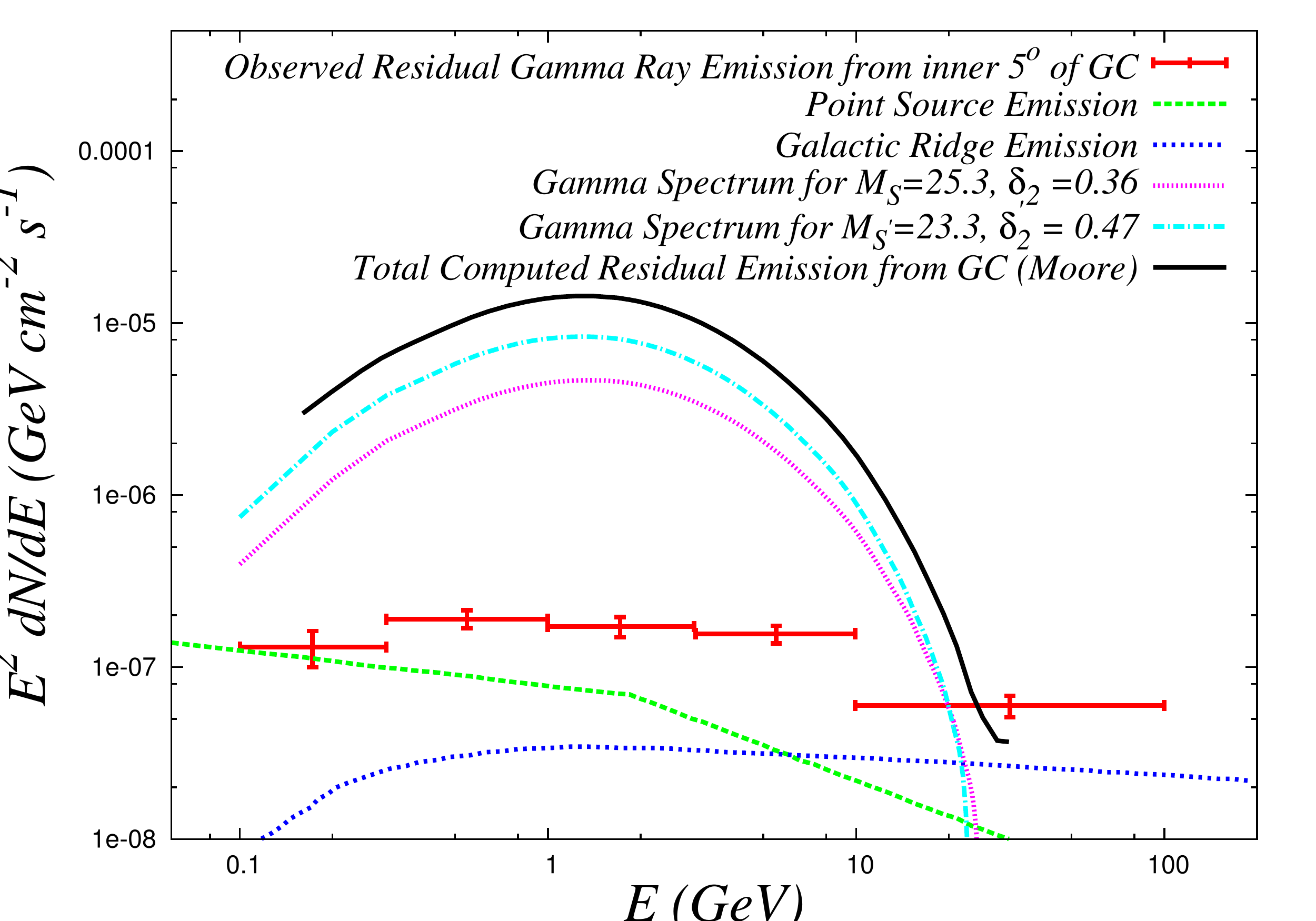}}
\caption{\label{fig:gammafromGC_cresst} \textit{Residual $\gamma$-ray flux from the inner 5$^o$ of
galactic centre.  DM annihilation is calculated for benchmark point 3 consistent with CRESST~II and Planck data (see Table~\ref{table_CRESST}). $SS$-annihilation is calculated for $M_S=25.3$~GeV and $\delta_2=0.36$.  For $S'S'$ annihilation we use $M_{S'}=23.3$~GeV and $\delta'_2=0.47$. Notations are same as in Fig.~\ref{fig:gammafromGC_cdms}.
} }
\end{center}
\end{figure}
\begin{figure}[h!]
\begin{center}
\subfigure[Isothermal]{
\includegraphics[width=2.1in,height=2.9in, angle=-90]{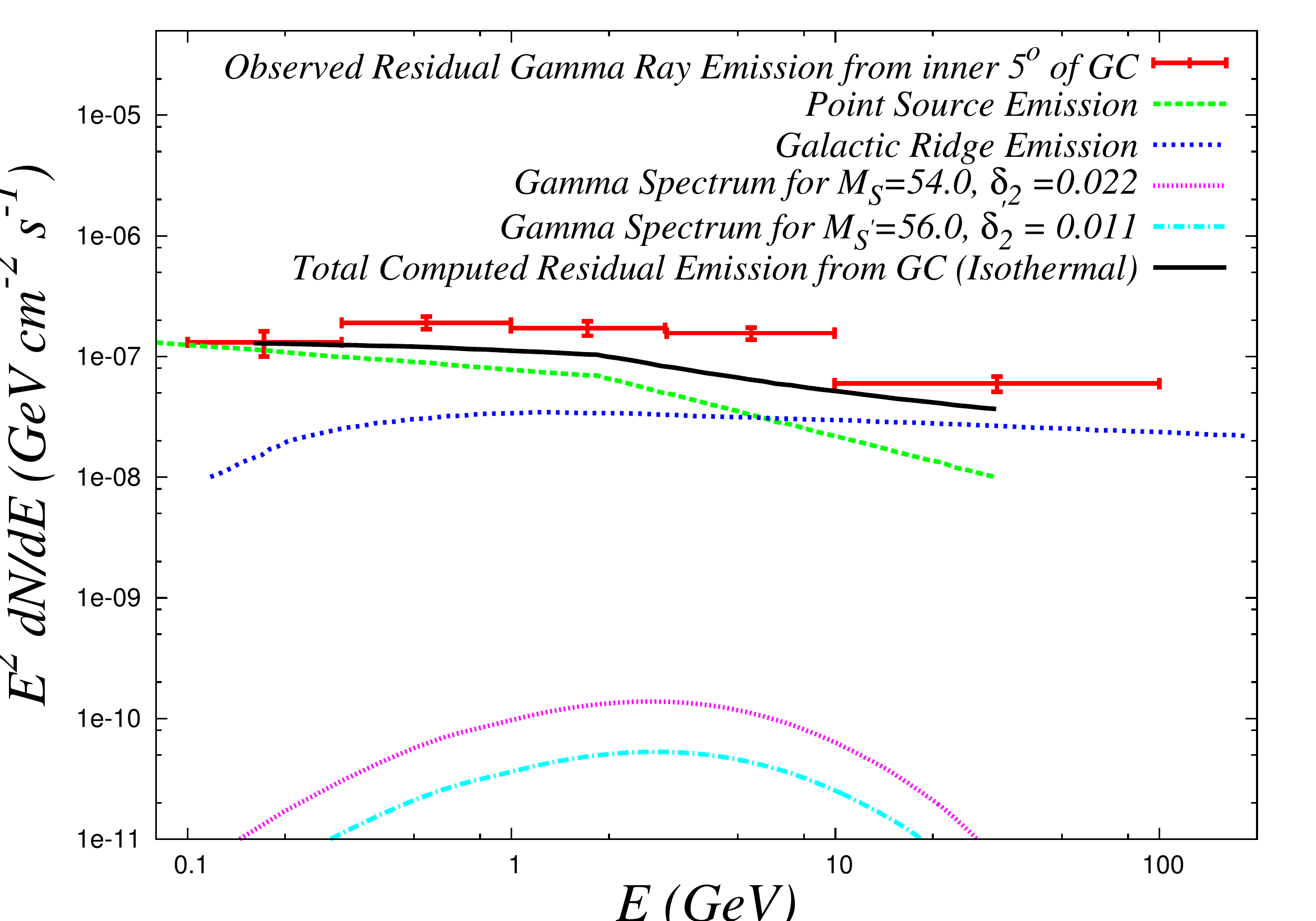}}
\subfigure[NFW]{
\includegraphics[width=2.1in,height=2.9in, angle=-90]{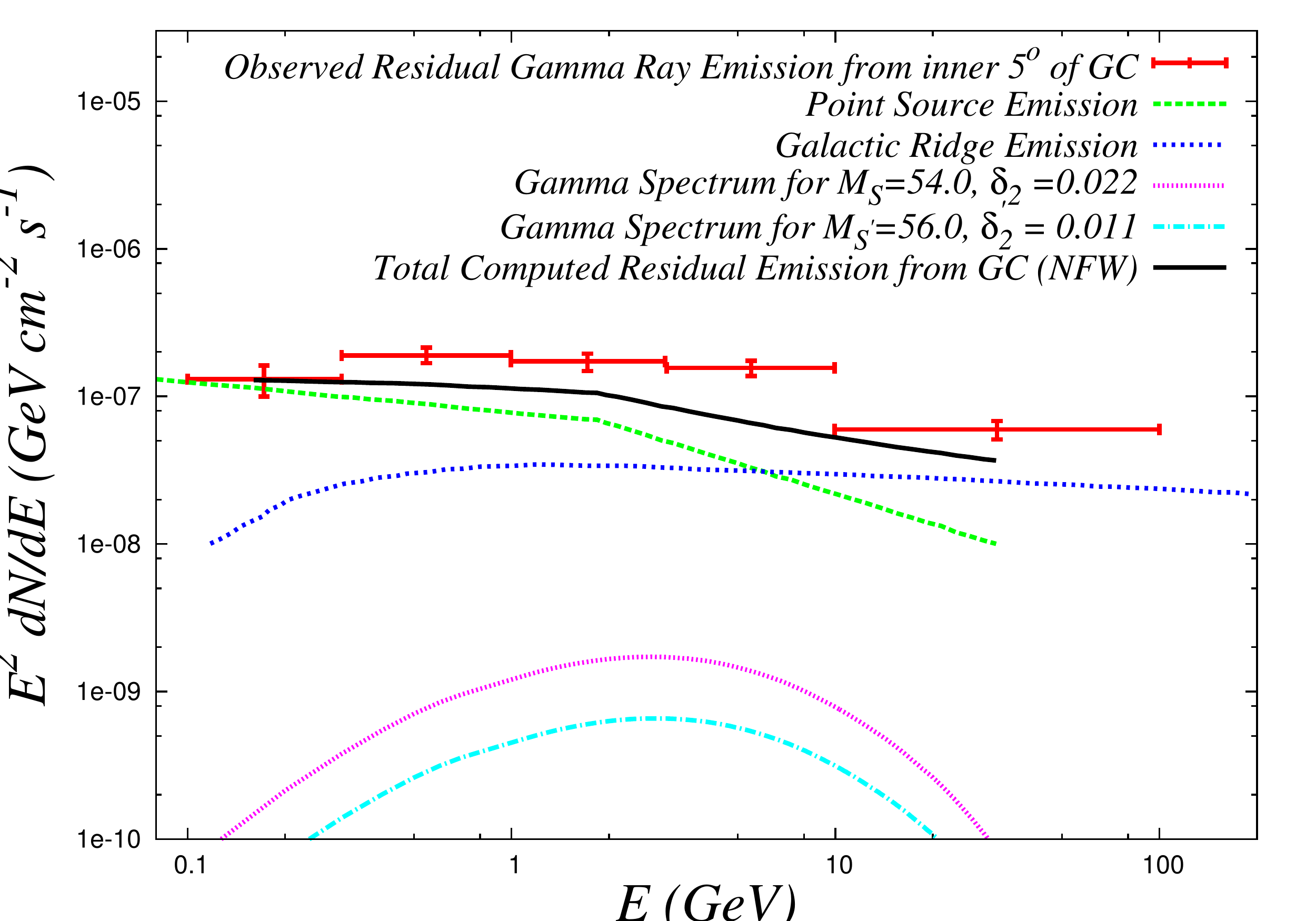}}
\subfigure[Einasto]{
\includegraphics[width=2.1in,height=2.9in, angle=-90]{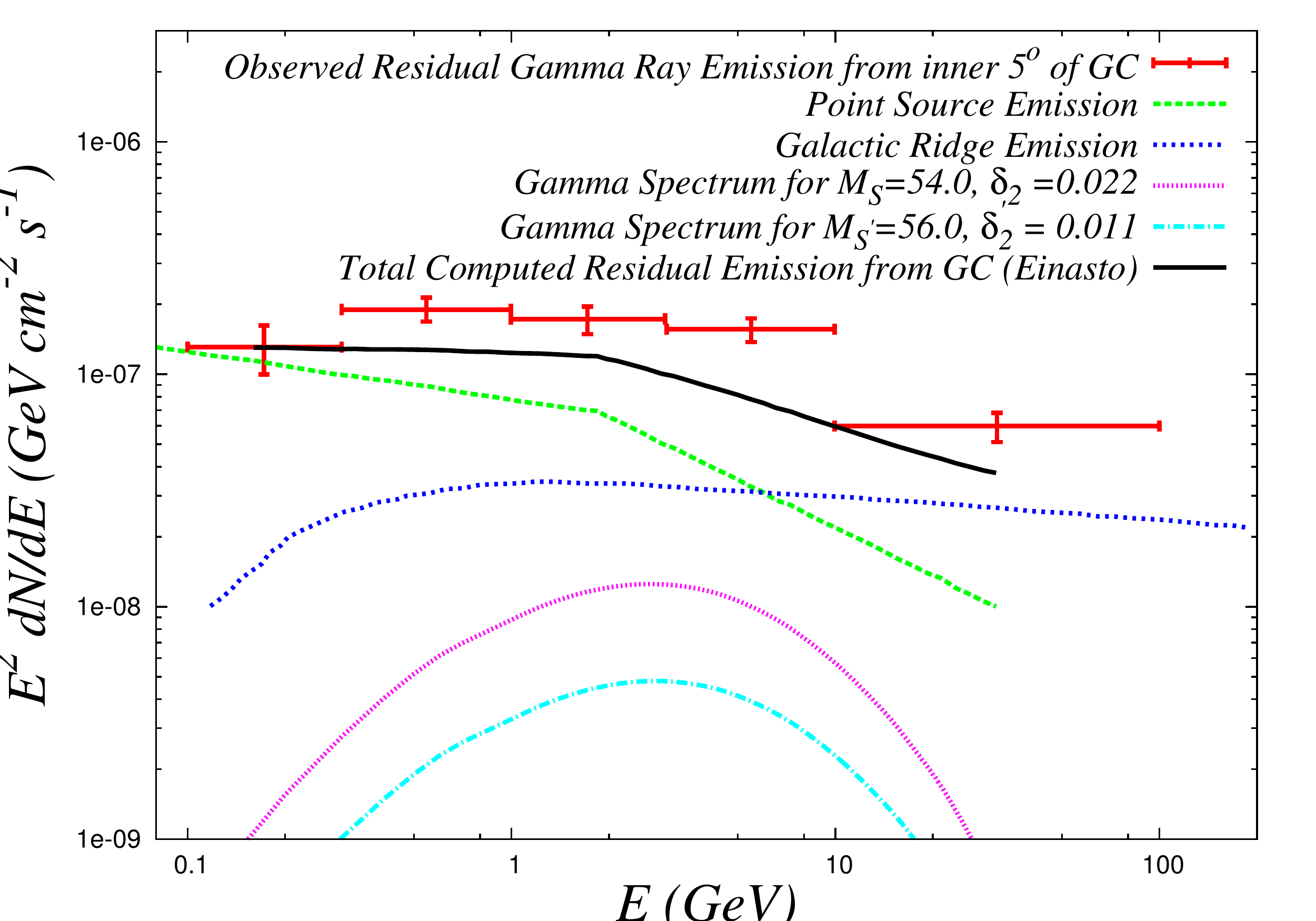}}
\subfigure[Moore]{
\includegraphics[width=2.1in,height=2.9in, angle=-90]{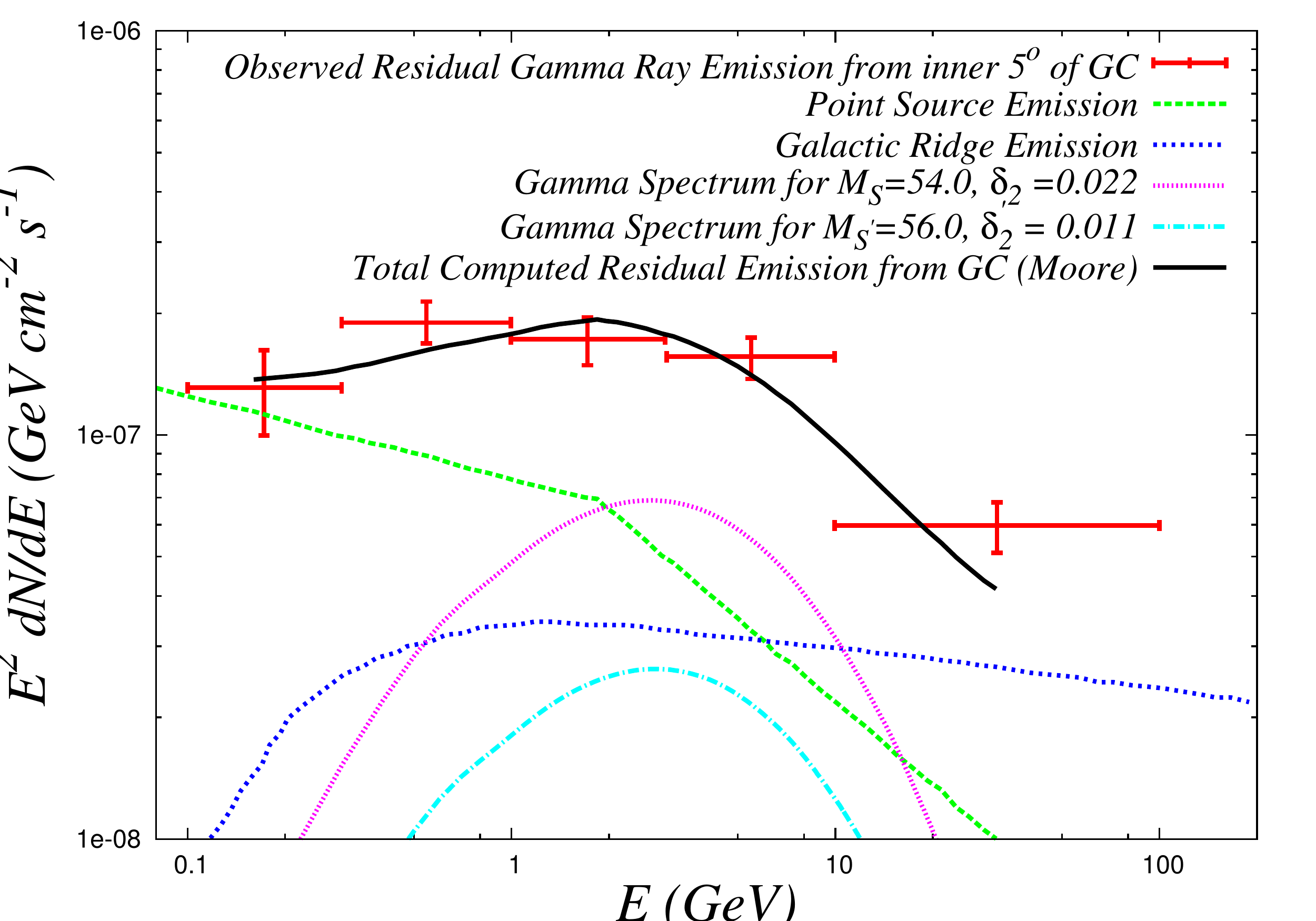}}
\caption{\label{fig:gammafromGC_xenon100A} \textit{Residual $\gamma$-ray flux from the inner 5$^o$ of
galactic centre.  DM annihilation is calculated for benchmark point 4A consistent with XENON~100 and Planck data (see Table~\ref{table_XENON100}). $SS$-annihilation is calculated for $M_S=54$~GeV and $\delta_2=0.022$.  For $S'S'$ annihilation we use $M_{S'}=56$~GeV and $\delta'_2=0.011$. Notations are same as in Fig.~\ref{fig:gammafromGC_cdms}.
} }
\end{center}
\end{figure}
\begin{figure}[h!]
\begin{center}
\subfigure[Isothermal]{
\includegraphics[width=2.1in,height=2.9in, angle=-90]{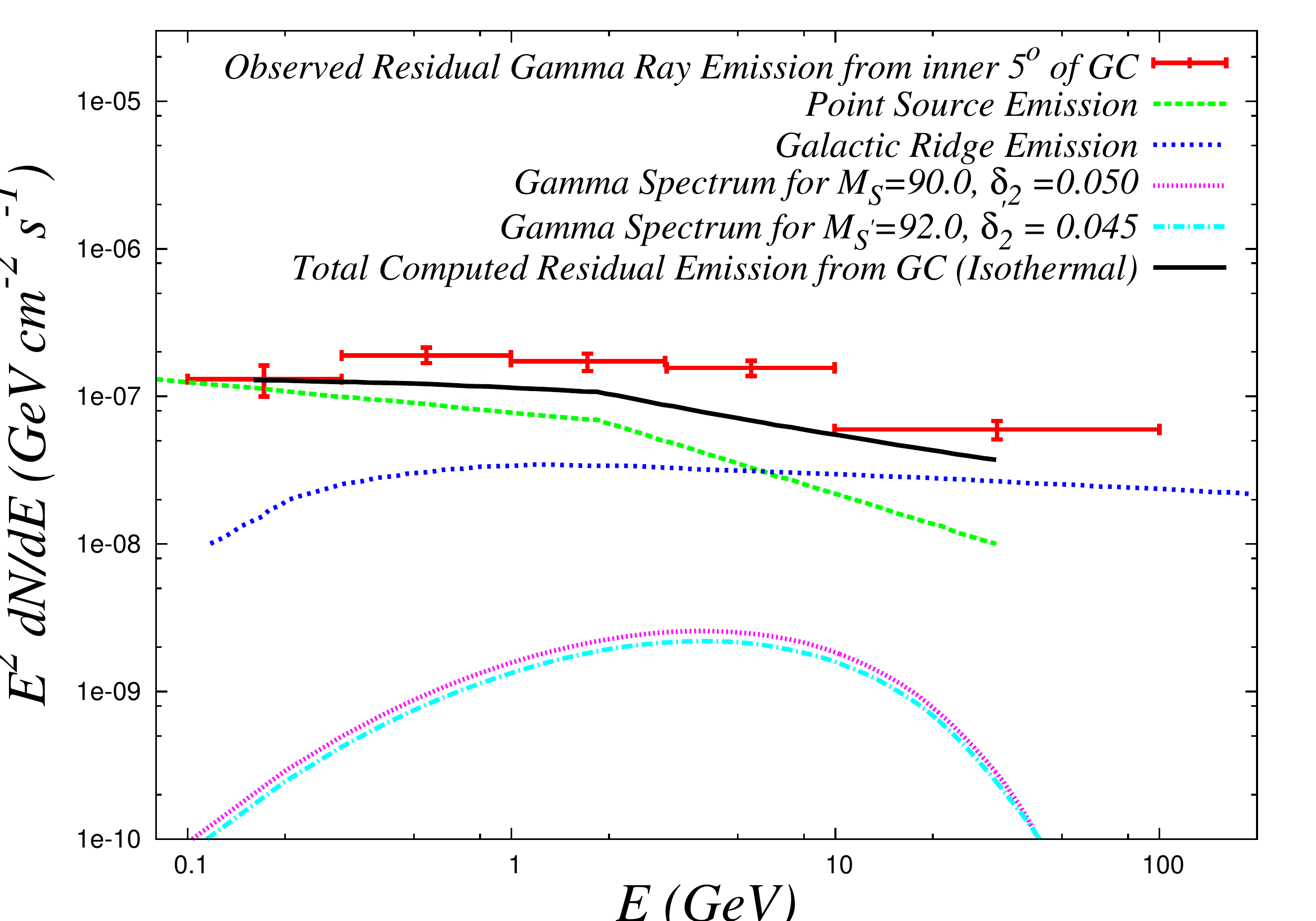}}
\subfigure[NFW]{
\includegraphics[width=2.1in,height=2.9in, angle=-90]{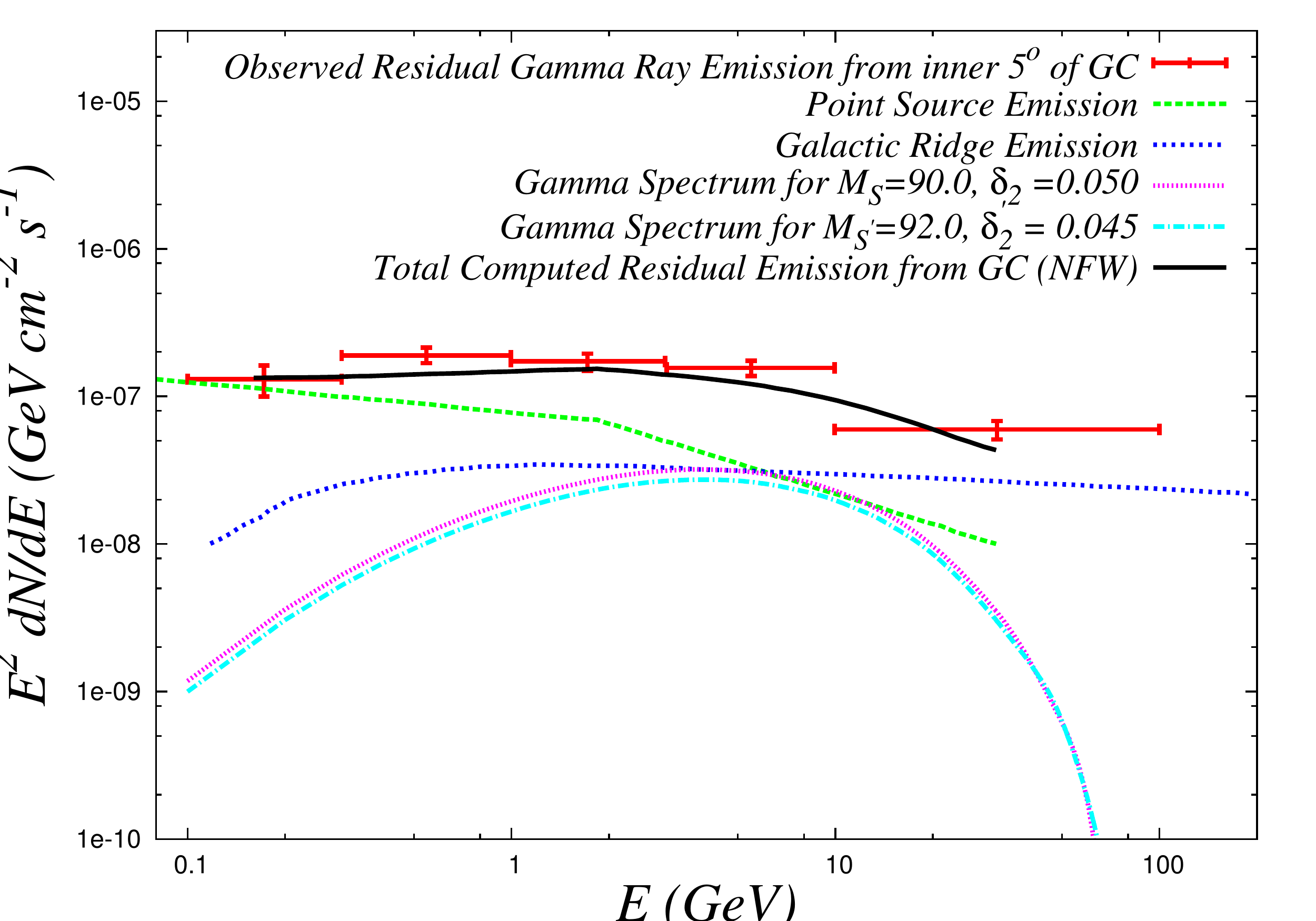}}
\subfigure[Einasto]{
\includegraphics[width=2.1in,height=2.9in, angle=-90]{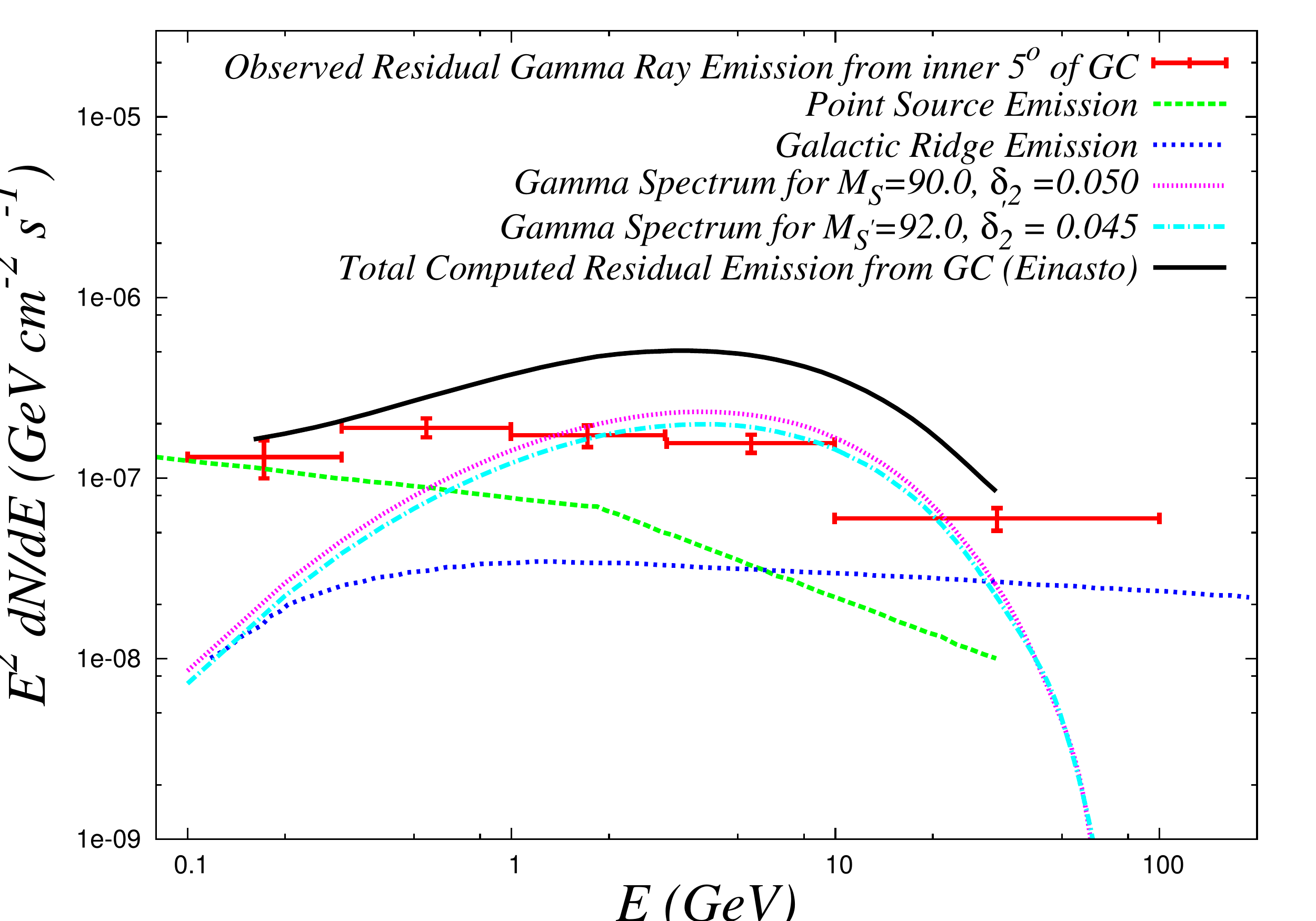}}
\subfigure[Moore]{
\includegraphics[width=2.1in,height=2.9in, angle=-90]{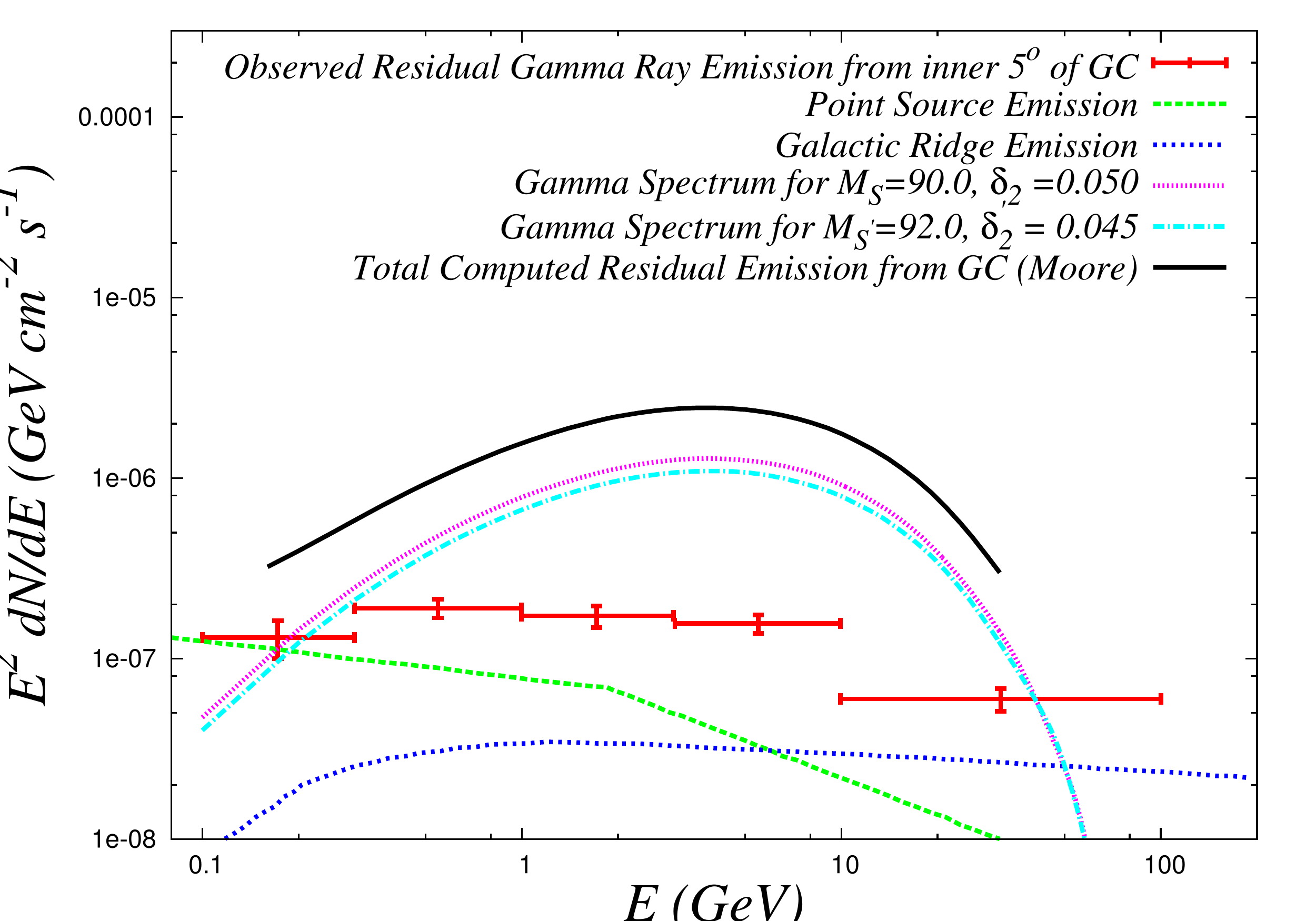}}
\caption{\label{fig:gammafromGC_xenon100B} \textit{Residual $\gamma$-ray flux from the inner 5$^o$ of
galactic centre.  DM annihilation is calculated for benchmark point 4B consistent with XENON~100 and Planck data (see Table~\ref{table_XENON100}). $SS$-annihilation is calculated for $M_S=90$~GeV and $\delta_2=0.05$.  For $S'S'$ annihilation we use $M_{S'}=92$~GeV and $\delta'_2=0.045$. Notations are same as in Fig.~\ref{fig:gammafromGC_cdms}.
} }
\end{center}
\end{figure}

 A potentially strong proposition about the nature of this bumpy
 feature of the residual gamma emission from GC is dark matter annihilation
 as indicated by Hooper \etal\cite{annihilating_dm, Hooper:2010im, Hooper:2012sr}.
 As in dark matter scenario, the angular width of the spectra is narrow since
 the astrophysical factor for flux calculation contains $\rho^2(r)$ which
 falls off very rapidly with radial distance from GC explaining the ``bump''. It also resolves the problem
 posed from pulsar explanation.

 In Ref.~\cite{annihilating_dm}
 it has been argued that by considering few annihilating dark matter scenarios
 with some standard dark matter halo profiles,
 low mass dark matter can fit the spectrum with good statistics. Few
 benchmark cases, such as 10 GeV dark matter
 annihilating to leptonic channels \cite{Hooper:2012ft} or $30$~GeV dark matter annihilating
 to $b\bar{b}$ channel with NFW halo profile have been shown to fit data~\cite{annihilating_dm}. In order to get an idea of where our specific model fits in such discussion of generic models, in Tables~\ref{table_CDMS}--\ref{table_XENON100} we have quoted branching ratios for different DM annihilation channels for different benchmark points. We see that although for $10-55$~GeV DM the $b\bar{b}$ channel has a branching ratio $\sim 80\%$, but for higher masses, when the $W^+W^-$ or $ZZ$ channel opens up, it drastically changes. However such a DM candidate is also compatible with data.

 In this section we extend the above discussion for the multi-component DM
 scenario as discussed in our model. As mentioned in the abstract, presence of more
 than one DM candidate helps enhance the total $\gamma$-ray emission due to DM annihilation.
 We work with benchmark points chosen from the model parameter space already constrained
 by direct detection experiments and Planck survey. For each such benchmark points we try to match
 the observed spectra from the theoretically calculated ones. We plot the emission from point sources and galactic ridge from 
 Ref.~\cite{annihilating_dm}. Then we add $SS$ and $S'S'$ annihilation spectra to get theoretically predicted residual flux for four DM halo profiles arranged in increasing order of ``cuspiness'': (1)~Isothermal~\cite{iso}, (2)~NFW~\cite{nfw}, (3)~Einesto~\cite{einasto} and (4)~Moore~\cite{moore}.

 We see that for low mass DM, the plots Fig.~\ref{fig:gammafromGC_cdms} and Fig.~\ref{fig:gammafromGC_cogent} corresponding to benchmark points 1 and 2 respectively, indicate that a flat DM halo profile like Isothermal profile offers a better agreement with the data. For benchmark point 3, Fig.~\ref{fig:gammafromGC_cresst} shows that Isothermal profile is still the promising one, whereas Moore profile overestimates the data. The XENON~100 benchmark point 4A is used for Fig.~\ref{fig:gammafromGC_xenon100A}, where we see that for DM masses $\sim$ 55~GeV, all DM profiles other than the cuspy Moore profile, the DM annihilation contribution is rather small compared to contributions from point sources and galactic ridge. NFW profile works better for XENON~100 benchmark point 4B, used for Fig.~\ref{fig:gammafromGC_xenon100B}.

\subsection[]{Explanation of Gamma Ray Bump from Fermi Bubble's Low Galactic Latitude}\label{ss:bubble}

 From the Fermi Gamma-Ray Space Telescope (FGST) data a pair of bilateral lobular structure
 that contain large amount of gamma-ray had been found
 in the upper and lower regions of galactic centre. These lobes,
 known as Fermi Bubbles, emit $\gamma$ rays between $\sim$ few GeV to
 $\sim$ 100 GeV range and they are extended almost $\sim 50^o$ ($r = \pm10$ kpc)
 up and down from the galactic plane.
 In Ref.~\cite{dobler}, the bubble emission has been studied as
 an extension of WMAP haze~\cite{wmap23_33} which is the non-thermal, microwave
 emission from the inner part of the galaxy confirmed from data of different
 ongoing experiments worldwide such as Planck~\cite{planck_haze} and ROSAT~\cite{rosat}
 X-ray emission data. Evidences~\cite{planck_haze} show that near the galactic
 plane, the $\gamma$-ray bubbles and the haze can have a strong correlation
 that attribute to the fact that they might have been a common origin.
 When we move far from the galactic plane along the Fermi bubble the
 gamma ray spectrum follows a power law, $E^{-\alpha}$ with
 spectral index, $\alpha$ = 2 over all the energy range observed by the FGST.
 This type of gamma ray spectrum can be well explained by approximate power
 law spectrum of electron distribution with spectral index, 3, i.e., $E^{-3}$
 where the inverse Compton scattering (ICS) is the mechanism of production
 of these types of gamma rays. Also, similar distribution can produce
 radio emission in the galaxy~\cite{dobler, mertsch} due to the synchrotron radiation
 effect with the interaction of microgauss galactic magnetic field.

\begin{figure}[h!]
\begin{center}
\subfigure[Benchmark point 1]{
\includegraphics[width=2.1in,height=2.9in, angle=-90]{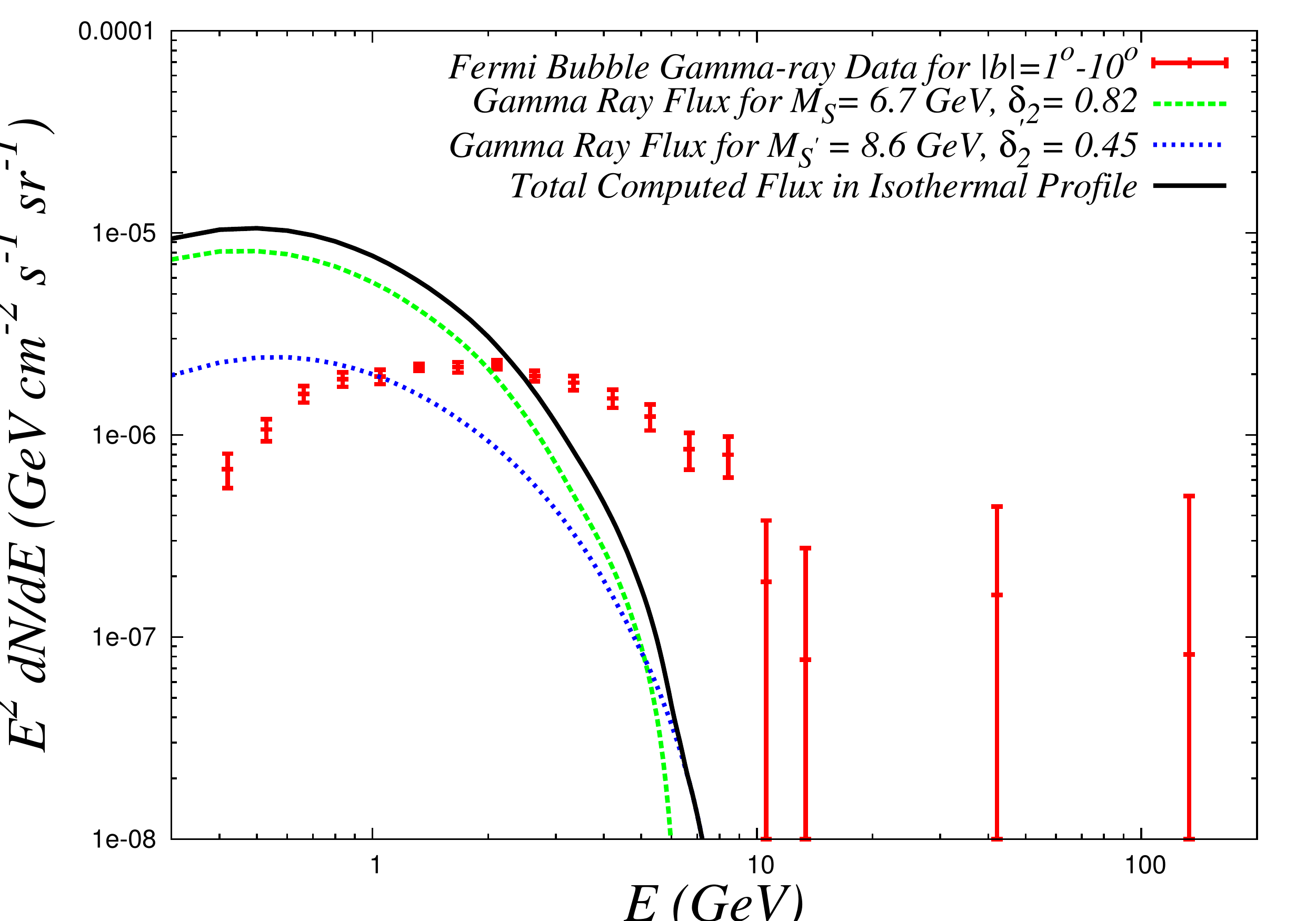}}
\subfigure[Benchmark point 2]{
\includegraphics[width=2.1in,height=2.9in, angle=-90]{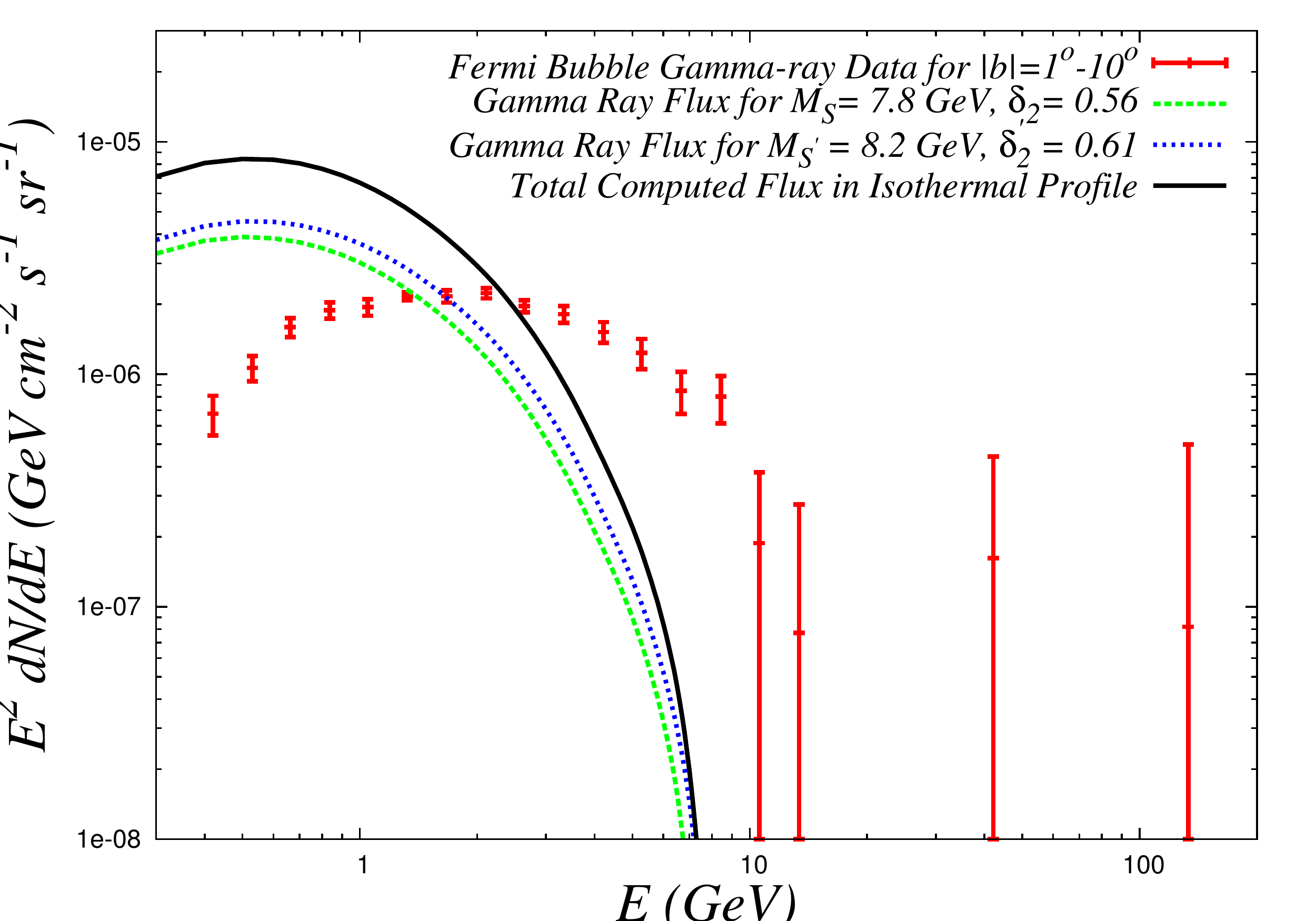}}
\subfigure[Benchmark point 3]{
\includegraphics[width=2.1in,height=2.9in, angle=-90]{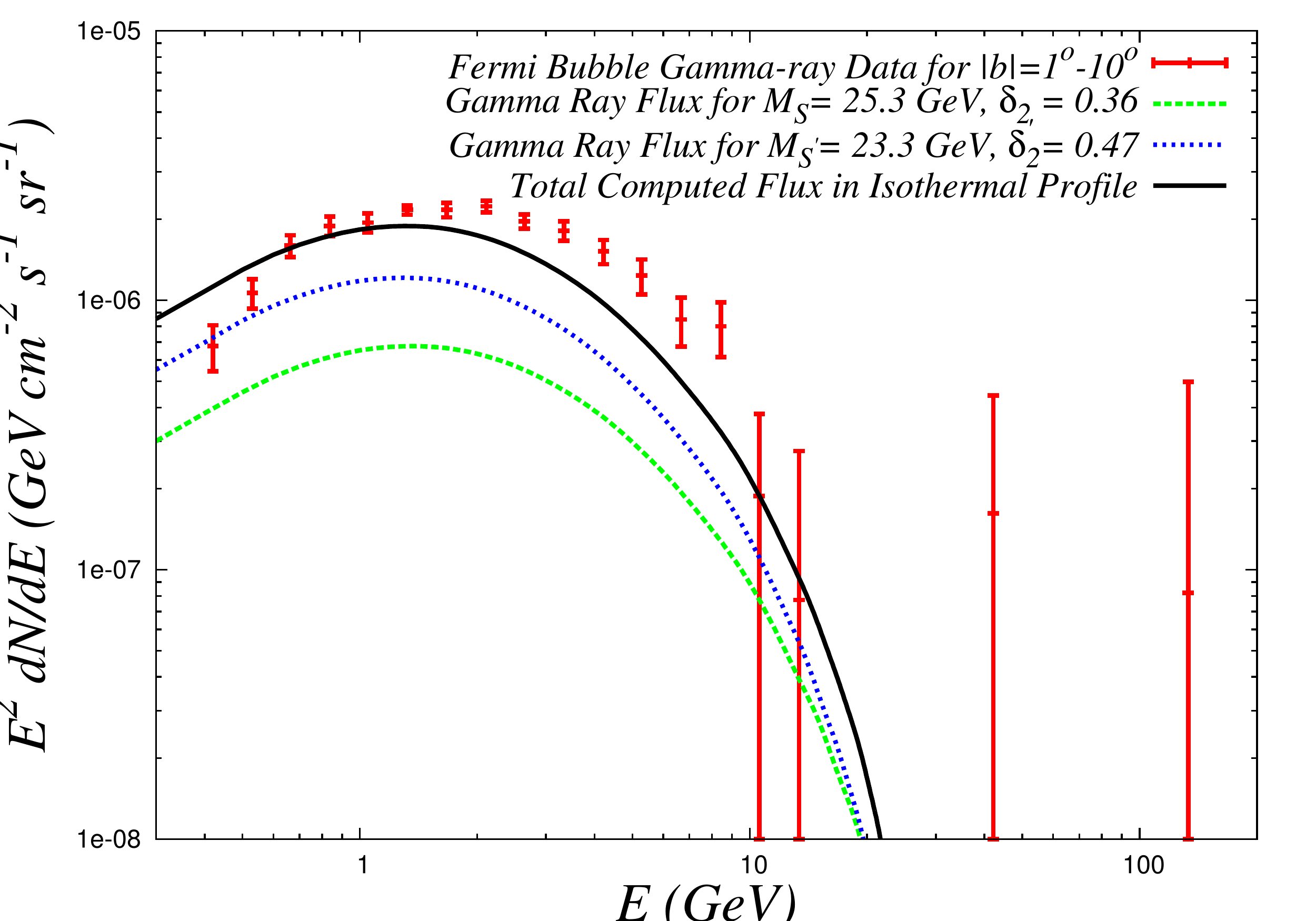}}\\
\subfigure[Benchmark point 4A]{
\includegraphics[width=2.1in,height=2.9in, angle=-90]{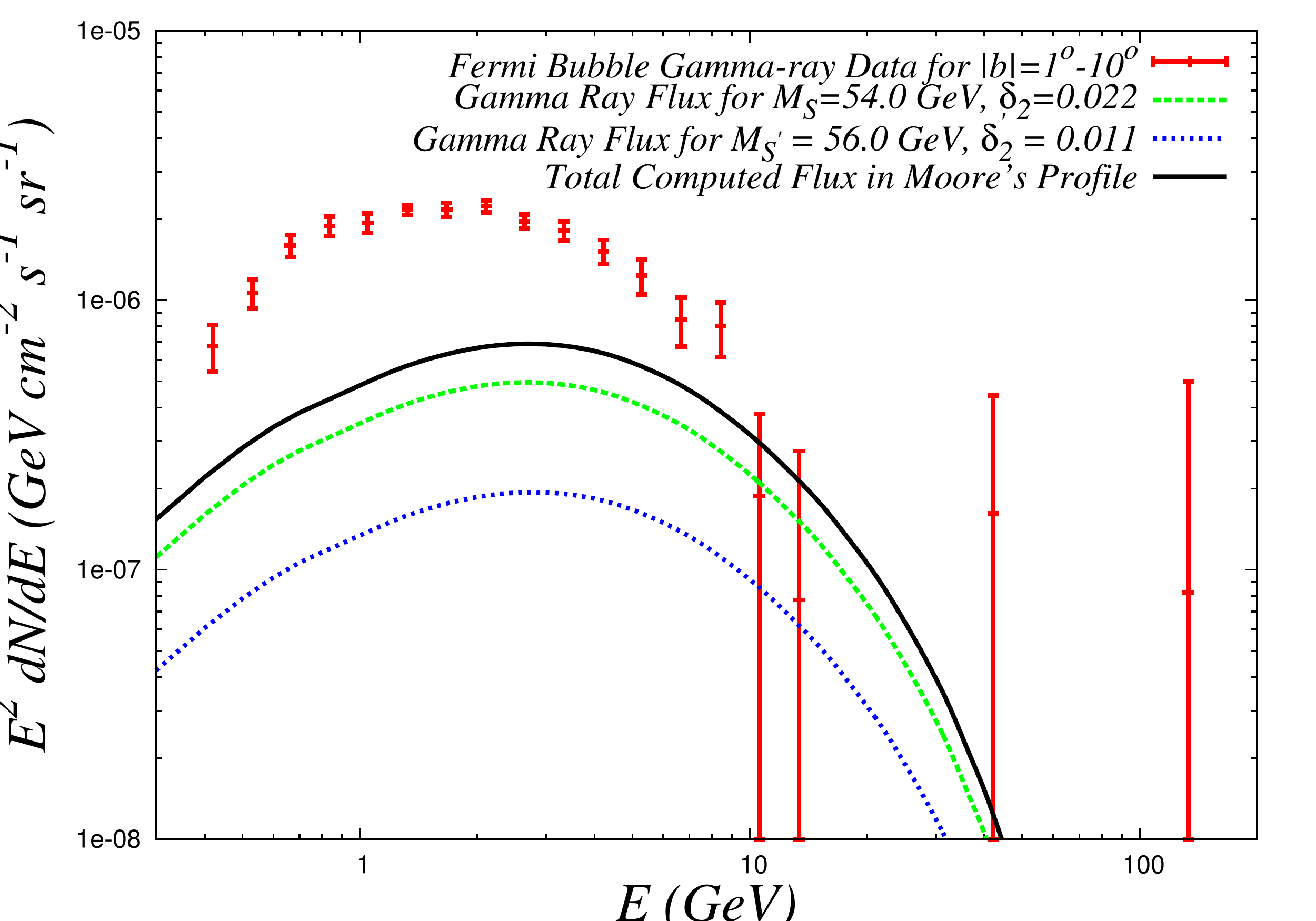}}
\subfigure[Benchmark point 4B]{
\includegraphics[width=2.1in,height=2.9in, angle=-90]{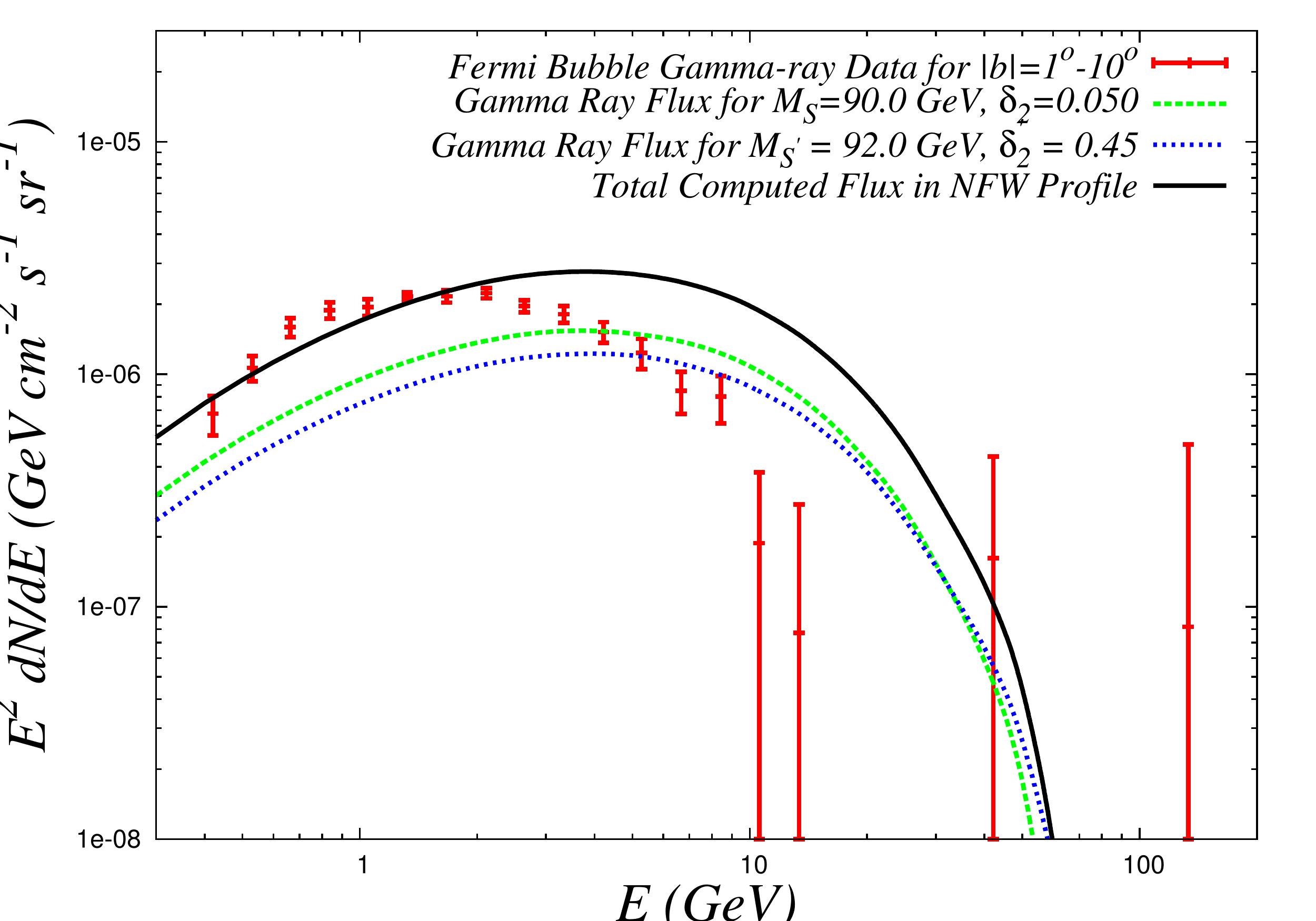}}
\caption{\label{fig:gammafromFB_1_10} \textit{
$\gamma$-ray emission spectrum from the Fermi Bubble's low-latitude ($|b|$=1$^o$-10$^o$) region.
The red points denote observed data after subtracting the ICS contribution. The green dashed line denote contribution from $SS$ annihilation. $S'S'$ annihilation is represented by the blue dotted line. The total DM annihilation contribution is shown by the solid black line. Each sub-figure is plotted for a different benchmark point with a DM halo profile which explains GC low energy $\gamma$-ray bump the best.
} }
\end{center}
\end{figure}

  The picture of the gamma ray emission from Fermi Bubble from the
  low galactic latitude is somewhat different. In this case the $\gamma$-spectrum has a peak
  at a few GeV energy range. This cannot be explained by the known
  astrophysical processes like inverse Compton
  scattering of light source
  by cosmic electrons in steady-state.
  This non-ICS nature of the spectrum has generated some interest in astrophysics
  community and different origins for this bumpy nature of $\gamma$-ray spectrum
  from bubble very near the galactic plane have been proposed. Population of
  millisecond pulsars \cite{millisecond_pulsar}, cosmic ray interaction with
  gas \cite{cosmic_gas} or an annihilating dark matter scenario
  \cite{hooperplb,annihilating_dm,Anchordoqui:2013pta,Hagiwara:2013qya} have been studied in great details.
  A detailed study on the morphology and spectral signature of Fermi Bubble is given in Hooper \etal \cite{hooperfermibubble}.

The explanation of this low latitude excess $\gamma$-ray emission as given
by the {\it diffuse emission mechanism} is due to the fact that the
cosmic ray protons are scattered with the gas present in the Milky Way region.
But the explanation cannot fully provide the observed phenomenology as the
gas distribution is merely correlated with the morphological structure of the
$\gamma$-emission and also the spectrum of the cosmic ray protons should follow 
a bumpy nature around $\sim$ 25 GeV or so to provide a good description of 
the observed emission. But this type of peak in the cosmic proton spectral
feature is not fully understood from the known astrophysical observation.

Another possibility of this $\gamma$-ray excess can be attributed to the
excess population of {\it millisecond pulsars} which have the advantage of 
producing $\gamma$-ray emission with very high luminosity over another 
types of pulsars. But the nature of $\gamma$-ray spectrum 
generated by such objects is not very well understood 
as a very few pulsars of this type have been discovered. 
Also the distribution of such objects
outside the galactic plane as proposed is much more constrained from various
astrophysical observations.

On the other hand the spectral nature of excess gamma
emission from the lower latitude of Fermi Bubble may be consistent with
the gamma spectrum calculated from the {\it annihilating dark matter} scenario at the
galactic halo region.

As the DM density is expected to be high for regions close to the GC, we concentrate on DM annihilation from low latitude $|\,b\,|=1^o$\,--\,10$^o$ zone of Fermi Bubble. Like we did for GC, we work with benchmark points consistent with direct detection experiments and Planck survey. However, rather than exploring for all DM halo profiles, we present the plot for that DM profile for which the GC data was better explained. These plots are presented in Fig.~\ref{fig:gammafromFB_1_10}. Here the observed flux is shown after deducting inverse Compton scattering contribution of best-fit steady state electron spectrum of the Bubble. We see that for very low DM mass $\sim 7-11$~GeV, as preferred by CDMS~II or CoGeNT, the spectrum peaks at a lower energy than that obtained from the data. The higher mass zone  $\sim 25$~GeV, as preferred by CRESST~II or $\sim 55$~GeV, as allowed by XENON~100 works better. For very high DM masses $\sim 90$~GeV allowed by XENON~100, the calculated spectra tend to peak at a bit higher energy than the observed spectrum. But overall we can conclude that the model holds some promise to explain the  morphological feature of the Fermi Bubble low latitude $\gamma$-ray excess.

\vskip 1cm
\section{Discussion and Conclusion}\label{S:conc}
We have explored a DM model adding two real scalar gauge singlets to the SM. The stability is ensured by a $\Z_2\times\Z_2'$ symmetry. We keep the symmetry unbroken to get a two component DM scenario. The annihilation of the heavier DM to the lighter ones is suppressed by considering a scenario where the DM candidates $S$ and $S'$ are almost degenerate in mass.  Such a two component real scalar DM model is better suited to explain both direct and indirect DM experiments compared to the DM models containing a single real scalar.

Detailed calculations for the vacuum stability, perturbative unitarity and triviality constraints on the model has been presented, which forms an integral part of the paper. The parameter space used to explain experimental results do respect these constraints. The model however brings forth the traits of any higgs-portal DM model. For low DM masses, the model predicts unacceptably high invisible Higgs decay width, which calls for adding lighter degrees of freedom to the model. In the present work we did not take up  that exercise as we feel that till the conflict of CDMS~II or CoGeNT observations with the XENON~100 or LUX observations are settled there is no pressing argument to believe that $7-11$~GeV DM do exist. So we considered 126~GeV Higgs mediated DM annihilation processes  throughout to reproduce indirect DM experimental results.

Guided by the direct detection experiments we considered three DM mass zones. The ``low" zone of $7-11$~GeV is indicated by CDMS~II, CoGeNT and CRESST~II experiments. CRESST~II also favours a ``mid" zone $\sim 25$~GeV. As XENON~100 and LUX seem to rule out these zones, the only DM masses consistent with both XENON~100 or LUX and Planck observations belong to a ``high" mass zone $> 50$~GeV. The advantage of dealing with this zone is that they do not give rise to unacceptable invisible branching ratio for Higgs. But a too high DM mass $> 100$~GeV predicts a photon flux from DM annihilations peaked at higher energies than what has been observed in the indirect detection experiments. This high DM mass zone will be probed by future XENON~1T~\cite{Aprile:2012zx} and LUX measurements. 

We have chosen some representative ``benchmark points" from the parameter space allowed by the direct detection experiments and Planck data. Now the question is how robust are they? DM annihilation cross-sections do depend on $M_S$ (or $M_{S'}$) and $\delta_2$ (or $\delta '_{2}$). But as it is proportional to $\delta_2^2$ (or ${\delta '}_2^2$), it is quite sensitive to the choice of  $\delta_2$ (or $\delta '_{2}$). For this reason we choose to show the allowed zones in the plots, so that the reader can roughly estimate the changes in the photon flux in the indirect detection experiments when we choose different benchmark points within the allowed parameter space. 

There is some advantage of addressing both direct and indirect detection experiments.  The allowed model parameter space is rather restricted by the direct DM detection experiments and relic density constraints as imposed by Planck. This makes indirect DM detection predictions quite sensitive to the assumed DM halo profiles. We would like to make a point that once some agreement in the direct DM sector is established and the background effects in the indirect detection experiments are better understood to delineate DM annihilation effects, in the framework of a given model, the experiments with the existing precision show some promise to identify the right DM halo profile. We have illustrated this with our proposed DM model. 

In conclusion, with the proposed model we do not intend to show that it can explain all the experimental results, which sometimes are contradictory in nature. Rather, in the framework of the model, we wanted to exploit the advantage of having a multi-component DM model satisfying both direct and indirect DM experiments, and in this process comment on the viability to choose the right DM halo profile. For completeness we also presented detailed calculations for theoretical constraints on this model as mentioned earlier.

\vskip 1cm
\noindent{\bf Acknowledgments:} SR acknowledges support of seed grant from IIT Indore.

\appendix
\appendixpage

\begin{appendices}
\section{Direct Detection Cross-section}\label{App-A}
The spin-independent singlet scalar -- nucleus elastic scattering cross-section in
the non-relativistic limit can be written as ~\cite{Burgess:2000yq}
\begin{eqnarray}
\sigma^{\rm SI}_{\rm nucleus}
&=&
\frac{\delta_2^2 v^2 |{\cal A}_N|^2}{4\pi}
\left( \frac{\mu^2_r}{{M_S}^2{M_H}^4}\right)
\label{eqcross1}
\end{eqnarray}
where, $\mu_r (N,S)= M_N M_S/(M_N + M_S)$ denote the reduced mass
for the system of singlet scalar and target nucleus
with individual masses $M_S$ and $M_N$ 
respectively.
${\cal A}_N$ represents the relevant
matrix element. The singlet scalar--nucleus
and singlet scalar--nucleon
elastic scattering cross-sections for the non-relativistic limit
are related as~\cite{Burgess:2000yq}
\begin{eqnarray}
\sigma^{\rm SI}_{\rm nucleus}
&=& \frac{A^2 \mu^2_r({\rm nucleus},S)}{ \mu^2_r({\rm nucleon},S)}
\sigma^{\rm SI}_{\rm nucleon}
\label{eqcross}
\end{eqnarray}
where $A$ is the atomic number of the nucleus.
$\sigma_{\rm nucleon}^{\rm SI}$ can be expressed as,
\begin{equation}
 \sigma_{p(n)}^{\rm SI}=\frac{4 m_{p(n)}^{2} M_{S}^{2}}{\pi\left(M_{S}+m_{p(n)}\right)^{2}}
    \left[f^{p(n)}\right]^{2}, \label{cs_nucleon}
\end{equation}
 where the expression for hadronic matrix element, $f_{Tq}^{(p,n)}$,
 are proportional to the matrix element, $\langle\bar q q \rangle$, of quarks in a nucleon
 and are given by
\begin{equation}
f^{p(n)}=\sum_{q=u,d,s}f_{T_{q}}^{p(n)}\mathcal{G}_{S
q}\frac{m_{p(n)}}{m_{q}}+\frac{2}{27}f_{T_{g}}^{p(n)}\sum_{q=c,b,t}\mathcal{G}_{S
q}\frac{m_{p(n)}}{m_{q}},\label{fpn}
\end{equation}
with suffix $p$ and $n$ denote proton and neutron respectively and $\mathcal{G}_{S
q}$ is the effective coupling between dark matter and nucleon~\cite{twin51},
\begin{eqnarray}
f^{p}_{Tu} &=& 0.020 \pm 0.004, \quad  f^{p}_{Td} = 0.026 \pm 0.005, \quad f^{p}_{Ts} = 0.118 \pm 0.062, \nonumber  \\
f^{n}_{Tu} &=& 0.014 \pm 0.003, \quad f^{n}_{Td} = 0.036 \pm 0.008,
\quad f^{n}_{Ts} = 0.118 \pm 0.062 \, .
\end{eqnarray}
where we have used the relation between $f_{Tg}^{(p,n)}$ and $f_{Tq}^{(p,n)}$
stated as,
\begin{equation}
f_{Tg}^{(p,n)} = 1 - \sum_{q=u,d,s} f_{Tq}^{(p,n)}.
\end{equation}
Thus $f^{p}_{TG} \approx 0.84$ and $f^{n}_{TG} \approx 0.83$~\cite{Ellis:2000ds}.
In fact, here $\sigma_{p}^{SI}\approx \sigma_{n}^{SI}$.

\section{Photon Flux Due to DM Annihilation}\label{App-B}
The differential flux of $\gamma-$ray due to dark matter annihilation in galactic halo
 in angular direction that produce a solid angle $d\Omega$
 is given by \cite{Cirelli}
\begin{eqnarray}
\frac{d\Phi_{\gamma}}{d\Omega dE_{\gamma}} = \frac{1}{8\pi\alpha}
\sum_f\frac{\langle \sigma v\rangle_f}{M^2_{S, S'}}
\frac{dN^{f}_{\gamma}}{dE_{\gamma}} r_{\odot} \rho^2_{\odot} J \,\, ,
\label{gammaflux1}
\end{eqnarray}
 $\alpha = 1$ for self-conjugated WIMP while $\alpha = 2$ when this is not the case.
 Here we consider $\alpha$ to be unity as the
 singlet scalars from
 the two scalar singlet model (the dark matter candidate chosen in the present work)
 are self-conjugated. In \Eqn{gammaflux1} $\frac{dN^{f}_{\gamma}}{dE_{\gamma}}$
 is the energy spectrum of photons produced in a single annihilation channel of dark
 matter with some specific final state, $f$.

The integrated $\gamma$-flux over a solid angle $\Delta\Omega$ can be expressed in terms of
averaged $J$ factor, $\bar{J}$ as
\begin{eqnarray}
\frac{d\Phi_{\gamma}}{dE_{\gamma}} = \frac{1}{8\pi\alpha}
\sum_f\frac{\langle \sigma v\rangle_f}{M^2_{S,S'}}
\frac{dN^{f}_{\gamma}}{dE_{\gamma}} r_{\odot} \rho^2_{\odot} \bar{J} \Delta\Omega \,\, ,
\label{gammaflux2}
\end{eqnarray}
with $l$ and $b$ denote
galactic longitude and latitude respectively.
\begin{eqnarray}
\bar{J} = \begin{cases}
            \frac{4}{\Delta\Omega}\int dl \int db \,\,\cos b \,\, J(l,b)  \,\,& (l, b\,\,\,\text{coordinate})\\
            \frac{2\pi}{\Delta\Omega} \int d\theta \sin\theta\,\, J(\theta)\,\, & (r, \theta\,\,\,\text{coordinate})
           \end{cases}
\label{jbar}
\end{eqnarray}
where the factor, $J$ can be written as,
\begin{eqnarray}
 J = \int_{l.o.s} \frac{ds}{r_\odot}
 \left(\frac{\rho(r)}{\rho_\odot}\right)^2 \,\,
 \label{j_lb_rt}
 \end{eqnarray}
and
\begin{eqnarray}
\Delta \Omega = \begin{cases}
                 4\int dl \int db\,\, \cos b \,\, & (l, b \,\,\,\text{coordinate})\\
                 2\pi \int d\theta\,\, \sin \theta \,\, & (r, \theta \,\,\,\text{coordinate})\,\, .
                 \end{cases}
\label{solidangle}
\end{eqnarray}
In the above $\rho(r)$ denote the DM halo profile.

The relation between radial distance $r$ from GC and line of sight $s$, can be given by,
\begin{eqnarray}
r = \begin{cases}
     \left( s^2 + r^2_{\odot} - 2sr_{\odot} {\rm cos}\,{l} \,{\rm cos}\,{b}\right)^{1/2}\,\, &  (l, b\,\,\,\text{coordinate})\\
     \left( s^2 + r^2_{\odot} - 2sr_{\odot} {\cos}\,{\theta}\right)^{1/2}\,\, & (r, \theta\,\,\,\text{coordinate})
    \end{cases}
\label{los}
\end{eqnarray}
In Eqs. (\ref{jbar}, \ref{solidangle}, \ref{los}) $\theta$ represents the angle
between the line of sight of an observer located at earth
while looking at some point $r$ from the galactic centre
and the line connecting the observer at earth to the Galactic Centre.

\section{DM Halo Profiles}\label{App-C}
The dark matter distribution is usually parametrised
 as a spherically symmetric profile,
 \begin{equation}
  \rho(r) = \rho_0 F_{\rm halo}(r) = \frac{\rho_0}{(r/r_c)^\gamma[1 + (r/r_c)^\gamma]^{(\beta - \gamma)/\alpha}} \,\,\, ,
 \end{equation}
 where 
 $\alpha$, $\beta$, $\gamma$ and $r_c$ are the parameters
 that represent some particular halo profile listed in Table ~\ref{table_7}. $\rho_0$ is the local dark matter
 halo density at solar location ($\rho (r_\odot)$) taken to be 0.4 GeV/cm$^3$ with $r_\odot$
 is the distance between sun to the galactic centre ($\sim$ 8.5 kpc).

 \begin{table}[h]
 \caption{Parameters used for widely used dark matter halo models}
 \centering
 \vskip 2 mm
 \begin{tabular}{| c | c | c | c | c |}
  \hline
 Halo Model & $\alpha$ & $\beta$ & $\gamma$ & $r_c$ (kpc) \\
\hline \hline
 Navarro, Frenk, White (NFW) \cite{nfw} & 1 & 3 & 1 & 20 \\
     \hline
 Moore \cite{moore} & 1.5 & 3 & 1.5 & 28 \\
    \hline
 Isothermal \cite{iso} & 2 & 2 & 0 & 3.5 \\
    \hline
\end{tabular}
  \label{table_7}
\end{table}

 Another halo profile, namely Einasto profile has also been involved for our study.
 A different kind of parametric form is adopted in this halo profile~\cite{einasto}
 which can be written as,
\begin{equation}
 F^{Ein}_{\rm halo}(r)=exp\left[\frac{-2}{\tilde \alpha}
\left(\left(\frac{r}{r_\odot}\right)^{\tilde \alpha}-1\right)\right]\,\,\, ,
\end{equation}
where $\tilde \alpha$ is a parameter of the halo profile. In our work
value of $\tilde \alpha$ is chosen to be 0.17.

\end{appendices}

\vskip 1cm

\bibliographystyle{JHEP}
\end{document}